\documentclass[12pt,a4paper]{article}

\usepackage{cite} 
\DeclareUnicodeCharacter{2212}{-}

\usepackage[utf8]{inputenc}
\usepackage{amsmath}
\usepackage{amsfonts} 
\usepackage{mathtools} 
\usepackage{amssymb} 
\usepackage{subfig} 
\usepackage{float} 
\usepackage{graphicx}
\usepackage{xcolor}
\usepackage{mathrsfs}
\usepackage{hyperref}
\usepackage{multirow}
\usepackage[normalem]{ulem}

\usepackage[page, toc]{appendix}

\graphicspath{{Figs/}}

\begin{document}
{
\renewcommand{\refeq}[1]{eq.\,\eqref{#1}} 
\newcommand{\refEq}[1]{Eq.\,\eqref{#1}} 
\newcommand{\refeqs}[1]{eqs.\,\eqref{#1}} 
\newcommand{\refEqs}[1]{Eqs.\,\eqref{#1}} 
\newcommand{\abs}[1]{|#1|} 
\newcommand{\Abs}[1]{\left|#1\right|} 
\newcommand{\re}[1]{\text{Re}\left(#1\right)}
\newcommand{\im}[1]{\text{Im}\left(#1\right)}
\newcommand{\Hc}{\text{H.c.}}

\newcommand{\COMM}[2]{\left[#1\,,\,#2\right]}
\newcommand{\comm}[2]{[#1,#2]}
\newcommand{\der}{\mathcal D}

\newcommand{\id}{\mathbf{1}}
\newcommand{\TR}[1]{\text{Tr}\left\{#1\right\}}
\newcommand{\tr}[1]{\text{Tr}\left(#1\right)}
\newcommand{\BR}[1]{\text{Br}\left(#1\right)}
\newcommand{\PL}{P_L}\newcommand{\PR}{P_R}
\newcommand{\Zn}[1]{\mathbb{Z}_{#1}}
\newcommand{\ZZ}{\Zn{2}}
\newcommand{\cb}{c_\beta}
\renewcommand{\sb}{s_\beta}
\newcommand{\cbb}{c_{2\beta}}
\newcommand{\sbb}{s_{2\beta}}
\newcommand{\cba}{c_{\alpha\beta}}
\newcommand{\sba}{s_{\alpha\beta}}
\newcommand{\tb}{t_\beta}
\newcommand{\tbb}{t_{2\beta}}
\newcommand{\tbinv}{\tb^{-1}}
\newcommand{\tti}{\tb+\tbinv}
\newcommand{\VEV}[1]{\langle #1 \rangle}
\newcommand{\vev}[1]{v_{#1}}
\newcommand{\ROTmat}{\mathcal R}
\newcommand{\ROTmatT}{\ROTmat^T}
\newcommand{\ROTmatinv}{\ROTmat^{-1}}
\newcommand{\ROT}[1]{\ROTmat_{#1}}
\newcommand{\HbROT}{\mathcal R_{\beta}^{\phantom{T}}}
\newcommand{\HbROTt}{\mathcal R_{\beta}^T}
\newcommand{\HbROTinv}{\mathcal R_{\beta}^{-1}}
\newcommand{\SMHD}{\Phi}
\newcommand{\SMHDd}{\Phi^\dagger}
\newcommand{\HD}[1]{\Phi_{#1}^{\phantom{\dagger}}}
\newcommand{\HDd}[1]{\Phi_{#1}^\dagger}
\newcommand{\HDC}[1]{\tilde\Phi_{#1}^{\phantom{\dagger}}}
\newcommand{\HDc}[1]{\Phi_{#1}^{\phantom{\dagger}\!\!\!\ast}}
\newcommand{\HHD}[1]{H_{#1}^{\phantom{\dagger}}}
\newcommand{\HHDd}[1]{H_{#1}^\dagger}
\newcommand{\HHDC}[1]{\tilde H_{#1}^{\phantom{\dagger}}}
\newcommand{\HHDc}[1]{H_{#1}^{\phantom{\dagger}\!\!\!\ast}}
\newcommand{\nHH}{{H}^0}
\newcommand{\nHR}{{R}^0}
\newcommand{\nHI}{{I}^0}
\newcommand{\nh}{\mathrm{h}}
\newcommand{\nhSM}{\mathrm{h_{SM}}}
\newcommand{\nH}{\mathrm{H}}
\newcommand{\nA}{\mathrm{A}}
\newcommand{\nS}{\mathrm{S}}
\newcommand{\cH}{\mathrm{H}^\pm}
\newcommand{\cHm}{\mathrm{H}^-}
\newcommand{\cHp}{\mathrm{H}^+}
\newcommand{\mNSc}{\mathcal M_0^2}
\newcommand{\mNScT}{{\mathcal M_0^{2}}^T}
\newcommand{\mh}{m_{\nh}}
\newcommand{\mhSM}{m_{\nhSM}}
\newcommand{\mH}{m_{\nH}}
\newcommand{\mA}{m_{\nA}}
\newcommand{\mcH}{m_{\cH}}
\newcommand{\mS}{m_{\mathrm{S}}}
\newcommand{\nl}[1]{n_{#1}}
\newcommand{\nrl}[1]{\re{n_{#1}}}
\newcommand{\nrle}{\nrl{e}}\newcommand{\nrlm}{\nrl{\mu}}\newcommand{\nrlt}{\nrl{\tau}}
\newcommand{\noml}[1]{\frac{\nrl{#1}}{m_{#1}}}

\newcommand{\Yd}[1]{\Gamma_{#1}}
\newcommand{\Yu}[1]{\Delta_{#1}}
\newcommand{\Ydc}[1]{\Gamma_{#1}^\ast}
\newcommand{\Yuc}[1]{\Delta_{#1}^\ast}
\newcommand{\Ydd}[1]{\Gamma_{#1}^\dagger}
\newcommand{\Yud}[1]{\Delta_{#1}^\dagger}
 \newcommand{\FQ}{Q}\newcommand{\FL}{L}
\newcommand{\Fu}{u}\newcommand{\Fd}{d}
\newcommand{\Fl}{\ell}\newcommand{\Fn}{\nu}
\newcommand{\ferX}[3]{{#1}_{#2#3}}\newcommand{\ferXb}[3]{\bar #1_{#2#3}}
\newcommand{\dL}[1]{\ferX{\Fd}{L}{#1}}\newcommand{\dLb}[1]{\ferXb{\Fd}{L}{#1}}
\newcommand{\dR}[1]{\ferX{\Fd}{R}{#1}}\newcommand{\dRb}[1]{\ferXb{\Fd}{R}{#1}}
\newcommand{\uL}[1]{\ferX{\Fu}{L}{#1}}\newcommand{\uLb}[1]{\ferXb{\Fu}{L}{#1}}
\newcommand{\uR}[1]{\ferX{\Fu}{R}{#1}}\newcommand{\uRb}[1]{\ferXb{\Fu}{R}{#1}}
\newcommand{\lL}[1]{\ferX{\Fl}{L}{#1}}\newcommand{\lLb}[1]{\ferXb{\Fl}{L}{#1}}
\newcommand{\lR}[1]{\ferX{\Fl}{R}{#1}}\newcommand{\lRb}[1]{\ferXb{\Fl}{R}{#1}}
\newcommand{\nL}[1]{\ferX{\Fn}{L}{#1}}\newcommand{\nLb}[1]{\ferXb{\Fn}{L}{#1}}
\newcommand{\nR}[1]{\ferX{\Fn}{R}{#1}}\newcommand{\nRb}[1]{\ferXb{\Fn}{R}{#1}}

\newcommand{\SD}[1]{\Phi_{#1}^{\phantom{\dagger}}}
\newcommand{\SDd}[1]{\Phi_{#1}^\dagger}
\newcommand{\SDc}[1]{\Phi_{#1}^\ast}
\newcommand{\SDti}[1]{\tilde\Phi_{#1}^{\phantom{\dagger}}}
\newcommand{\bilSD}[2]{\SDd{#1}\SD{#2}}
\newcommand{\pbilSD}[2]{\big(\bilSD{#1}{#2}\big)}

\newcommand{\Hv}{H_1}\newcommand{\Hvd}{H_1^\dagger}\newcommand{\Hvti}{\tilde H_1}
\newcommand{\Ho}{H_2}\newcommand{\Hod}{H_2^\dagger}\newcommand{\Hoti}{\tilde H_2}

\newcommand{\weakferX}[3]{{#1}_{#2#3}^0}\newcommand{\weakferXb}[3]{\bar #1_{#2#3}^0}

\newcommand{\wQL}[1]{\weakferX{\FQ}{L}{#1}}\newcommand{\wQLb}[1]{\weakferXb{\FQ}{L}{#1}}
\newcommand{\wdL}[1]{\weakferX{\Fd}{L}{#1}}\newcommand{\wdLb}[1]{\weakferXb{\Fd}{L}{#1}}
\newcommand{\wdR}[1]{\weakferX{\Fd}{R}{#1}}\newcommand{\wdRb}[1]{\weakferXb{\Fd}{R}{#1}}
\newcommand{\wuL}[1]{\weakferX{\Fu}{L}{#1}}\newcommand{\wuLb}[1]{\weakferXb{\Fu}{L}{#1}}
\newcommand{\wuR}[1]{\weakferX{\Fu}{R}{#1}}\newcommand{\wuRb}[1]{\weakferXb{\Fu}{R}{#1}}
\newcommand{\wLL}[1]{\weakferX{\FL}{L}{#1}}\newcommand{\wLLb}[1]{\weakferXb{\FL}{L}{#1}}
\newcommand{\wlL}[1]{\weakferX{\Fl}{L}{#1}}\newcommand{\wlLb}[1]{\weakferXb{\Fl}{L}{#1}}
\newcommand{\wlR}[1]{\weakferX{\Fl}{R}{#1}}\newcommand{\wlRb}[1]{\weakferXb{\Fl}{R}{#1}}
\newcommand{\wnL}[1]{\weakferX{\Fn}{L}{#1}}\newcommand{\wnLb}[1]{\weakferXb{\Fn}{L}{#1}}
\newcommand{\QL}[1]{\ferX{\FQ}{L}{#1}}\newcommand{\QLb}[1]{\ferXb{\FQ}{L}{#1}}
\newcommand{\LL}[1]{\ferX{\FL}{L}{#1}}\newcommand{\LLb}[1]{\ferXb{\FL}{L}{#1}}

\newcommand{\CKM}{V}\newcommand{\CKMdag}{V^\dagger}
\newcommand{\V}[1]{\CKM^{\phantom{\ast}}_{#1}}
\newcommand{\Vc}[1]{\CKM^{\ast}_{#1}}
\newcommand{\Vd}[1]{\CKM^{\dagger}_{#1}}\newcommand{\Vt}[1]{\CKM^{t}_{#1}}
\newcommand{\PMNS}{U}\newcommand{\PMNSdag}{U^\dagger}
\newcommand{\U}[1]{\PMNS^{\phantom{\ast}}_{#1}}
\newcommand{\Uc}[1]{\PMNS^{\ast}_{#1}}
\newcommand{\Ud}[1]{\PMNS^{\dagger}_{#1}}\newcommand{\Ut}[1]{\PMNS^{t}_{#1}}
\newcommand{\matXF}[2]{{\rm #1}_{#2}}\newcommand{\matXFd}[2]{{\rm #1}_{#2}^\dagger}
\newcommand{\wmatXF}[2]{{\rm #1}_{#2}^{0}}\newcommand{\wmatXFd}[2]{{\rm #1}_{#2}^{0\dagger}}
\newcommand{\basematM}{M}\newcommand{\basematN}{N}
\newcommand{\matNf}[1]{\matXF{\basematN}{#1}}\newcommand{\matNfd}[1]{\matXFd{\basematN}{#1}}
\newcommand{\matMf}[1]{\matXF{\basematM}{#1}}\newcommand{\matMfd}[1]{\matXFd{\basematM}{#1}}
\newcommand{\matND}{\matXF{\basematN}{d}}\newcommand{\matNDd}{\matXFd{\basematN}{d}}
\newcommand{\matNU}{\matXF{\basematN}{u}}\newcommand{\matNUd}{\matXFd{\basematN}{u}}
\newcommand{\matNL}{\matXF{\basematN}{\ell}}\newcommand{\matNLd}{\matXFd{\basematN}{\ell}}
\newcommand{\matNN}{\matXF{\basematN}{\nu}}\newcommand{\matNNd}{\matXFd{\basematN}{\nu}}
\newcommand{\matMD}{\matXF{\basematM}{d}}\newcommand{\matMDd}{\matXFd{\basematM}{d}}
\newcommand{\matMU}{\matXF{\basematM}{u}}\newcommand{\matMUd}{\matXFd{\basematM}{u}}
\newcommand{\matML}{\matXF{\basematM}{\ell}}\newcommand{\matMLd}{\matXFd{\basematM}{\ell}}
\newcommand{\matMN}{\matXF{\basematM}{\nu}}\newcommand{\matMNd}{\matXFd{\basematM}{\nu}}
\newcommand{\wmatNf}{\wmatXF{\basematN}{f}}\newcommand{\wmatNDf}{\wmatXF{\basematN}{f}}
\newcommand{\wmatMf}{\wmatXF{\basematM}{f}}\newcommand{\wmatMDf}{\wmatXF{\basematM}{f}}
\newcommand{\wmatND}{\wmatXF{\basematN}{d}}\newcommand{\wmatNDd}{\wmatXFd{\basematN}{d}}
\newcommand{\wmatNU}{\wmatXF{\basematN}{u}}\newcommand{\wmatNUd}{\wmatXFd{\basematN}{u}}
\newcommand{\wmatNL}{\wmatXF{\basematN}{\ell}}\newcommand{\wmatNLd}{\wmatXFd{\basematN}{\ell}}
\newcommand{\wmatNN}{\wmatXF{\basematN}{\nu}}\newcommand{\wmatNNd}{\wmatXFd{\basematN}{\nu}}
\newcommand{\wmatMD}{\wmatXF{\basematM}{d}}\newcommand{\wmatMDd}{\wmatXFd{\basematM}{d}}
\newcommand{\wmatMU}{\wmatXF{\basematM}{u}}\newcommand{\wmatMUd}{\wmatXFd{\basematM}{u}}
\newcommand{\wmatML}{\wmatXF{\basematM}{\ell}}\newcommand{\wmatMLd}{\wmatXFd{\basematM}{\ell}}
\newcommand{\wmatMN}{\wmatXF{\basematM}{\nu}}\newcommand{\wmatMNd}{\wmatXFd{\basematM}{\nu}}

\newcommand{\UfX}[2]{\mathcal U_{#1_{#2}}}\newcommand{\UfXd}[2]{\mathcal U_{#1_{#2}}^\dagger}
\newcommand{\UuL}{\UfX{u}{L}}\newcommand{\UuLd}{\UfXd{u}{L}}\newcommand{\UuR}{\UfX{u}{R}}\newcommand{\UuRd}{\UfXd{u}{R}}
\newcommand{\UdL}{\UfX{d}{L}}\newcommand{\UdLd}{\UfXd{d}{L}}\newcommand{\UdR}{\UfX{d}{R}}\newcommand{\UdRd}{\UfXd{d}{R}}
\newcommand{\UlL}{\UfX{\ell}{L}}\newcommand{\UlLd}{\UfXd{\ell}{L}}\newcommand{\UlR}{\UfX{\ell}{R}}\newcommand{\UlRd}{\UfXd{\ell}{R}}
\newcommand{\UnL}{\UfX{\nu}{L}}\newcommand{\UnLd}{\UfXd{\nu}{L}}\newcommand{\UnR}{\UfX{\nu}{R}}\newcommand{\UnRd}{\UfXd{\nu}{R}}

\newcommand{\matYukF}[2]{Y_{#1#2}}\newcommand{\matYukFd}[2]{Y_{#1#2}^\dagger}\newcommand{\matYukFt}[2]{Y_{#1#2}^t}\newcommand{\matYukFc}[2]{Y_{#1#2}^\ast}
\newcommand{\YukF}[3]{(Y_{#1#2})_{#3}}\newcommand{\YukFd}[3]{(Y_{#1#2}^\dagger)_{#3}}\newcommand{\YukFt}[3]{(Y_{#1#2}^t)_{#3}}\newcommand{\YukFc}[3]{(Y_{#1#2}^\ast)_{#3}}
\newcommand{\matYukU}[1]{\matYukF{u}{#1}}\newcommand{\matYukUd}[1]{\matYukFd{u}{#1}}\newcommand{\matYukUt}[1]{\matYukFt{u}{#1}}\newcommand{\matYukUc}[1]{\matYukFc{u}{#1}}
\newcommand{\YukU}[2]{\YukF{u}{#1}{#2}}\newcommand{\YukUd}[2]{\YukFd{u}{#1}{#2}}\newcommand{\YukUt}[2]{\YukFt{u}{#1}{#2}}\newcommand{\YukUc}[2]{\YukFc{u}{#1}{#2}}
\newcommand{\matYukD}[1]{\matYukF{d}{#1}}\newcommand{\matYukDd}[1]{\matYukFd{d}{#1}}\newcommand{\matYukDt}[1]{\matYukFt{d}{#1}}\newcommand{\matYukDc}[1]{\matYukFc{d}{#1}}
\newcommand{\YukD}[2]{\YukF{d}{#1}{#2}}\newcommand{\YukDd}[2]{\YukFd{d}{#1}{#2}}\newcommand{\YukDt}[2]{\YukFt{d}{#1}{#2}}\newcommand{\YukDc}[2]{\YukFc{d}{#1}{#2}}
\newcommand{\matYukL}[1]{\matYukF{\ell}{#1}}\newcommand{\matYukLd}[1]{\matYukFd{\ell}{#1}}\newcommand{\matYukLt}[1]{\matYukFt{\ell}{#1}}\newcommand{\matYukLc}[1]{\matYukFc{\ell}{#1}}
\newcommand{\YukL}[2]{\YukF{\ell}{#1}{#2}}\newcommand{\YukLd}[2]{\YukFd{\ell}{#1}{#2}}\newcommand{\YukLt}[2]{\YukFt{\ell}{#1}{#2}}\newcommand{\YukLc}[2]{\YukFc{\ell}{#1}{#2}}
\newcommand{\matYukN}[1]{\matYukF{\nu}{#1}}\newcommand{\matYukNd}[1]{\matYukFd{\nu}{#1}}\newcommand{\matYukNt}[1]{\matYukFt{\nu}{#1}}\newcommand{\matYukNc}[1]{\matYukFc{\nu}{#1}}
\newcommand{\YukN}[2]{\YukF{\nu}{#1}{#2}}\newcommand{\YukNd}[2]{\YukFd{\nu}{#1}{#2}}\newcommand{\YukNt}[2]{\YukFt{\nu}{#1}{#2}}\newcommand{\YukNc}[2]{\YukFc{\nu}{#1}{#2}}
\newcommand{\YuH}[1]{Y_{u}^{#1}}
\newcommand{\YdH}[1]{Y_{d}^{#1}}
\newcommand{\YlH}[1]{Y_{\ell}^{#1}}

\newcommand{\glFC}[1]{#1-g$\ell$FC}
\newcommand{\solA}{[A]}
\newcommand{\solB}{[B]}
\newcommand{\solBpm}{[B$_\pm$]}
\newcommand{\solBp}{[B$_+$]}
\newcommand{\solBm}{[B$_-$]}

 \hfill\begin{minipage}[r]{0.3\textwidth}\begin{flushright}  IFIC/22-17 \end{flushright} \end{minipage}

\begin{center}

\vspace{0.50cm}
{\large\bf {Muon and electron $g-2$ anomalies in a flavor conserving 2HDM with an oblique view on the CDF $M_W$ value}}\\

\vspace{0.50cm}

\hbox{Francisco J. Botella  $^{a,}$\footnote{\texttt{Francisco.J.Botella@uv.es}}, 
Fernando Cornet-Gomez  $^{a,b,}$\footnote{\texttt{Fernando.CornetGomez@case.edu}}, 
Carlos Miró  $^{a,}$\footnote{\texttt{Carlos.Miro@uv.es}}, 
Miguel Nebot $^{a,}$\footnote{\texttt{Miguel.Nebot@uv.es}}}
\end{center}
\vspace{0.50cm}
\begin{flushleft}
\emph{$^a$ Departament de F\' \i sica Te\`orica and IFIC, Universitat de Val\`encia-CSIC,\\ \quad E-46100, Burjassot, Spain.} \\
\emph{$^b$ Physics Department and Center for Education and Research in Cosmology and Astrophysics (CERCA), Case Western Reserve University, Cleveland, OH 44106, USA.}
\end{flushleft}
\vspace{0.5cm}
\date{\today}
\begin{abstract}
We consider a type I or type X two Higgs doublets model with a modified lepton sector. The generalized lepton sector is also flavor conserving but with the new Yukawa couplings completely decoupled from lepton mass proportionality. The model is one loop stable under renormalization group evolution and it allows to reproduce the $g-2$ muon anomaly together with the different scenarios one can consider for the electron $g-2$ anomaly, related to the Cesium and/or to the Rubidium recoil measurements of the fine structure constant. Thorough parameter space analyses are performed to constrain all the model parameters in the different scenarios, either including or not including the recent CDF measurement of the W boson mass. For light new scalars with masses in the $0.2$-$1.0$ TeV range, the muon anomaly receives dominant one loop contributions; it is for heavy new scalars with masses above $1.2$ TeV that two loop Barr-Zee diagrams are needed. The electron $g-2$ anomaly, if any, must always be obtained with the two loop contributions. The final allowed regions are quite sensitive to the assumptions about perturbativity of Yukawa couplings, which influence unexpected observables like the allowed scalar mass ranges. On that respect, intermediate scalar masses, highly constrained by direct LHC searches, are allowed provided that the new lepton Yukawa couplings are fully scrutinized, including values up to 250 GeV. In the framework of a complete model, fully numerically analysed, we show the implications of the recent $M_{W}$ measurement.
\end{abstract}
\clearpage
\section{Introduction}\label{Sec:Intro}

In the search of Physics beyond the Standard Model (SM), disagreement between measurements and theoretical expectations, that is ``anomalies'', can play the role of beacons to guide our explorations. One longstanding anomaly concerns the anomalous magnetic moment of the muon $a_\mu=\frac{g_\mu-2}{2}$. The \emph{Muon g-2} experiment at Brookhaven \cite{Muong-2:2006rrc} and its successor at Fermilab \cite{Muong-2:2021ojo,Muong-2:2021vma} have produced the following result
\begin{equation}\label{eq:damu:Exp}
 \delta a_\mu^{\rm Exp}=a_\mu^{\rm Exp}-a_\mu^{\rm SM}=(2.5\pm 0.6)\times 10^{-9}\,.
\end{equation}
where $a_\mu^{\rm Exp}$ is the experimental observation and $a_\mu^{\rm SM}$ the SM theoretical expectation\cite{Aoyama:2020ynm,Aoyama:2012wk,Aoyama:2019ryr,Czarnecki:2002nt,Gnendiger:2013pva,Davier:2017zfy,Keshavarzi:2018mgv,Colangelo:2018mtw,Hoferichter:2019mqg,Davier:2019can,Keshavarzi:2019abf,Kurz:2014wya,Melnikov:2003xd,Masjuan:2017tvw,Colangelo:2017fiz,Hoferichter:2018kwz,Gerardin:2019vio,Bijnens:2019ghy,Colangelo:2019uex,Blum:2019ugy,Colangelo:2014qya}. 
Although there are unsettled discrepancies concerning Hadronic Vacuum Polarization (HVP) contributions to $a_\mu^{\rm SM}$ \cite{Borsanyi:2020mff,Ce:2022kxy,Alexandrou:2022amy}, we interpret $\delta a_\mu^{\rm Exp}$ in \refeq{eq:damu:Exp} as a signal of New Physics (NP).\footnote{Solving the anomaly in \refeq{eq:damu:Exp} by enhancing the HVP contribution could generate other tensions in electroweak precision fits \cite{Crivellin:2020zul,Keshavarzi:2020bfy,Colangelo:2020lcg,Passera:2008jk}.}
\\
Besides the muon, recent results concerning the anomalous magnetic moment of the electron might also be interpreted as NP hints \cite{Davoudiasl:2018fbb}. On the one hand, perturbative calculations of $a_e=\frac{g_e-2}{2}$, which have reached impressive levels \cite{Aoyama:2012wj,Aoyama:2012wk,Laporta:2017okg,Aoyama:2017uqe,Volkov:2019phy}, yield $a_e^{\rm SM}$ as a series in powers of the fine structure constant $\alpha$. On the other hand, we have precise measurements of $a_e^{\rm Exp}$ such as \cite{Hanneke:2008tm}. In the past, such measurements were indeed used to infer values of $\alpha$. On the contrary, measurements of atomic recoils \cite{Tiesinga:2021myr} provide now more precise determinations of $\alpha$, which give values of $a_e^{\rm SM}$ such that 
\begin{equation}\label{eq:dae:ExpCs}
 \delta a_e^{\rm Exp,Cs}=-(8.7\pm 3.6)\times 10^{-13}\, ,
\end{equation}
from measurements with $^{133}{\rm Cs}$ \cite{Parker:2018vye}, and
\begin{equation}\label{eq:dae:ExpRb}
 \delta a_e^{\rm Exp,Rb}=(4.8\pm 3.0)\times 10^{-13}\, ,
\end{equation}
from measurements with $^{87}{\rm Rb}$ \cite{Morel:2020dww}.

In reference \cite{Botella:2020xzf} the possibility to explain the values of $\delta a_\mu^{\rm Exp}$ from the \emph{Muon g-2} Brookhaven experiment \cite{Muong-2:2006rrc} together with $\delta a_e^{\rm Exp,Cs}$ in \refeq{eq:dae:ExpCs} was successfully addressed within a subclass of Two Higgs Doublets Models (2HDMs) with general flavor conservation \cite{Penuelas:2017ikk,Botella:2018gzy}. This was achieved, of course, without conflicting with a large set of high and low energy constraints. The specific model considered, the so-called \glFC{I} 2HDM is a 2HDM without tree level scalar flavor changing neutral couplings (SFCNC): in the quark sector it is a type I 2HDM while in the lepton sector it is a general flavor conserving model. The existence of these two anomalies has been addressed in a variety of scenarios, including models with extra Higgs doublets \cite{Broggio:2014mna,Han:2018znu,Haba:2020gkr,Jana:2020pxx,Dutta:2020scq,Sabatta:2019nfg,Chun:2020uzw,Li:2020dbg,DelleRose:2020oaa,Hernandez:2021tii,Keung:2021rps,Han:2021gfu,Hernandez:2021iss,
Jueid:2021avn,CarcamoHernandez:2021iat,De:2021crr,Bharadwaj:2021tgp,Hue:2021xzl,Barman:2021xeq}, models with other scalar extensions \cite{Liu:2018xkx,Hiller:2019mou,Endo:2020mev,Hati:2020fzp,Chua:2020dya,Arbelaez:2020rbq,Escribano:2020wua,Jana:2020joi,Chen:2021rnl,Biswas:2021dan,Chowdhury:2022jde} and supersymmetric models \cite{Endo:2019bcj,Badziak:2019gaf,Cao:2021lmj,Li:2021koa,Li:2021xmw}. There are also plenty of studies with other approaches such as leptoquarks, vector-like fermions or extra gauge bosons, among others \cite{Bauer:2019gfk,CarcamoHernandez:2020pxw,Bigaran:2020jil,Calibbi:2020emz,Chen:2020jvl,Dorsner:2020aaz,Bodas:2021fsy,Fajfer:2021cxa,Lee:2021jdr,Bhattacharya:2021shk, Cadeddu:2021dqx,Borah:2021khc,Bigaran:2021kmn,Hue:2021zyw, Li:2021wzv,Julio:2022ton,Julio:2022bue}.

The present work extends and improves several aspects of \cite{Botella:2020xzf}.
\begin{itemize}
 \item An improved numerical exploration of the parameter space shows that some unexpected regions of interest can be appropriately covered.
 \item Some theoretical assumptions like the perturbativity limits on Yukawa couplings had a significant impact on the analysis and were not fully considered.
 \item The latest \emph{Muon g-2} Fermilab result \cite{Muong-2:2021ojo,Muong-2:2021vma} consolidates the need of NP brought by the previous Brookhaven result.
 \item For $a_e^{\rm Exp}$ the situation is rather unclear: within the present scenario, accommodating the values in \refeq{eq:dae:ExpCs} or in \refeq{eq:dae:ExpRb} may have non-trivial consequences in the model, since they differ in size and in sign.
 \item The recent measurement of the W boson mass by the CDF collaboration \cite{CDF:2022hxs}, which disagrees with SM expectations \cite{ParticleDataGroup:2020ssz}, can also be addressed in this context.
\end{itemize}
All in all, we are entering an era of exclusion or discovery at the LHC and improved analyses of such potential NP hints are necessary.

The manuscript is organized as follows. In section \ref{Sec:Model}, the model is presented. Section \ref{Sec:General} is devoted to a discussion of general constraints which apply regardless of $\delta a_\ell$. The new contributions to $\delta a_\ell$ are analysed in section \ref{Sec:dal}. The main aspects of the numerical analysis are introduced in section \ref{Sec:Analysis}. Next, section \ref{Sec:Results} contains the results of the different analyses together with the corresponding discussions. Finally, the conclusions are presented in section \ref{Sec:Conclusions}. We relegate to the appendices some aspects concerning different sections.
\section{Model}\label{Sec:Model}

The 2HDM is based on the SM gauge group with identical fermion matter content\footnote{As in the SM, we do not include right-handed neutrinos.} and an additional complex scalar doublet. Hence, we have $\SD{j}$ ($j=1,2$) and their corresponding $\mathcal{C}$-conjugate fields defined as $\SDti{j}\equiv i\sigma_2\SDc{j}\,$, with opposite sign hypercharge.

The most general scalar potential of 2HDMs can be written as
\begin{equation}\label{eq:ScalarPotential}
\begin{split}
\mathcal{V}(\SD{1}, \SD{2}) =\ &\mu_{11}^2 \SDd{1} \SD{1} + \mu_{22}^2 \SDd{2} \SD{2} + (\mu_{12}^2 \SDd{1} \SD{2} + \Hc) \\
&+ \lambda_1 (\SDd{1} \SD{1})^2 + \lambda_2 (\SDd{2} \SD{2})^2 + 2 \lambda_3 (\SDd{1} \SD{1})(\SDd{2} \SD{2}) + 2 \lambda_4 (\SDd{1} \SD{2})(\SDd{2} \SD{1}) \\
&+ [\lambda_{5} (\SDd{1} \SD{2})^2 + \Hc] + [\lambda_{6} (\SDd{1} \SD{1}) (\SDd{1} \SD{2}) + \lambda_{7} (\SDd{2} \SD{2}) (\SDd{1} \SD{2}) + \Hc]\, ,
\end{split}
\end{equation}
with real $\mu_{11}^2$, $\mu_{22}^2$ and $\lambda_i$ ($i = 1\ \mathrm{to}\ 4$), whereas $\mu_{12}^2$ and $\lambda_j$ ($j = 5\ \mathrm{to}\ 7$) are complex in general. We assume that $\mathcal{V}(\langle \SD{1} \rangle,\langle \SD{2} \rangle)$ has an appropriate minimum at
\begin{equation}
\left\langle 0\middle|\SD{j}\middle|0 \right\rangle=\frac{1}{\sqrt{2}}\begin{pmatrix}0\\v_j  e^{i\theta_j}\end{pmatrix}\, ,
\end{equation}
being $\theta_j$ and $v_j$ ($v_j \geq 0$) real numbers. Taking this into account, the Higgs doublets can be parametrized around the vacuum as
\begin{equation}\label{eq:SD:expansion}
\SD{j} = e^{i\theta_j} \begin{pmatrix} \varphi_j^{+} \\ (v_j + \rho_j + i\eta_j)/\sqrt{2}\end{pmatrix}.
\end{equation}

Introducing\footnote{From now on, $\tbinv = \cot{\beta}$.} $\cb \equiv \cos{\beta} \equiv v_1/v$, $\sb \equiv \sin{\beta} \equiv v_2/v$, $\tb \equiv \tan{\beta} = v_2/v_1$, with $\beta \in [0;\pi/2]$ and $v^2 = v_1^2 + v_2^2 = (\sqrt{2} G_F)^{-1} \simeq (246\ \mathrm{GeV})^2$, one can perform a global $SU(2)$ rotation in the scalar space and express the scalar doublets in the so-called Higgs basis \cite{GEORGI197995,PhysRevD.19.945,PhysRevD.51.3870}
\begin{equation}\label{eq:HiggsBasis:01}
\begin{pmatrix}\Hv\\ \Ho\end{pmatrix}=\HbROT\,
\begin{pmatrix}e^{-i\theta_1}\SD{1}\\ e^{-i\theta_2}\SD{2}\end{pmatrix},\quad \text{with}\quad 
\HbROT=\begin{pmatrix}\phantom{-}\cb & \sb\\ -\sb & \cb \end{pmatrix}\quad \text{and} \quad \HbROTt = \HbROTinv,
\end{equation}
where only one linear combination of the scalar doublets, namely $\Hv$, has a non-zero vacuum expectation value (vev):
\begin{equation}
\langle \Hv \rangle = \frac{v}{\sqrt{2}}\begin{pmatrix} 0 \\ 1 \end{pmatrix}, \quad \langle \Ho \rangle= \begin{pmatrix} 0 \\ 0 \end{pmatrix}.
\end{equation}
The explicit degrees of freedom in this basis are defined by
\begin{equation}
\Hv = \begin{pmatrix} G^{+} \\ \frac{v+\nHH+iG^{0}}{\sqrt{2}} \end{pmatrix}, \quad \Ho = \begin{pmatrix} \cHp \\ \frac{\nHR+i\nHI}{\sqrt{2}} \end{pmatrix},
\end{equation}
where
\begin{equation}
\begin{pmatrix} G^{+}\\ \cHp \end{pmatrix} = \ROT \beta \begin{pmatrix} \varphi_1^{+}\\ \varphi_2^{+} \end{pmatrix}, \quad \begin{pmatrix} \nHH \\ \nHR \end{pmatrix} = \ROT \beta \begin{pmatrix} \rho_1\\ \rho_2 \end{pmatrix}, \quad \begin{pmatrix} G^{0}\\ \nHI \end{pmatrix} = \ROT \beta \begin{pmatrix} \eta_1\\ \eta_2 \end{pmatrix}.
\end{equation}
As we can check, the would-be Goldstone bosons $G^0$ and $G^\pm$ get isolated as components of the first Higgs doublet. Likewise, we already identify two charged physical scalars $\cH$ and three neutral fields $\{\nHH,\nHR,\nHI\}$ that are not, in general, the mass eigenstates. The latter are determined by the scalar potential, which generates their mass matrix $\mNSc$. This can be diagonalized by a $3 \times 3$ real orthogonal transformation, $\ROTmat$, as
\begin{equation}
\ROTmatT \mNSc \ROTmat = \mathrm{diag}(\mh^2,\mH^2,\mA^2), \quad \ROTmatT = \ROTmatinv,
\end{equation}
and thus the physical scalars $\{\nh,\nH,\nA\}$ are given by
\begin{equation}\label{eq:PhysNeutralScalars}
\begin{pmatrix} \nh \\ \nH \\ \nA \end{pmatrix} = \ROTmatT \begin{pmatrix} \nHH \\ \nHR \\ \nHI \end{pmatrix}.
\end{equation}
Neglecting CP violation in the scalar sector, one has
\begin{equation}\label{eq:ScalarRot}
\ROTmat = \begin{pmatrix} \sba & -\cba & 0 \\ \cba & \sba & 0 \\ 0 & 0 & 1 \end{pmatrix},
\end{equation}
where $\sba \equiv \sin(\alpha + \beta)$ and $\cba \equiv \cos(\alpha + \beta)$, with $\pi/2 - \alpha$ being the mixing angle that parametrizes the change of basis from the fields in \refeq{eq:SD:expansion} to the mass eigenstates in \refeq{eq:PhysNeutralScalars}. We should point out that different conventions for \refeq{eq:ScalarRot} can be found in the literature.

Regarding the Yukawa sector, it is extended to
\begin{equation}\label{eq:2HDMLagrYuk}
\begin{multlined}[c][.9\displaywidth]
\mathscr{L_{\rm Y}}=
 -\wQLb{}\left(\SD{1}\matYukD{1}+\SD{2}\matYukD{2}\right)\wdR{}
 -\wQLb{}\left(\SDti{1}\matYukU{1}+\SDti{2}\matYukU{2}\right)\wuR{} \\
 -\wLLb{}\left(\SD{1}\matYukL{1}+\SD{2}\matYukL{2}\right)\wlR{}
 +\Hc\, ,
\end{multlined}
\end{equation}
where the couplings $\matYukD{j}$, $\matYukU{j}$ and $\matYukL{j}$ $(j = 1,2)$ are $3 \times 3$ complex matrices in flavor space. One should notice that there are only two flavor structures in the leptonic sector because we are not considering right-handed neutrinos. In the Higgs basis, the Yukawa Lagrangian takes the form
\begin{equation}\label{eq:2HDM:YukLag:HiggsBasis}
\begin{multlined}[c][.9\displaywidth]
\mathscr{L_{\rm Y}}=
 -\frac{\sqrt{2}}{v}\wQLb{}\left(\Hv\wmatMD+\Ho\wmatND\right)\wdR{}
 -\frac{\sqrt{2}}{v}\wQLb{}\left(\Hvti\wmatMU+\Hoti\wmatNU\right)\wuR{}\\
 -\frac{\sqrt{2}}{v}\wLLb{}\left(\Hv\wmatML+\Ho\wmatNL\right)\wlR{}+\Hc \ .
\end{multlined}
\end{equation}

It is then clear that the matrices $\wmatMf$ ($f = d, u, \ell$) are the non-diagonal fermion mass matrices since they are coupled to the only Higgs doublet that acquires a non-vanishing vev, i.e., $\Hv$. 

The model we are considering in the quark sector is defined by
\begin{equation}
\matYukD{2} = d \matYukD{1}\, , \quad \matYukU{2} = d^{*} \matYukU{1}\, ,
\end{equation}
which is equivalent to
\begin{equation}
\wmatND=\tbinv\wmatMD\, , \quad \wmatNU=\tbinv\wmatMU\, .
\end{equation}
In the leptonic sector, there exist two unitary matrices $W_L$ and $W_R$ such that both $W_L^\dagger \matYukL{i} W_R$ ($i=1,2$) get simultaneously diagonalized. It is well-known that the structure of the quark sector can be enforced through a $\ZZ$ symmetry, but this is not the case in the lepton sector. Nevertheless, as it is shown in appendix \ref{appendix:RGE:stability}, the entire Yukawa structure is stable under one loop renormalization group evolution (RGE) and, therefore, the model is free from unwanted SFCNC. 

Going to the fermion mass bases for our \glFC{I} model --type I in the quark sector and general flavor conserving in the lepton sector--, we get the relevant new Yukawa structures:
\begin{equation}\label{eq:2HDM:YukLag:HiggsBasis:Diagonal} 
\begin{multlined}[c][.9\displaywidth]
\mathscr{L_{\rm Y}}=
 -\frac{\sqrt{2}}{v}\QLb{}\left(\Hv\matMD+\Ho\matND\right)\dR{}
 -\frac{\sqrt{2}}{v}\QLb{}\left(\Hvti\matMU+\Hoti\matNU\right)\uR{}\\
 -\frac{\sqrt{2}}{v}\LLb{}\left(\Hv\matML+\Ho\matNL\right)\lR{}+\Hc \, ,
\end{multlined}
\end{equation}
with
\begin{equation}
\matND=\tbinv\matMD\, , \qquad \matNU=\tbinv\matMU\, , \qquad \matNL=\mathrm{diag}(\nl{e},\nl{\mu},\nl{\tau})\, ,
\end{equation}
and $\matMf{f}$ ($f = u,d,\ell$) the corresponding diagonal fermion mass matrices. Note that the quark couplings $\matNU$ and $\matND$ are those from 2HDMs of type I or X. On the other hand, the matrices $\matNL$ correspond to a general flavor conserving lepton sector. Therefore, they are diagonal, arbitrary and one loop stable under RGE, as it was shown in \cite{Botella:2018gzy}, meaning that they remain diagonal.

We must stress that it is the fact that $\nl{e}$ and $\nl{\mu}$ are completely independent what implements the desired decoupling between electron and muon NP couplings in order to have enough freedom to address the corresponding $(g-2)_\ell$ anomalies. We assume that these couplings are real, i.e., $\mathrm{Im}(\nl{\ell}) = 0$. This prevents us from dangerous contributions to electric dipole moments (EDMs), that are tightly constrained: $|d_e| < 1.1 \times 10^{−29}\ \text{e}\cdot\text{cm}$ \cite{ACME:2018yjb}.

Furthermore, we consider an scalar potential shaped by a $\ZZ$ symmetry that is softly broken by the term $\mu_{12}^2 \neq 0$. Hence, we have to take $\lambda_6 = \lambda_7 = 0$ in \refeq{eq:ScalarPotential}. We also assume that there is no CP violation in the scalar sector, so \refeq{eq:ScalarRot} is fulfilled.

Under these assumptions, the flavor conserving Yukawa interactions of neutral scalars read
\begin{equation}\label{eq:neutral:couplings}
\begin{split}
\mathscr{L_{\rm N}} =& -\frac{m_{u_j}}{v} \left(\sba + \cba \tbinv \right) \nh\, \bar{u}_j u_j -\frac{m_{d_j}}{v} \left(\sba + \cba \tbinv \right) \nh\, \bar{d}_j d_j \\
&-\frac{m_{\ell_j}}{v} \left(\sba + \cba \frac{\nrl{\ell_j}}{m_{\ell_j}} \right)  \nh\, \bar{\ell}_j \ell_j \\
& -\frac{m_{u_j}}{v} \left(-\cba + \sba \tbinv \right) \nH\, \bar{u}_j u_j -\frac{m_{d_j}}{v} \left(-\cba + \sba \tbinv \right) \nH\, \bar{d}_j d_j \\
&-\frac{m_{\ell_j}}{v} \left(-\cba + \sba \frac{\nrl{\ell_j}}{m_{\ell_j}} \right)  \nH\, \bar{\ell}_j \ell_j \\
&+i\frac{m_{u_j}}{v} \tbinv \nA\, \bar{u}_j \gamma_5 u_j -i\frac{m_{d_j}}{v} \tbinv \nA\, \bar{d}_j \gamma_5 d_j -i\frac{\nrl{\ell_j}}{v} \nA\, \bar{\ell}_j \gamma_5 \ell_j\, , \\
\end{split}
\end{equation}
and those involving charged scalars are
\begin{equation}\label{eq:charged:couplings}
\begin{split}
\mathscr{L_{\rm Ch}} =&\ \frac{\cHm}{\sqrt{2}v}\, \bar{d}_i\, V_{ji}^{*} \tbinv \left[(m_{u_j}-m_{d_i}) + (m_{u_j}+m_{d_i}) \gamma_5\right] u_j \\
&+ \frac{\cHp}{\sqrt{2}v}\, \bar{u}_j\, V_{ji} \tbinv \left[(m_{u_j}-m_{d_i}) - (m_{u_j}+m_{d_i}) \gamma_5\right] d_i \\
&- \frac{\cHm}{\sqrt{2}v}\, \bar{\ell}_j\, \nrl{\ell_j} \left(1-\gamma_5\right) \nu_j - \frac{\cHp}{\sqrt{2}v}\, \bar{\nu}_j\, \nrl{\ell_j} \left(1+\gamma_5\right) \ell_j\, , \\
\end{split}
\end{equation}
with $i,j = 1,2,3$ summing over generations. It is easy to check that $\nh$ presents the same couplings as the SM Higgs boson when we take the scalar alignment limit, i.e., $\sba \rightarrow 1$.

\section{General constraints}\label{Sec:General}

Before addressing the different contributions to the anomalous magnetic moments $\delta a_\ell$, we discuss in this section some general constraints which are relevant in the scenario under consideration. By ``general'' we mean that they do not depend specifically on the values of $\nrle$, $\nrlm$, $\delta a_e$ and $\delta a_\mu$. Furthermore, their effects can be understood in simple terms. 
\begin{itemize}
 \item \underline{Alignment}. The couplings of the scalar $\nh$, assumed to be the SM-Higgs-like particle with $\mh=125$ GeV, deviate from SM values through the scalar mixing in \refeq{eq:ScalarRot}. Measurements of the signal strengths in the usual set of production mechanisms and decay channels impose $\cba\ll 1$. Concerning the scalar sector, we are thus in the \emph{alignment limit}.

 \item \underline{Oblique parameters and $M_W$}. Electroweak precision measurements constrain deviations in the oblique parameters $S$ and $T$ \cite{ParticleDataGroup:2020ssz,Grimus:2008nb}:
\begin{equation}\label{eq:obliqueval}
 \Delta S=0.00\pm 0.07,\quad \Delta T=0.05\pm 0.06,\quad \rho=0.92 \ \text{(correlation)}.
\end{equation}
In 2HDMs, in the alignment limit mentioned above, one can observe that the corrections to $S$ and $T$ are kept under control when either $\mcH\simeq\mA$ or $\mcH\simeq\mH$, as shown in figure \ref{sfig:DSDT:0}. Recently, the CDF collaboration announced a measurement of the W boson mass which disagrees with SM expectations \cite{CDF:2022hxs}. In fits of electroweak precision observables this disagreement can be translated into values of the oblique parameters $(\Delta S,\Delta T)\neq (0,0)$ \cite{deBlas:2022hdk,Lu:2022bgw} (although fits including $\Delta U$ have also been considered, we focus on the case $\Delta U=0$, appropriate here). In order to ``explain'' the CDF $M_W$ ``anomaly'' one can thus consider  $(\Delta S,\Delta T)$ constraints from \cite{deBlas:2022hdk,Lu:2022bgw} instead of \refeq{eq:obliqueval}. We can consider, in particular,
\begin{itemize}
 \item[(i)] the ``conservative scenario'' in \cite{deBlas:2022hdk} which combines the CDF with previous measurements and gives
 \begin{equation}\label{eq:MWST:cons}
  \Delta S=0.086 \pm 0.077\,,\quad \Delta T=0.177\pm 0.070\,,\quad \rho=0.89\,,
 \end{equation}
\item[(ii)] the results in \cite{Lu:2022bgw} which solely use the CDF measurement and give
 \begin{equation}\label{eq:MWST:nocons}
  \Delta S=0.15 \pm 0.08\,,\quad \Delta T=0.27\pm 0.06\,,\quad \rho=0.93\,.
 \end{equation}
\end{itemize}
In the alignment limit, for $\mcH=1$ TeV, \refeqs{eq:MWST:cons} and \eqref{eq:MWST:nocons} give the allowed regions represented in figures \ref{sfig:DSDT:W2} and \ref{sfig:DSDT:W1} respectively. In sharp contrast with figure \ref{sfig:DSDT:0}, notice in figures \ref{sfig:DSDT:W2} and \ref{sfig:DSDT:W1} how near degeneracy of the three new scalars is excluded, and how even near degeneracies $\mcH\simeq\mA$ or $\mcH\simeq\mH$ are quite disfavored. Furthermore, notice that the $1\sigma$ region (2D-$\Delta\chi^2\leq 2.23$) does not appear in figure \ref{sfig:DSDT:W1}: contrary to \refeq{eq:MWST:cons}, with \refeq{eq:MWST:nocons} one cannot obtain the minimum $\chi^2_{\rm Min}$ with $\mcH=1$ TeV.
\begin{figure}[H]
\begin{center}
\subfloat[$(\Delta S,\Delta T)$ in \refeq{eq:obliqueval}.\label{sfig:DSDT:0}]{\includegraphics[width=0.3\textwidth]{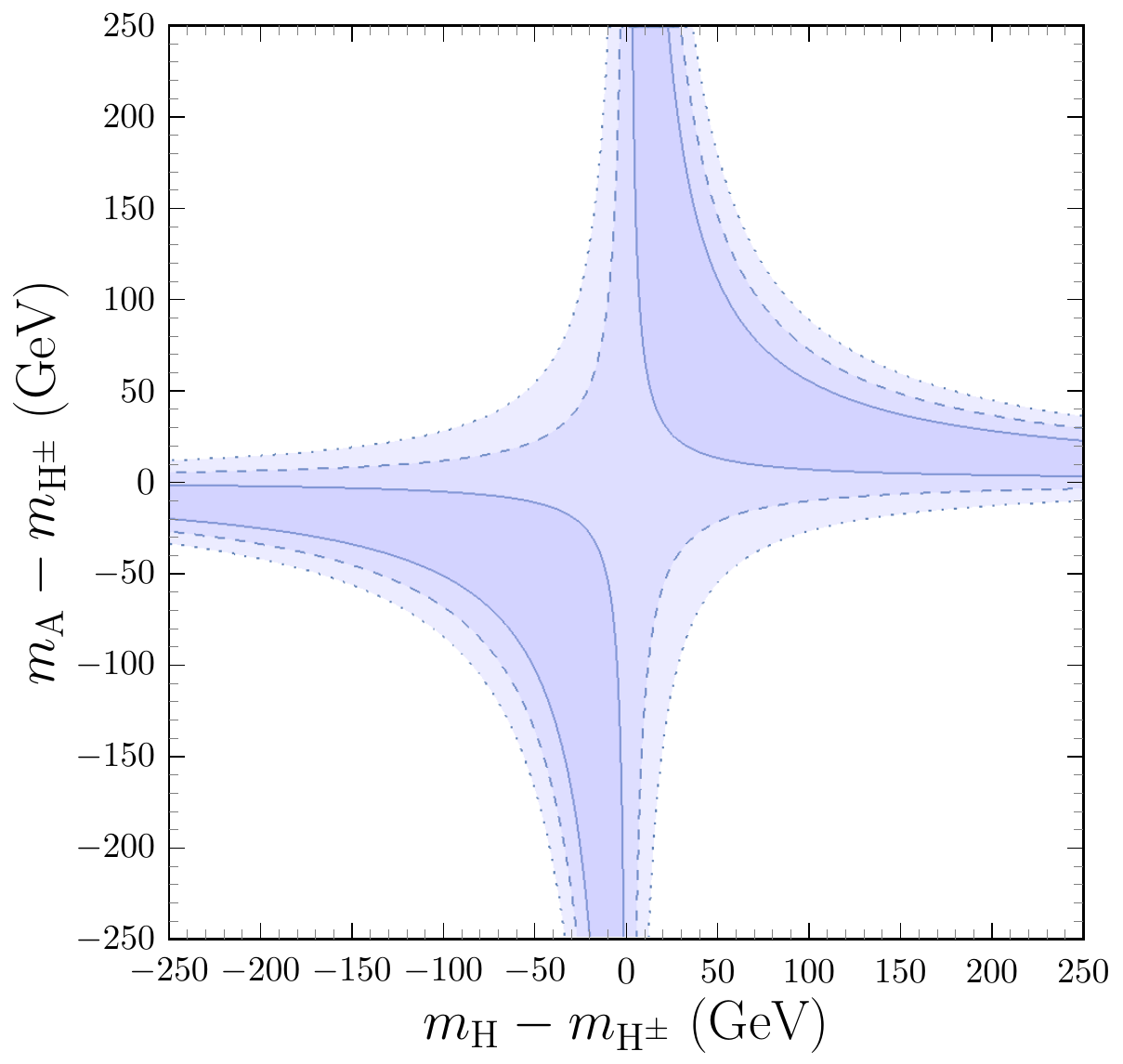}}\quad 
\subfloat[$(\Delta S,\Delta T)$ in \refeq{eq:MWST:cons}.\label{sfig:DSDT:W2}]{\includegraphics[width=0.3\textwidth]{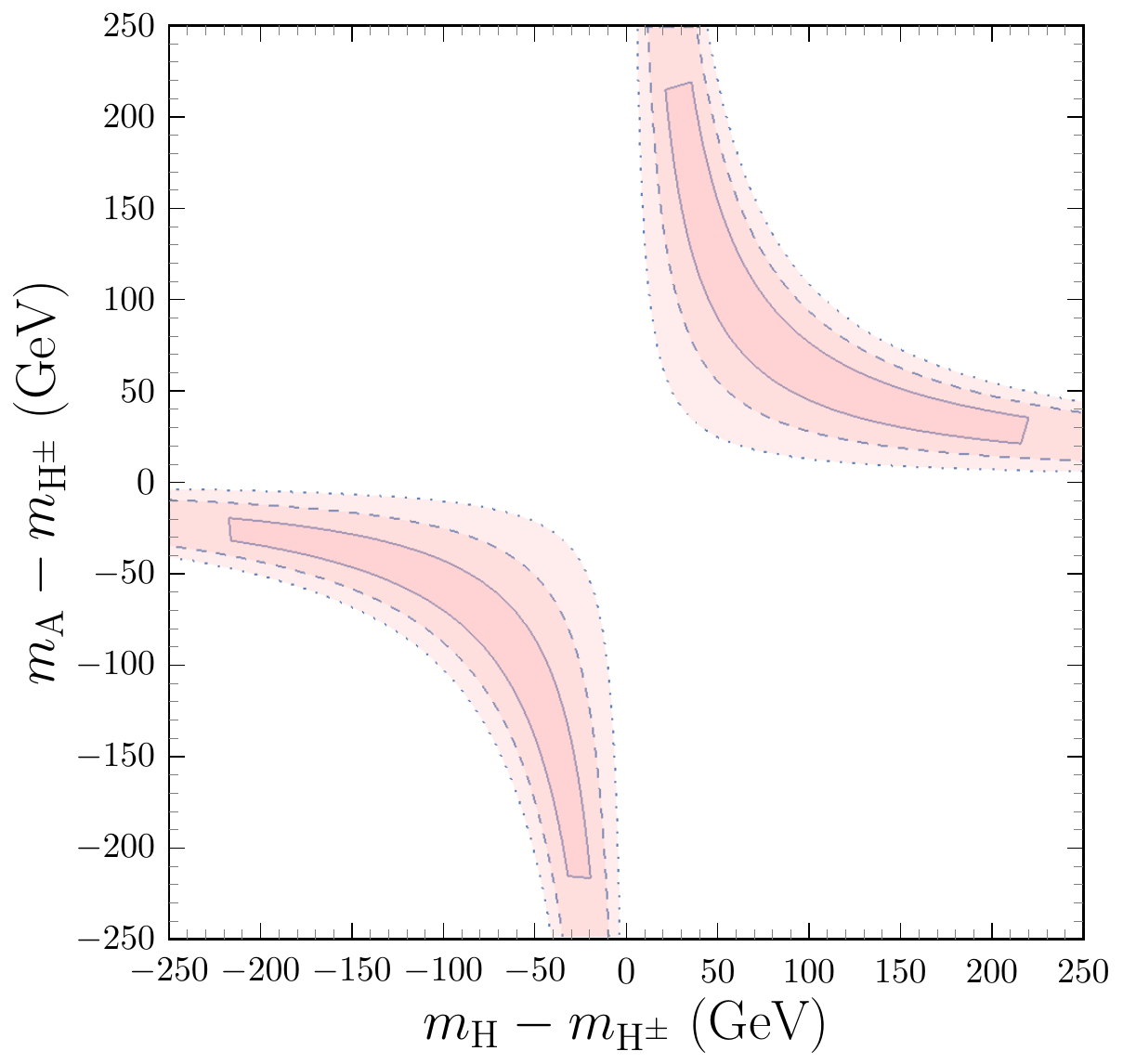}}\quad 
\subfloat[$(\Delta S,\Delta T)$ in \refeq{eq:MWST:nocons}.\label{sfig:DSDT:W1}]{\includegraphics[width=0.3\textwidth]{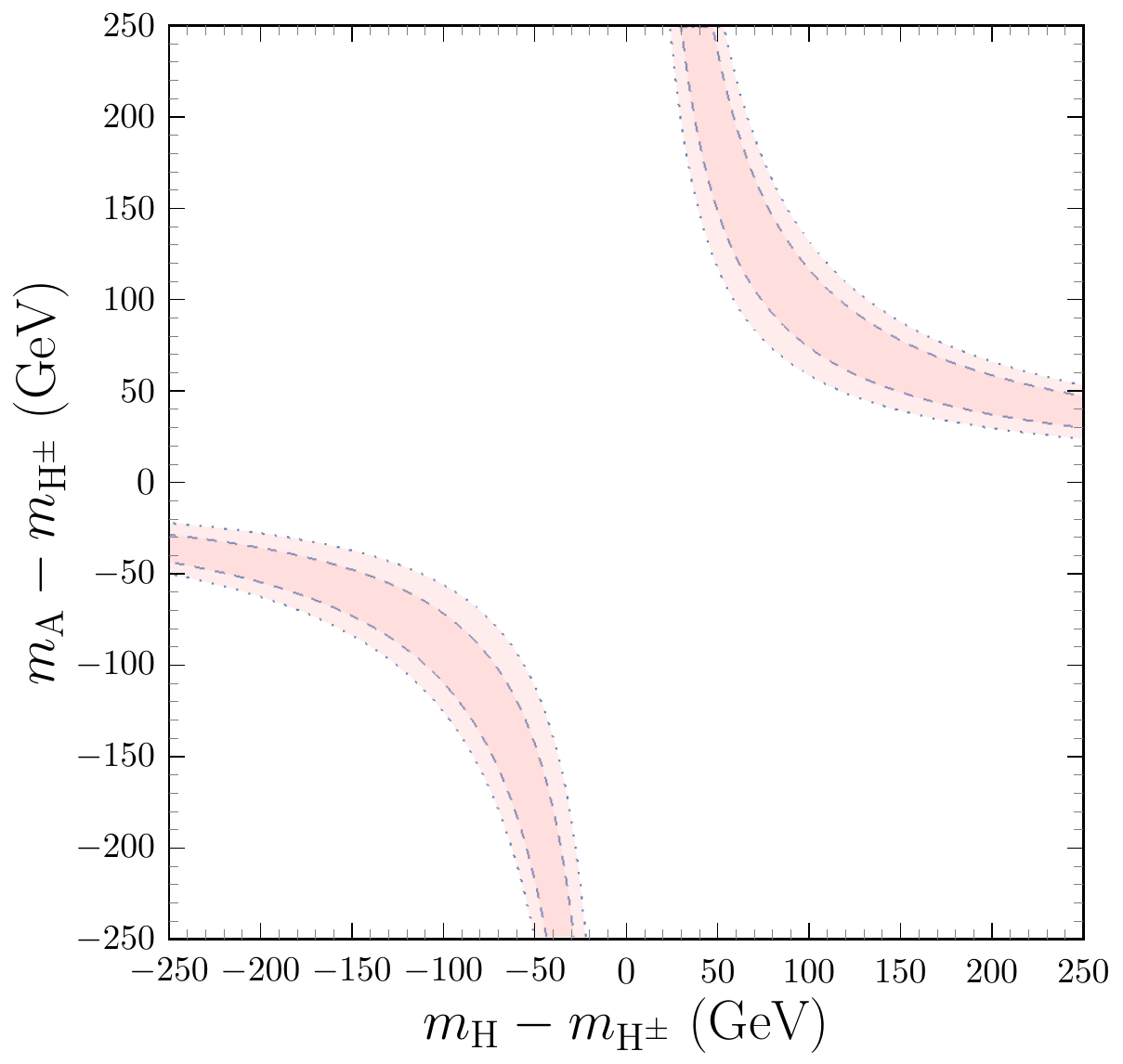}}
\caption{Oblique parameters: allowed regions in $\mA-\mcH$ vs. $\mH-\mcH$. Darker to lighter colors correspond to 2D-$\Delta\chi^2$ 1, 2 and 3$\sigma$ regions. The plot corresponds to $\mcH=1$ TeV and scalar alignment.\label{fig:oblique}}
\end{center}
\end{figure}
\item \underline{$\cH$-induced FCNC}.
The charged scalar $\cH$ can contribute to $\Delta F=1$ and $\Delta F=2$ FCNC processes like $b\to s\gamma$ and $B_q$--$\bar B_q$ mixings (for example, through SM-like box diagrams for $B_q$--$\bar B_q$ in which $W^\pm$ are replaced with $\cH$). The dominant contributions involve virtual top quarks as in the SM, with couplings including now $\tbinv$ factors. Keeping those contributions within experimental bounds only allows, roughly, the colored region in figure \ref{fig:box}. For each value of $\mcH$ there is a lower bound on $\tb$. See \cite{Misiak:2006zs,Crivellin:2013wna,Botella:2014ska} for further details.
\begin{figure}[H]
\begin{center}
\includegraphics[width=0.3\textwidth]{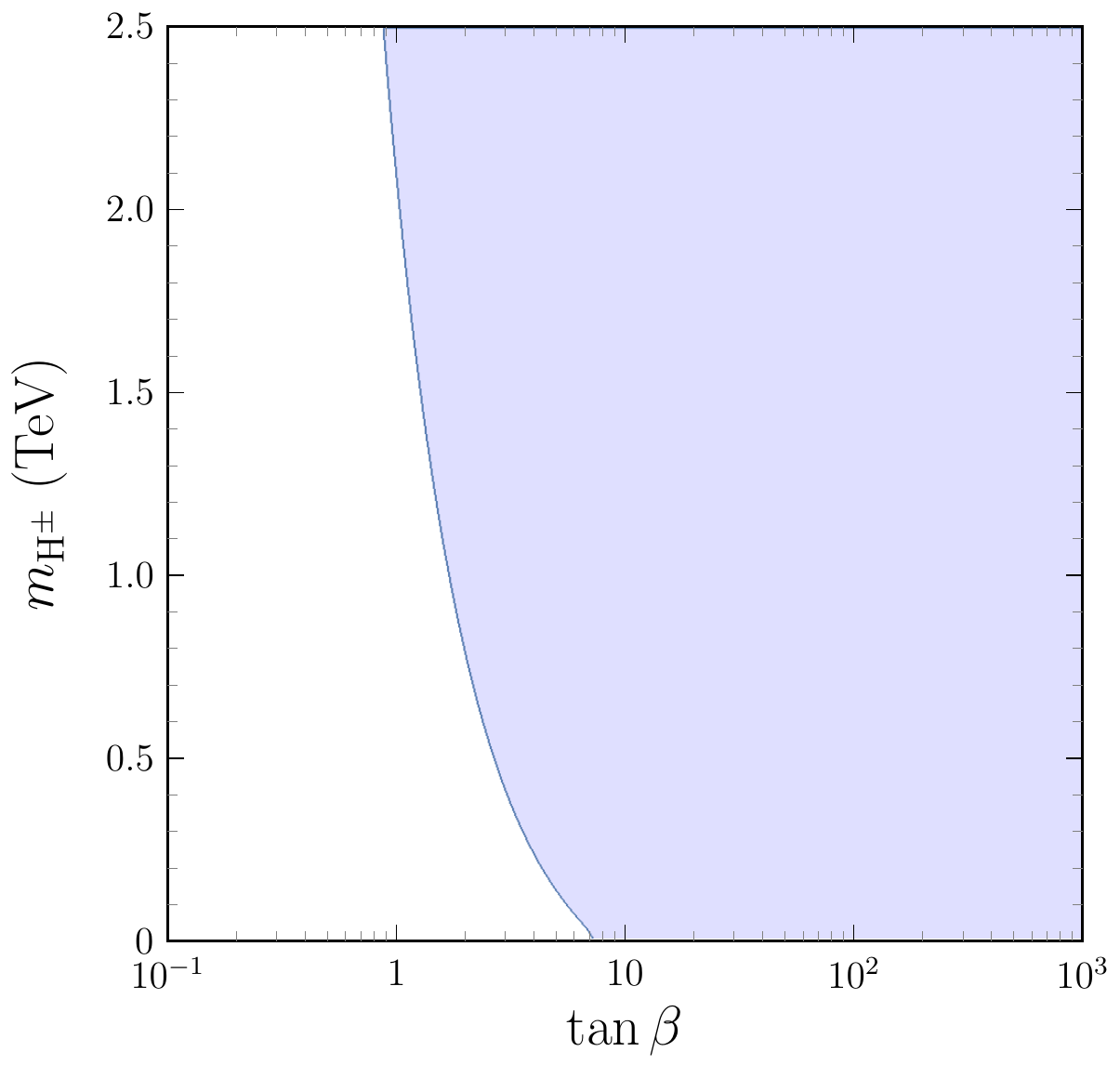}
\caption{$\cH$ FCNC: $\mcH$ vs. $\tb$ allowed region when contributions of $\cH$ to $B_q$--$\bar B_q$ are below experimental uncertainty in $\Delta M_{B_q}$. \label{fig:box}}
\end{center}
\end{figure}
\item \underline{Scalar sector perturbativity}. Additional constraints on scalar masses vs. $\tb$ arise from perturbativity requirements on the quartic coefficients of the scalar potential and from perturbative unitarity of $2\to 2$ scattering amplitudes \cite{Kanemura:1993hm,Akeroyd:2000wc,Ginzburg:2005dt,Horejsi:2005da,Kanemura:2015ska,Grinstein:2015rtl,Nebot:2019qvr}. With a $\ZZ$ symmetric potential, it is difficult to obtain masses above 1 TeV and values of $\tb$ larger than 8. Larger values of the masses and larger values of $\tb$ can be nevertheless obtained with the introduction of a soft symmetry breaking term $\mu_{12}^2\neq 0$ in \refeq{eq:ScalarPotential} \cite{Das:2015qva,Nebot:2019qvr}. 
\item \underline{Gluon-gluon fusion production cross section}. Let us consider the production cross section of $\nH$ and $\nA$ through the one loop gluon-gluon fusion process. In the scalar alignment limit, one can read from \refeq{eq:neutral:couplings} that the same $\tbinv$ factor applies to both pure scalar $\nH$ and pure pseudoscalar $\nA$ couplings with the top quark in the triangle loop:
\begin{equation}
\sigma(pp\to S)_{\rm ggF}\propto \tb^{-2}|F_{S}(x)|^2,\ x=4m_t^2/m_S^2,\quad S=\nH,\nA.
\end{equation}
The corresponding loop functions $F_{\nH}$ and $F_{\nA}$ \cite{LHCHiggsCrossSectionWorkingGroup:2016ypw,Harlander:2002wh,Ravindran:2003um,Pak:2011hs,Harlander:2002vv,Anastasiou:2002wq} are different due to the scalar or pseudoscalar character:
\begin{equation}
\begin{aligned}
& F_{\nH}(x)=-2x(1+(1-x)f(x)),\\ & F_{\nA}(x)=-2xf(x),\\ 
& \quad f(x)=
\left\{\begin{aligned}
&\arcsin^2(1/\sqrt{x}),\quad && x\geq 1\\ &-\frac{1}{4}\left(\ln\left(\frac{1+\sqrt{1-x}}{1-\sqrt{1-x}}\right)-i\pi\right)^2, &&x<1
\end{aligned}\right\}.
\end{aligned}
\end{equation}
Figure \ref{fig:ggxS} shows $|F_{\nH}(x)|^2$, $|F_{\nA}(x)|^2$ and the ratio $|F_{\nA}(x)|^2/|F_{\nH}(x)|^2$ as a function of the scalar mass. It is clear that the pseudoscalar $\nA$ has a larger gluon-gluon production cross section than the scalar $\nH$ for $\mA=\mH$ (up to a factor of 6 for $\mA=\mH=2m_t$). Since dimuon searches $[pp]_{\rm ggF}\to S\to\mu^+\mu^-$ at the LHC can be rather constraining for scalar masses $m_{\rm S}<1$ TeV, one can expect that in that low mass region $\mA > \mH$. One could have worried about the validity of this expectation in case $\BR{\nA\to\mu^+\mu^-}\ll \BR{\nH\to\mu^+\mu^-}$, but the only way to achieve a suppression of $\BR{\nA\to\mu^+\mu^-}$ relative to $\BR{\nH\to\mu^+\mu^-}$ is through the existence of $\nA\to\nH Z$ decays, which are only available if $\mA > \mH$, and thus cannot change that expectation.
\begin{figure}[H]
\begin{center}
 \includegraphics[width=0.4\textwidth]{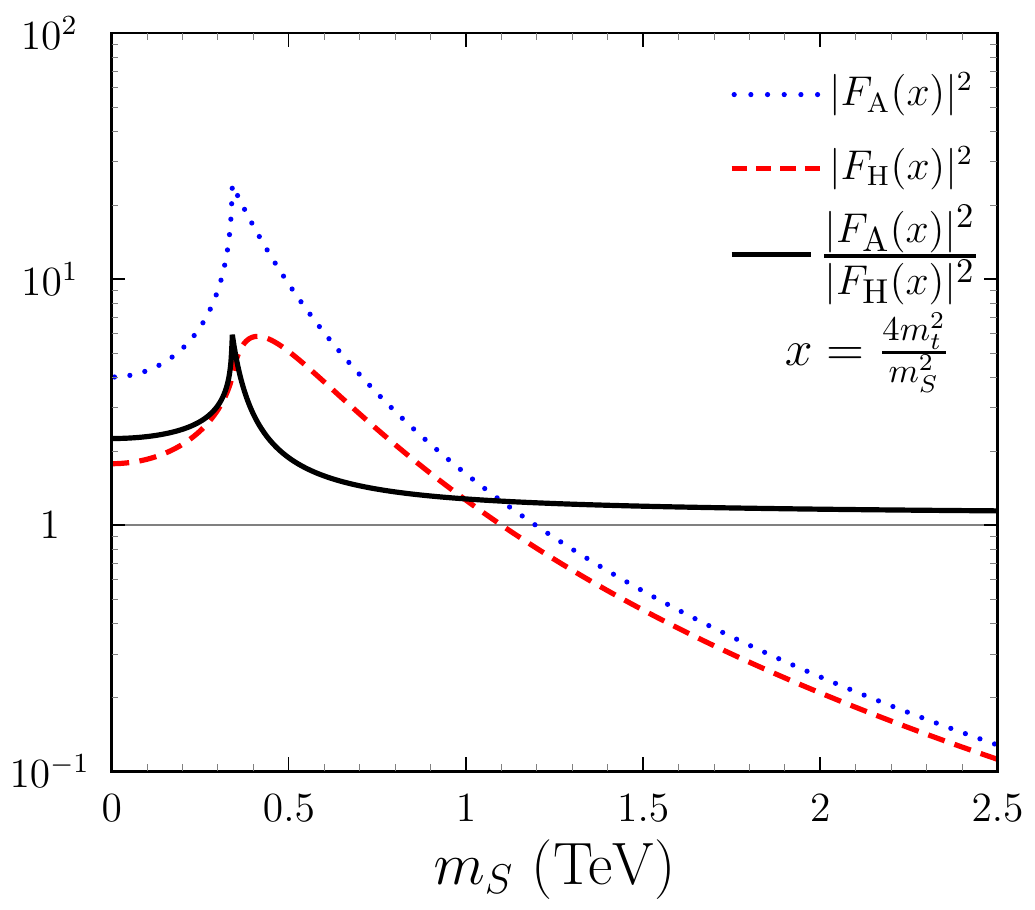}
 \caption{Loop functions controlling gluon-gluon production cross sections of scalars.\label{fig:ggxS}}
\end{center}
\end{figure}
\item \underline{$e^+e^-\to\mu^+\mu^-$ at LEP}. Sizable $n_e$ and $n_\mu$ are necessary ingredients for the contributions to $a_e$ and $a_\mu$ involving the new scalars $\nH$, $\nA$ and $\cH$. Data from LEP \cite{ALEPH:2006jhv} on $e^+e^-\to\mu^+\mu^-$ with $\sqrt{s}$ up to 210 GeV are sensitive to $s$-channel $\nH$ and $\nA$ mediated contributions (contrary to the LHC gluon-gluon fusion process, being scalar or pseudoscalar does not change the sensitivity of LEP data). One can roughly expect that agreement with LEP data imposes $\mA,\mH\geq 210$ GeV.
\end{itemize}
\section{Contributions to $\delta a_\ell$}\label{Sec:dal}

The complete prediction of the anomalous magnetic moment $a_\ell^{\rm Th}$, $\ell=e,\mu$, is
\begin{equation}
a_{\ell}^{\rm Th}=a_{\ell}^{\rm SM}+\delta a_{\ell}\, ,
\end{equation}
where $a_{\ell}^{\rm SM}$ is the SM contribution and $\delta a_\ell$ the NP correction. The anomalies in \refeqs{eq:damu:Exp}--\eqref{eq:dae:ExpRb} are ``solved'' for $\delta a_e=\delta a_e^{\rm Exp}$ and $\delta a_\mu=\delta a_\mu^{\rm Exp}$. We introduce for convenience $\Delta_\ell$ such that
\begin{equation}
 \delta a_\ell=K_\ell\,\Delta_\ell,\qquad K_\ell=\frac{1}{8\pi^2}\left(\frac{m_\ell}{v}\right)^2.
\end{equation}
For $\delta a_\mu$ one needs
\begin{equation}
 \Delta_\mu\simeq 1\, ,
\end{equation}
while for $\delta a_e$ one needs
\begin{equation}
 \Delta_e^{\rm Cs}\simeq -16\, ,\qquad \Delta_e^{\rm Rb}\simeq 9\,,
\end{equation}
where the superscript corresponds to the different values in \refeqs{eq:dae:ExpCs} and \eqref{eq:dae:ExpRb}.\\

In the model considered here, it is well known that both one loop and two loop (of Barr-Zee type) contributions can be dominant. In this section we analyse both types of contributions in the scalar alignment limit $\sba\to 1$ and keeping only leading terms in $\frac{m_\ell^2}{m_{S}^2}$, ${S}=\nH,\nA,\cH$. Full results, used for instance in the numerical analyses, can be found in appendix \ref{appendix:dal:loops}.

\subsection{One loop contributions to $\delta a_\ell$}\label{sSec:dal:1loop}
%
The one loop result $\Delta_\ell^{(1)}$ has contributions from $\nH$, $\nA$ and $\cH$. With the approximations mentioned above and the couplings in \refeqs{eq:neutral:couplings} and \eqref{eq:charged:couplings}, we have
\begin{equation}\label{eq:deltal:1loop}
 \Delta_\ell^{(1)}\simeq |\nl{\ell}|^2\left(\frac{I_{\ell \nH}}{\mH^2}-\frac{I_{\ell \nA}-2/3}{\mA^2}-\frac{1}{6\mcH^2}\right)\, ,
\end{equation}
where
\begin{equation}
 I_{\ell S}=-\frac{7}{6}-2\ln\left(\frac{m_\ell}{m_S}\right)\,.
\end{equation}
The range of interest in our analyses will be $m_S\in[0.2;2.5]$ TeV, in which case
\begin{equation}
 I_{\mu S}\in [13.9;18.9]\,,
\end{equation}
while
\begin{equation}
 I_{eS}=I_{\mu S}+2\ln \left(\frac{m_\mu}{m_e}\right)=I_{\mu S}+10.7\,.
\end{equation}
In \refeq{eq:deltal:1loop}, the $\nH$ contribution is positive, the $\nA$ contribution is negative and the $\cH$ contribution is negligible. One can then anticipate the following.
\begin{itemize}
\item The muon anomaly $\Delta_\mu\simeq 1$ can only be explained with the one loop $\nH$ contribution and provided
  \begin{equation}\label{eq:nmvsmH}
  1 \simeq \frac{|\nl{\mu}|^2}{\mH^2} I_{\mu\nH}\ \Rightarrow\ |\nl{\mu}|\sim \frac{1}{4}\mH\,.
 \end{equation}
 Considering $|\nl{\mu}|<250$ GeV, a priori there could be a one loop explanation of $\delta a_\mu$ for $\mH < 1$ TeV. Since the $\nA$ contribution has opposite sign, if $\mA\sim\mH$ a substantial cancellation would occur. As discussed in section \ref{Sec:General}, it is precisely for light $\nH$ that one expects $\mA>\mH$, in which case that cancellation is largely avoided and a one loop $\nH$ explanation viable. For heavier $\mH$, the muon anomaly needs other contributions.
\item For the electron ${\rm Cs}$ anomaly, $\Delta_e^{\rm Cs}\simeq -16$ can only be explained with the one loop $\nA$ contribution provided
\begin{equation}\label{eq:nevsmA:Cs}
-16\simeq -\frac{|\nl{e}|^2}{\mA^2} I_{e\nA}\ \Rightarrow\ |\nl{e}|\sim \frac{4}{5}\mA\,.
\end{equation}
For $|\nl{e}|<250$ GeV, this would require the pseudoscalar $\nA$ to be rather light, $\mA<300$ GeV. On the other hand, $\mA>200$ GeV would require $|\nl{e}|>160$ GeV: besides perturbativity concerns, such values of $|n_e|$ might be hard to reconcile with other constraints. More importantly, since we expect $\mH<\mA$ for light $\nA$, we also expect a sizable cancellation among $\nH$ and $\nA$ contributions. From this simple analysis, obtaining $\Delta_e^{\rm Cs}\simeq -16$ with one loop contributions does not appear to be feasible.
\item For the electron ${\rm Rb}$ anomaly, $\Delta_e^{\rm Rb}\simeq 9$ can only be explained with the one loop $\nH$ contribution and provided
\begin{equation}\label{eq:nevsmH:Rb}
 9\simeq \frac{|\nl{e}|^2}{\mH^2} I_{e\nH}\ \Rightarrow\ |\nl{e}|\sim \frac{3}{5}\mH\,.
\end{equation}
For $\mH>200$ GeV, this would require $|\nl{e}|>120$ GeV. If the same concerns on the values of $|\nl{e}|$ mentioned for $\Delta_e^{\rm Cs}\simeq -16$ apply here, obtaining  $\Delta_e^{\rm Rb}\simeq 9$ does not seem to be feasible neither; otherwise $\Delta_e^{\rm Rb}\simeq 9$ would be ``easier'' to accommodate with one loop contributions than $\Delta_e^{\rm Cs}\simeq -16$ because of the sign difference and the smaller absolute value.
\end{itemize}

\subsection{Two loop contributions to $\delta a_\ell$}\label{sSec:dal:2loop}
%
The dominant two loop contributions are the Barr-Zee ones. Diagrammatically they correspond to contributions where a closed fermion loop is attached to the external lepton through two propagators: one photon and one of the new scalars $\nH$, $\nA$. In the scalar alignment limit, 
\begin{equation}\label{eq:deltal:2loop}
 \Delta_\ell^{(2)}=-\frac{2\alpha}{\pi}\,\frac{\nrl{\ell}}{m_\ell}\,F\,.
\end{equation}
It is important to notice that these contributions are linear in $\nl{\ell}$. Detailed expressions are provided in appendix \ref{appendix:dal:loops}. In \refeq{eq:deltal:2loop} we have
\begin{equation}
 \frac{2\alpha}{\pi m_e}\simeq 9.1\,\text{GeV}^{-1},\qquad \frac{2\alpha}{\pi m_\mu}\simeq 0.044\,\text{GeV}^{-1}.
\end{equation}
The function $F$ depends on the masses of the fermions in the closed loop, their couplings to $\nH$ and $\nA$, and on $\mA$ and $\mH$. Considering the dominant contributions from top and bottom quarks, and also from tau and muon leptons since $\nl{\tau}$ and $\nl{\mu}$ are free parameters, 
\begin{equation}\label{eq:F:2loop}
 F=\frac{\tbinv}{3}\left[4(f_{t\nH}+g_{t\nA})+(f_{b\nH}-g_{b\nA})\right]
 +\frac{\nrlt}{m_\tau}(f_{\tau\nH}-g_{\tau\nA})+\frac{\nrlm}{m_\mu}(f_{\mu\nH}-g_{\mu\nA})\,,
\end{equation}
with 
\begin{equation}
 f_{xS}=f\left(\frac{m^2_{x}}{m^2_{S}}\right),\quad g_{xS}=g\left(\frac{m^2_{x}}{m^2_{S}}\right)\,.
\end{equation}
The functions $f(z)$ and $g(z)$ are defined in appendix \ref{appendix:dal:loops}. It is to be noticed that (i) $f(z)\sim g(z)$ in the range of interest, (ii) larger values correspond to heavier fermions, (iii) for the top quark loop, $f$ and $g$ vary between $0.08$ and $1$ in the relevant range of scalar masses, $m_S\in[0.2;2.5]$ TeV. 
\begin{itemize}
 \item If the electron anomaly is to be obtained through the two loop contributions, 
 \begin{equation}\label{eq:Delta:2loop}
  \Delta_e\simeq -9.1\,F\,\nrle/\text{GeV}\,,
 \end{equation}
and thus
\begin{equation}\label{eq:neF}
\begin{aligned}
&\text{from } \Delta_e^{\rm Cs}\,,\quad \nrle\,F\simeq 1.8\,\text{GeV}\,,\\
&\text{from } \Delta_e^{\rm Rb}\,,\quad \nrle\,F\simeq -1.0\,\text{GeV}\,.
\end{aligned}
\end{equation}
The sign and the magnitude of $F$ is fixed by the $\nrle$ value to fix $\delta a_e$.
\item For $\mH>1$ TeV, two loop contributions are necessary to explain the muon anomaly, in which case
\begin{equation}
 \Delta_\mu\simeq -0.044\,F\,\nrlm/\text{GeV}\ \Rightarrow\ \nrlm\,F\simeq -23\,\text{GeV}\,.
\end{equation}
If follows that, for $\mH>1$ TeV, 
\begin{equation}\label{eq:nmu:ne}
\begin{aligned}
&\text{for } \Delta_e^{\rm Cs}\text{ and }\Delta_\mu\,,\quad \nrlm\sim -13\nrle\,,\\
&\text{for } \Delta_e^{\rm Rb}\text{ and }\Delta_\mu\,,\quad \nrlm\sim 23\nrle\,.
\end{aligned}
\end{equation}
\end{itemize}
These correlations show that, in the present framework, the independence of $\nl{e}$ and $\nl{\mu}$ is essential to explain the different sign of $\Delta_e^{\rm Cs}$ and $\Delta_\mu$. This sign difference is challenging for many scenarios addressing simultaneously both anomalies. In this sense, addressing $\Delta_e^{\rm Rb}$ and $\Delta_\mu$ is less challenging.

\section{Analysis}\label{Sec:Analysis}

In section \ref{Sec:General} we have discussed some general constraints that apply without regard to the values of $\nl{e}$ and $\nl{\mu}$ of interest to reproduce the $\delta a_\ell$ anomalies; in section \ref{Sec:dal} we have explored the obtention of the $\delta a_\ell$ anomalies through one and two loop contributions. It is now time to present the main aspects of our detailed numerical analyses. The goal of the numerical analyses is to explore the parameter space of the model and map the different regions where a chosen set of relevant constraints is satisfied and the $\delta a_\ell$ anomalies are explained in terms of the new contributions. The independent parameters of the model are $\{\tb,\mH,\mA,\mcH,\mu_{12}^2,\cba,\nrle,\nrlm,\nrlt\}$: $\{\tb,\mH,\mA,\mcH,\mu_{12}^2,\cba\}$ control the scalar sector (together with $v$ and $\mh$) while $\{\nrle,\nrlm,\nrlt\}$ give the lepton Yukawa couplings (quark Yukawa couplings are fixed by $\tb$). The set of relevant constraints includes the following.
\begin{itemize}
 \item Boundedness from below of the scalar potential \cite{Ivanov:2015nea}, perturbativity of quartic couplings and perturbative unitarity of high energy $2\to 2$ scattering in the scalar sector \cite{Ginzburg:2005dt}.
 \item Corrections to the oblique parameters $S$ and $T$ in agreement with electroweak precision data \cite{ParticleDataGroup:2020ssz,Grimus:2008nb}.
 \item ``Production $\times$ decay" predictions for $\nh$ in agreement with Higgs signal strengths \cite{Aad:2020ago,Aad:2020jym,ATLAS:2020syy,ATLAS:2020pvn,Aad:2020xfq,ATLAS:2020qdt,Aad:2019lpq,Aad:2020mkp,CMS:2016mmc,CMS:2020gsy, Sirunyan:2020two, Sirunyan:2021ybb, Sirunyan:2021rug}. As already mentioned, this constraint forces the alignment limit in the scalar sector: in the analyses of section \ref{Sec:Results} one obtains indeed $\cba<3\times 10^{-3}$.
 \item Lepton flavor universality in leptonic and semileptonic decays \cite{ParticleDataGroup:2020ssz,Cirigliano:2007xi,Pich:2013lsa}.
 \item $b\to s\gamma$ and $B_q^0$--$\bar B_q^0$ data \cite{ParticleDataGroup:2020ssz,Misiak:2006zs,Crivellin:2013wna}.
 \item $e^+e^-\to \mu^+\mu^-$, $\tau^+\tau^-$ data from LEP (with center of mass energies up to 210 GeV) \cite{ALEPH:2006jhv}.
 \item LHC searches: resonant $pp\to S\to \mu^+\mu^-,\tau^+\tau^-$ searches with gluon-gluon fusion $pp\to S$ production \cite{ATLAS:2017fih,CMS:2019mij,ATLAS:2017eiz,CMS:2016xbv,CMS:2018rmh} and $\cH$ searches in $pp\to \cH tb$, $\cH\to \tau\nu,tb$ \cite{ATLAS:2018ntn,CMS:2019bfg,CMS:2019rlz,CMS:2020imj}.
\end{itemize}
For additional details on the different constraints we refer to \cite{Botella:2020xzf}. The constraints are typically modelled with a gaussian likelihood or an equivalent $\chi^2$ term, the overall likelihood is sampled over parameter space using Markov chain Monte Carlo techniques in order to obtain the regions where (best) agreement with the constraints is obtained. There are two final aspects of central importance which require a specific discussion: (i) how are the anomalies included in the analyses, (ii) what ranges are considered for the $n_\ell$ parameters.\\
Concerning the $a_\ell$ anomalies, the situation for $\delta a_\mu^{\rm Exp}$ is clear: one should consider \refeq{eq:damu:Exp}. On the contrary, for $\delta a_e^{\rm Exp}$ the situation is not settled: we have \refeq{eq:dae:ExpCs} and \refeq{eq:dae:ExpRb}, which are rather incompatible. In order to have a complete picture, we analyse both cases separately. Furthermore, we also consider two additional possibilities concerning $\delta a_e^{\rm Exp}$: 
\begin{itemize}
 \item despite the marginal compatibility of $\delta a_e^{\rm Exp,Cs}$ and $\delta a_e^{\rm Exp,Rb}$, we combine them into
 \begin{equation}\label{eq:dae:ExpAvg}
  \delta a_e^{\rm Exp,Avg}=-(2.0\pm 2.2)\times 10^{-13},
 \end{equation}
 which has the same sign as $\delta a_e^{\rm Exp,Cs}$, i.e. opposite to $\delta a_\mu^{\rm Exp}$, but a size close to 4 times smaller;
 \item a conservative approach in which we only assume that $|\delta a_e|\leq 20\times 10^{-13}$. Rather than targeting a specific value, this analysis may help to single out regions of parameter space where one cannot reproduce $\delta a_\mu^{\rm Exp}$ together with any value of $\delta a_e$ compatible with $\delta a_e^{\rm Exp,Cs}$ or $\delta a_e^{\rm Exp,Rb}$.
\end{itemize}
We will refer to these separate analyses as ``$a_e^{\rm Cs}$'', ``$a_e^{\rm Rb}$'', ``$a_e^{\rm Avg}$'', ``$a_e^{\rm Bound}$''. For their implementation in the analyses, we assign a joint $\chi^2$ contribution (corresponding to a gaussian factor in the likelihood)
\begin{equation}\label{eq:chi2:al}
 \chi^2_{g-2}(\delta a_e,\delta a_\mu)=\left(\frac{\delta a_e-c_e}{s_e}\right)^2+\left(\frac{\delta a_\mu-c_\mu}{s_\mu}\right)^2\,,
\end{equation}
where $c_\ell$ is the experimental central value and $s_\ell$ is the experimental uncertainty \emph{divided by 4}. The scope of this choice -- dividing the experimental uncertainty by 4 instead of simply using the experimental uncertainty -- is to show clearly that the model can reproduce easily and simultaneously both the muon and the electron anomalies, and to guarantee that we are definitely reproducing a sizable deviation from the SM both in $a_\mu$ and in all cases for $a_e$, except the ``$a_e^{\rm Bound}$'' analysis where there is no $\delta a_e$ term in \refeq{eq:chi2:al} and $|\delta a_e|\leq 20\times 10^{-13}$ is imposed. As a summary, all four selected cases of $\delta a_\mu^{\rm Exp}$ vs. $\delta a_e^{\rm Exp}$ are represented in figure \ref{fig:damu:dae}. 
\begin{figure}[h!tb]
\begin{center}
 \includegraphics[width=0.4\textwidth]{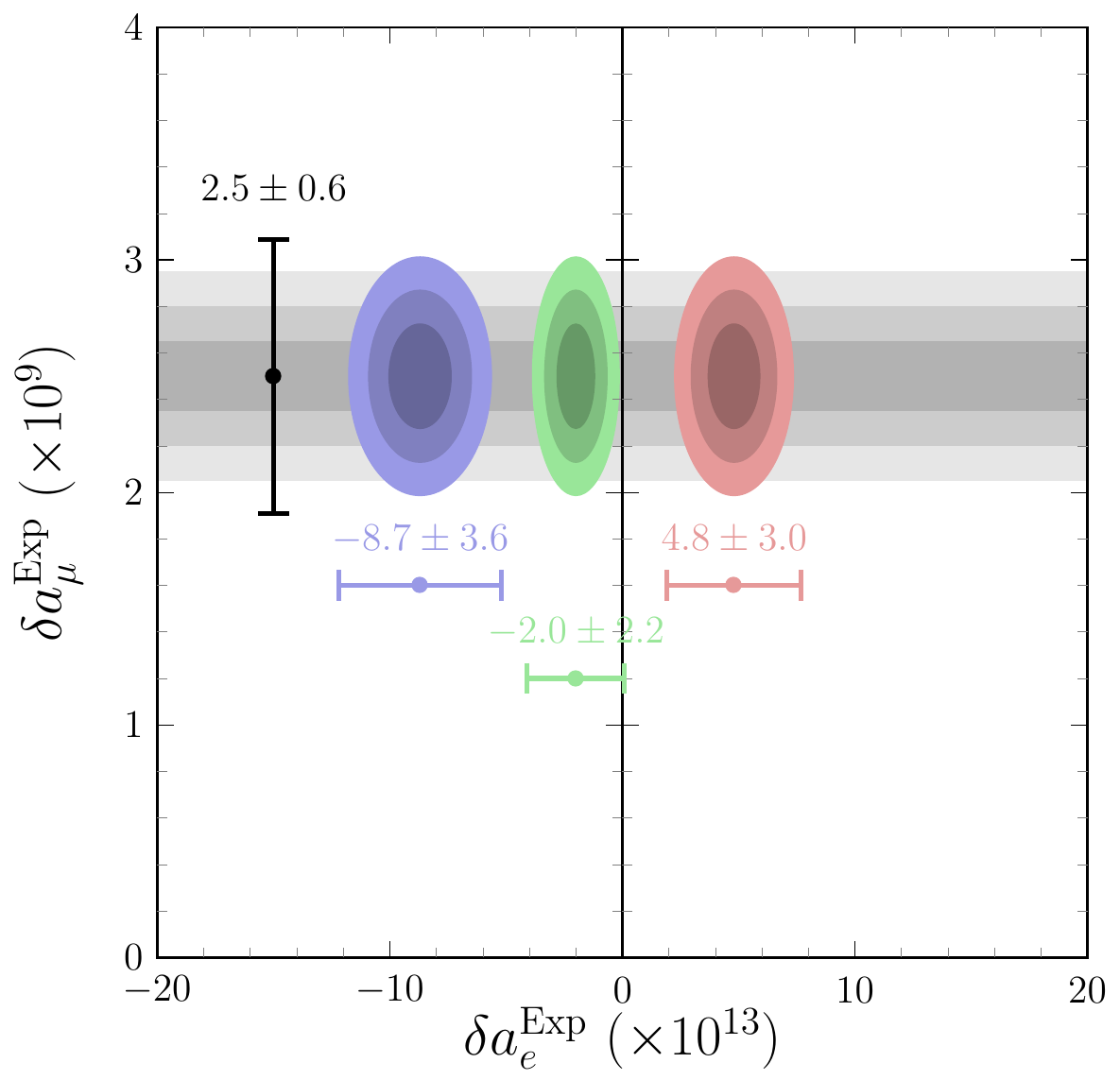}
 \caption{Allowed $\delta a_\mu^{\rm Exp}$ vs. $\delta a_e^{\rm Exp}$ regions in the different analyses.\label{fig:damu:dae}}
\end{center}
\end{figure}
The different colored regions in figure \ref{fig:damu:dae} represent three contours in the joint $\Delta \chi^2 = \chi^2 - \chi^2_{\rm{Min}}$ considering only \refeq{eq:chi2:al}. In a 2D-$\Delta \chi^2$ distribution they correspond, darker to lighter, to 1, 2 and 3$\sigma$ regions with 68.2\% C.L., 95.4\% C.L. and 99.7\% C.L., respectively. The same color coding is used in the figures below illustrating the final results of the analyses, where all observables and constraints have been included.\\ 
Finally, in \cite{Botella:2020xzf}, only $|\nl{\ell}|\leq 100$ GeV were considered. Although for $|\nl{e}|,|\nl{\mu}|\sim 100$ GeV lepton couplings to the new scalars are hugely enhanced with respect to $\nh$ couplings, it is true that $\frac{|\nl{\ell}|}{v}\sim \frac{100\text{ GeV}}{v}\sim 0.4$ does not appear to pose a perturbativity challenge. In fact, the one loop correction to the imaginary part of the $\mH$ mass is controlled by $\Gamma(\nH\to\ell\bar\ell)$ and the relevant ratio is $\frac{\Gamma}{\mH}=\frac{1}{8\pi}\frac{|\nl{\ell}|^2}{\vev{}^2}$, therefore arriving to $|\nl{\ell}|=v\sim 250$ GeV represents one loop corrections at the 4\% level. For this reason the analyses have been done with $|\nl{\ell}|\leq 250$ GeV $\sim v$; furthermore, the analysis ``$a_e^{\rm Cs}$'' has been conducted both with $|\nl{\ell}|\leq 100$ GeV (since this case is the closest one to \cite{Botella:2020xzf}) and with $|\nl{\ell}|\leq 250$ GeV.

\section{Results\label{Sec:Results}}

In the next subsections we discuss the most relevant results of the analyses done following the lines of the previous section. In subsection \ref{sSec:Results:Cs100} we consider the scenario ``$a_e^{\rm Cs}$'' when $|\nl{\ell}|\leq100$ GeV is imposed. The implications of changing this last assumption to $|\nl{\ell}|\leq250$ GeV are addressed in subsection \ref{sSec:Results:Cs250}. The implications of the different assumptions for the electron anomaly, that is scenarios ``$a_e^{\rm Rb}$'', ``$a_e^{\rm Avg}$'' and ``$a_e^{\rm Bound}$'' are explored in subsection \ref{sSec:Results:diff:dae}. The impact of the recent measurement of $M_W$ by the CDF collaboration is considered in subsection \ref{sSec:Results:CDF}. Finally, to further illustrate these discussions, a few complete example cases are shown in subsection \ref{sSec:Results:examples}.

\subsection{$|\nl{\ell}|\leq 100$ GeV}\label{sSec:Results:Cs100}
Here we present the results of the analysis ``$a_e^{\rm Cs}$'' with the perturbativity constraint $|\nl{\ell}|\leq 100$ GeV. This serves to revisit the main results of \cite{Botella:2020xzf} and as a reference for the analysis with $|\nl{\ell}|\leq 250$ GeV addressed in the following subsection.
\begin{figure}[h!tb]
\begin{center}
\subfloat[$\nrlm$ vs. $\mH$.\label{sfig:nmu:mH:Cs100}]{\includegraphics[width=0.3\textwidth]{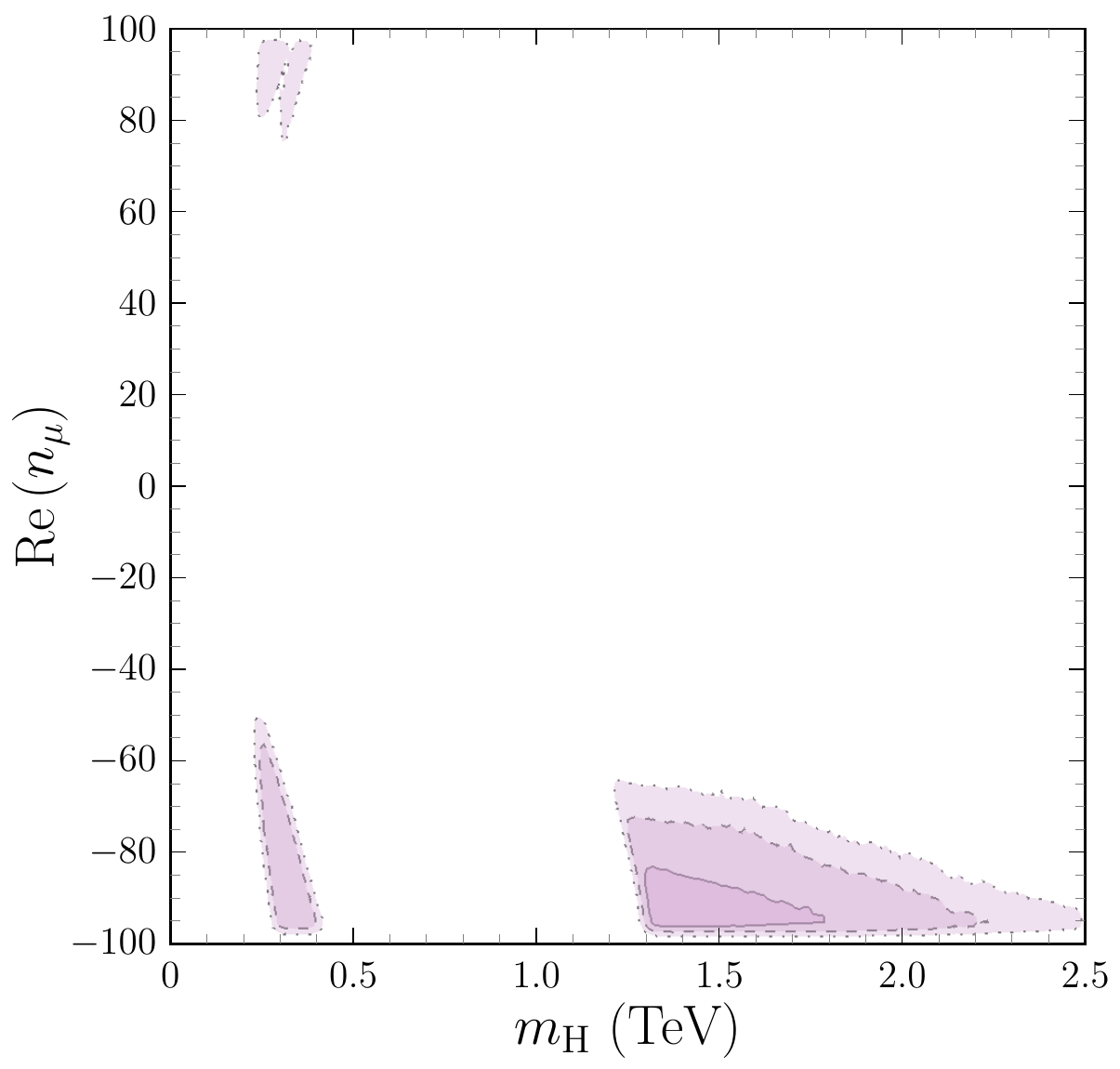}}\quad 
\subfloat[$\mH$ vs. $\tb$.\label{sfig:mH:tb:Cs100}]{\includegraphics[width=0.3\textwidth]{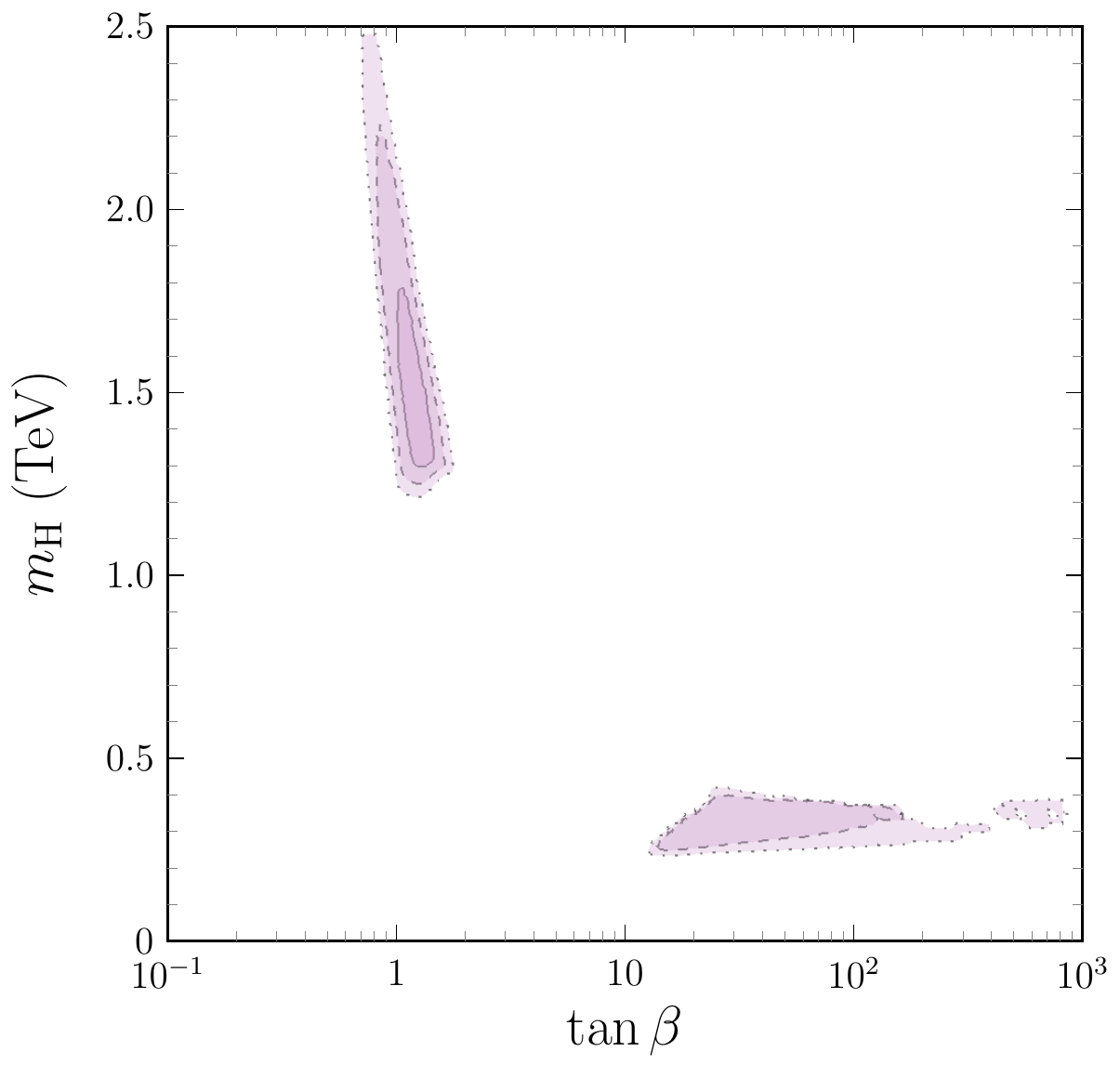}}\quad 
\subfloat[$\mH$ vs. $\mcH$.\label{sfig:mH:mCH:Cs100}]{\includegraphics[width=0.3\textwidth]{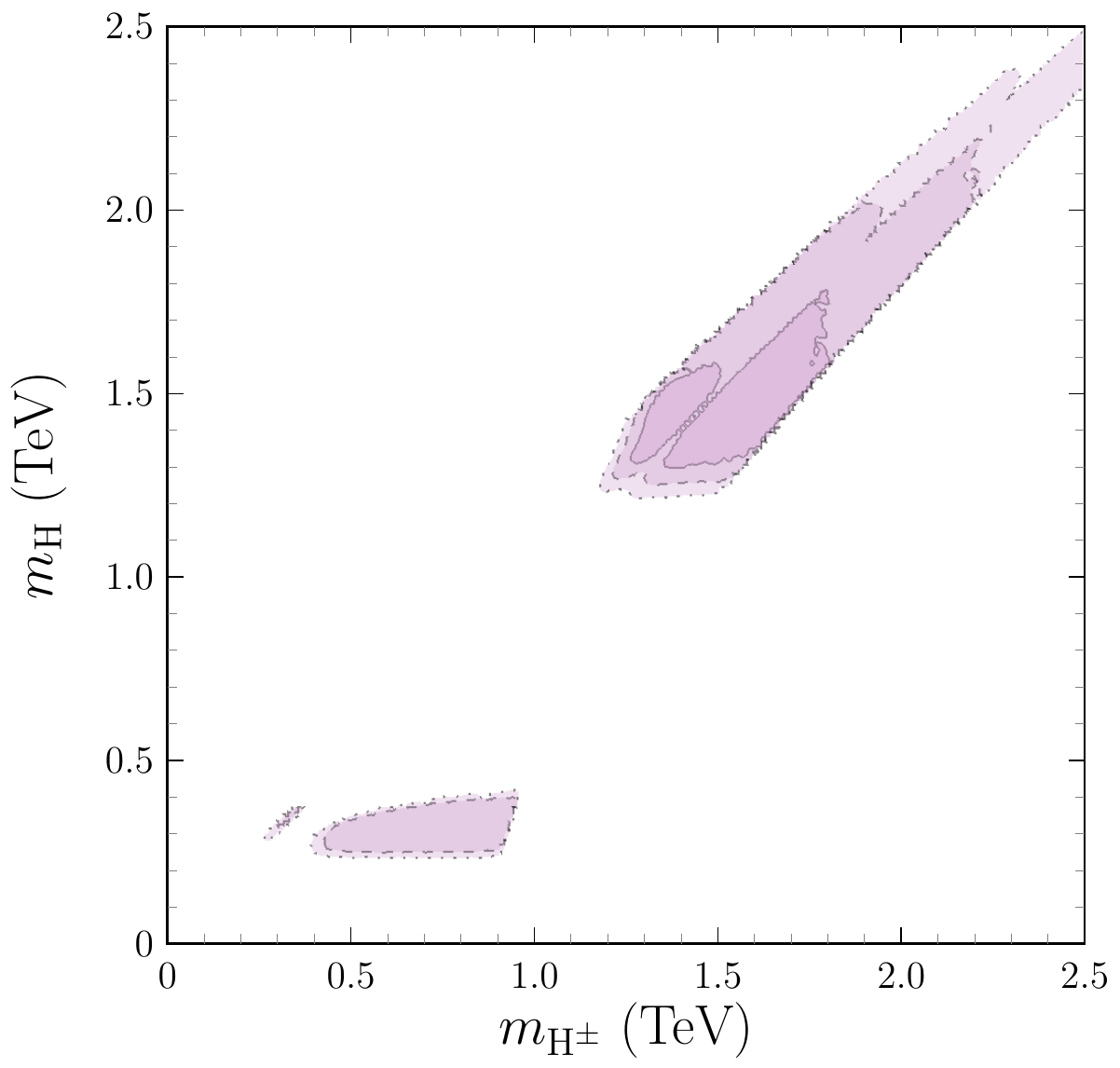}}
\caption{Allowed regions: relevant correlations involving $\mH$ with $|\nl{\ell}|\leq100$ GeV.\label{fig:01:Cs100}}
\end{center}
\end{figure}
The perturbativity constraint limits the possibility of explaining $\delta a_e^{\rm Exp,Cs}$ via the one loop contribution, since it requires  $\mA\leq125$ GeV for $|\nl{e}|\leq100$ GeV (see \refeq{eq:nevsmA:Cs}) which is not allowed by $e^+e^-\to\mu^+\mu^-$ LEP data. 
On that respect, lepton flavor universality constraints also limit the possibility of a one loop explanation for the electron anomaly, as discussed later. 
This leaves us with two scenarios, one where both anomalies are explained via the two loop contribution, following the scaling law in \refeq{eq:nmu:ne}, and another where the muon anomaly is one loop dominated while the electron one is still generated at two loops.\\
In figure \ref{sfig:nmu:mH:Cs100} the allowed regions for $\nrlm$ are presented as a function of $\mH$. Three disjoint regions in the scalar mass can be seen: two in the 200-400 GeV range and the other above 1.2 TeV. The low mass regions belong to the scenario where the muon anomaly is obtained through the one loop contribution in agreement with the relation in \refeq{eq:nmvsmH}. Note that this contribution depends on the absolute value of the coupling, so both signs are allowed for $\nrlm$. In the large mass region both leptonic anomalies are two loop dominated.\\
Figure \ref{sfig:mH:tb:Cs100} shows $\mH$ vs. $\tb$. It contains two separate allowed regions again: in the $\tb\sim 1$ regime only scalar masses above 1.2 TeV are allowed; conversely for $\tb$ larger than 10, $\mH$ lies in the 200-400 GeV interval.\\
To complement the previous two plots, in figure \ref{sfig:mH:mCH:Cs100} the relation between the masses $\mH$ and $\mcH$ is shown. In the low mass region we can clearly distinguish two scenarios. One where $\mcH\simeq\mH$ and another where $\mcH>\mH$; in the latter, $\mcH\simeq\mA$.The degeneracy of $\cH$ with either $\nH$ or $\nA$ arises from the oblique parameters constraint, as mentioned in section \ref{Sec:General}. In the large mass region the mass differences do not exceed $\pm 300$ GeV.\\
\begin{figure}[htb]
\begin{center}
\subfloat[For $S=\nH$.\label{sfig:ppHmumu:mH:Cs100}]{\includegraphics[width=0.3\textwidth]{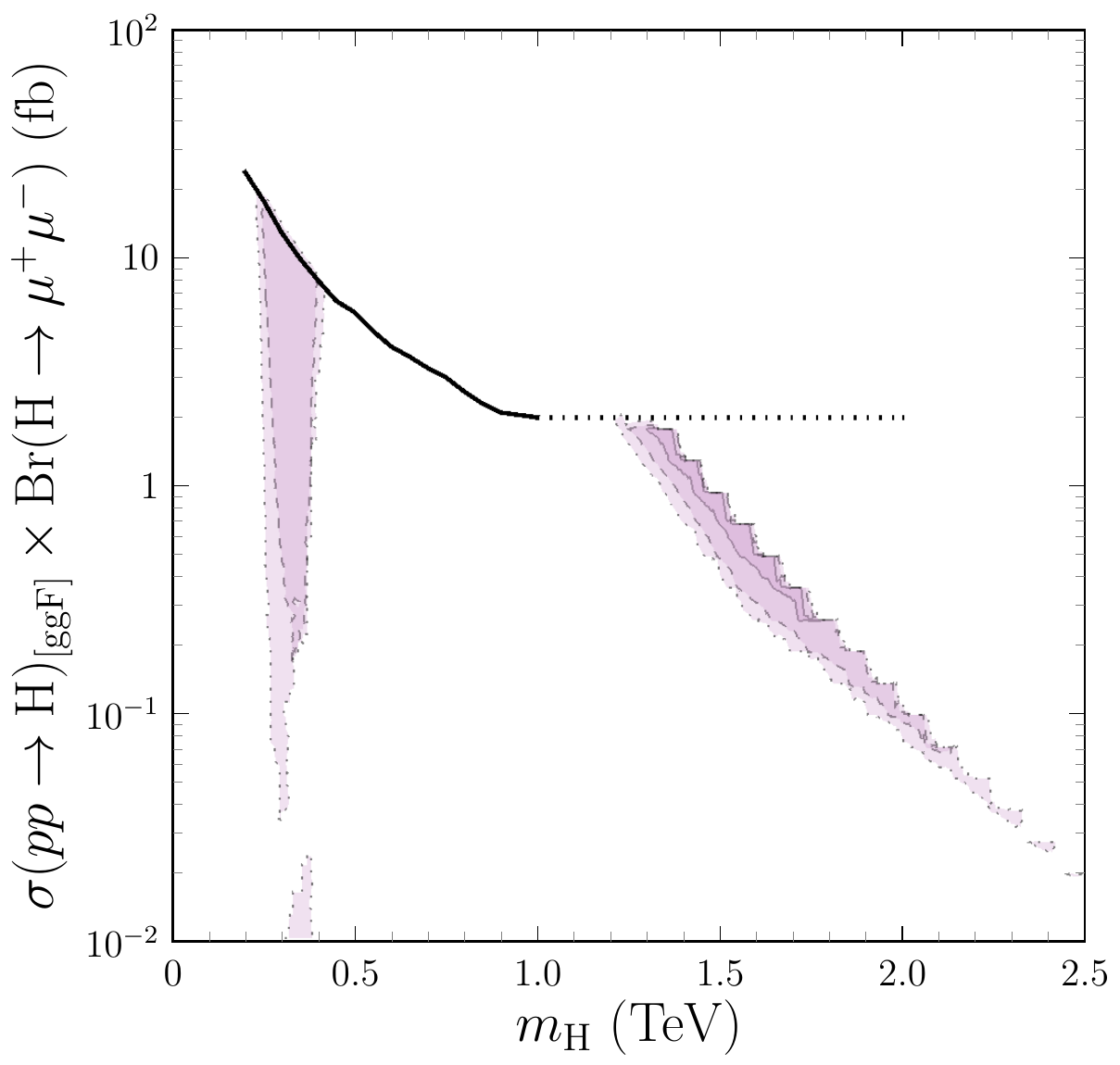}}\qquad 
\subfloat[For $S=\nA$.\label{sfig:ppAmumu:mA:Cs100}]{\includegraphics[width=0.3\textwidth]{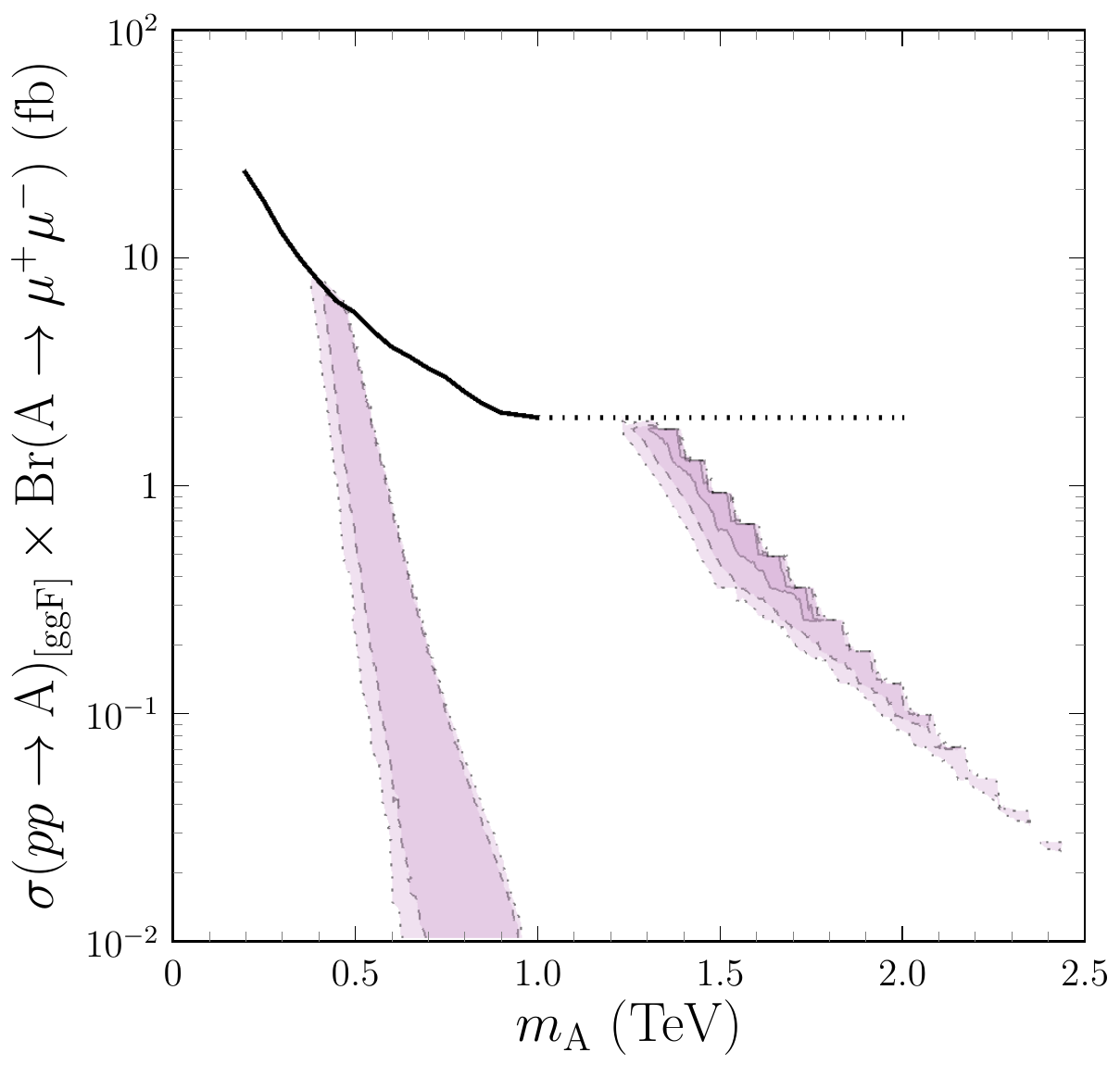}}
\caption{$\sigma(pp\to S)_{\rm [ggF]}\times\BR{S\to\mu^+\mu^-}$ vs. $m_S$ allowed regions with $|\nl{\ell}|\leq100$ GeV.\label{fig:02:Cs100}}
\end{center}
\end{figure}
Figure \ref{fig:02:Cs100} illustrates the allowed regions for the resonant process $[pp]_{\rm{ggF}} \to S \to \mu^{+}\mu^{-}$ with respect to the scalar mass $m_S$ for $S = \nH, \nA$. The black line corresponds to the limit observed by CMS \cite{CMS:2019mij}. 
Although LHC direct searches are already constraining the allowed regions, there is 
ample room for extra scalars that can explain both $g-2$ anomalies simultaneously.

\begin{figure}[h!tb]
\begin{center}
\subfloat[$\nrle$ vs. $\mH$.\label{sfig:ne:mH:Cs100}]{\includegraphics[width=0.3\textwidth]{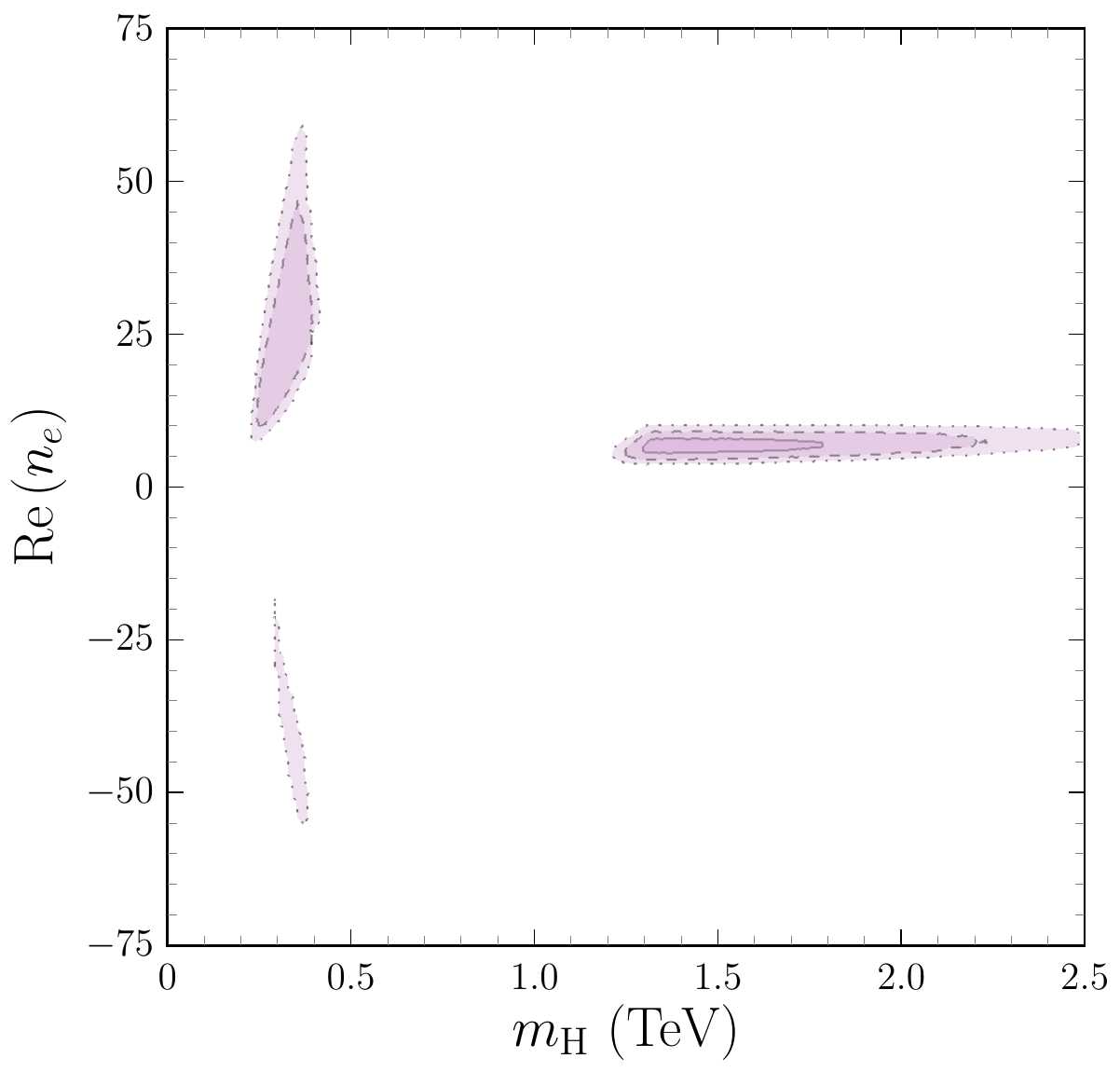}}\qquad 
\subfloat[$\nrlt$ vs. $\nrle$.\label{sfig:nt:ne:Cs100}]{\includegraphics[width=0.3\textwidth]{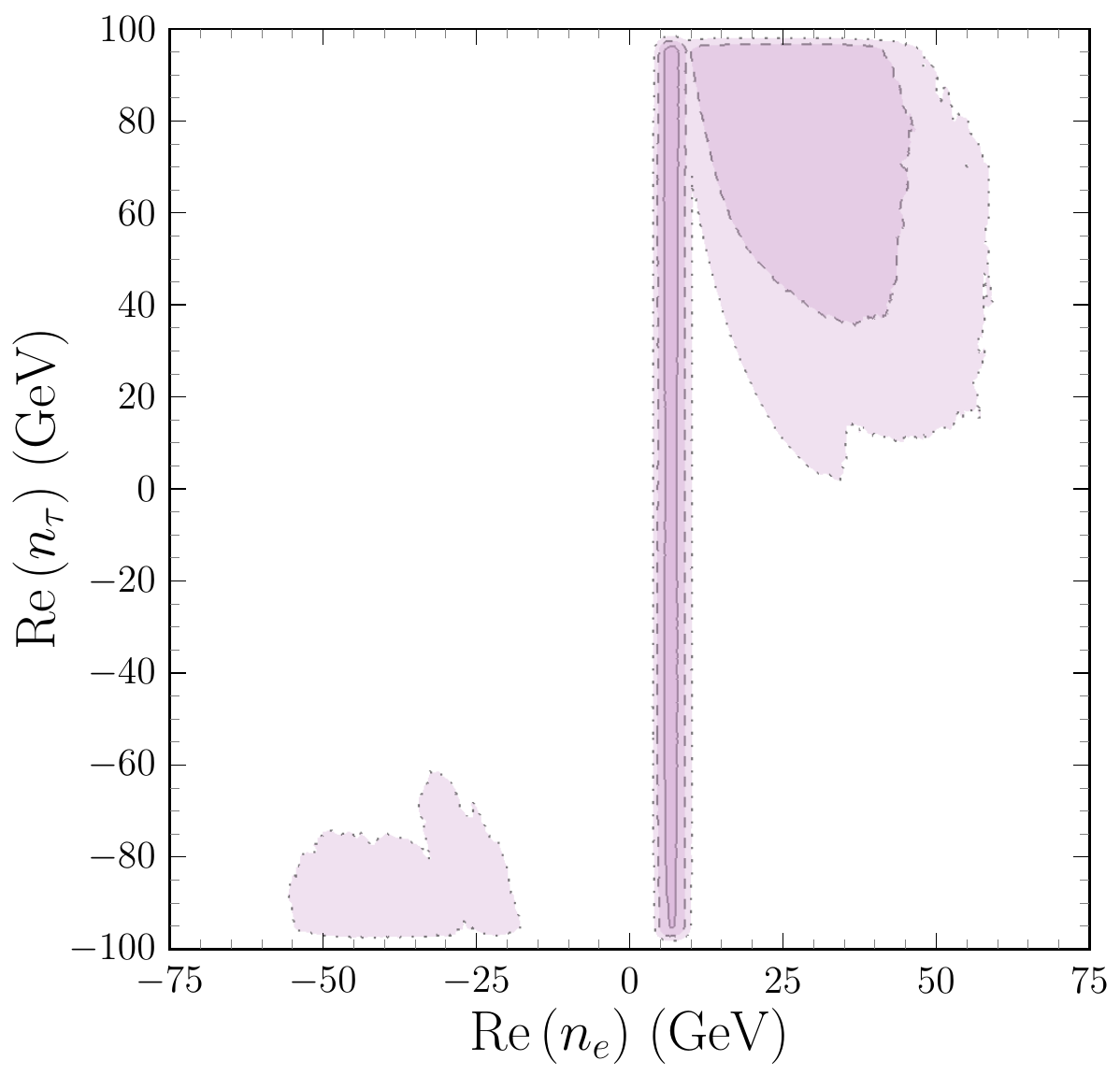}}
\caption{Allowed regions for $\nrle$ with $|\nl{\ell}|\leq 100$ GeV.\label{fig:03:Cs100}}
\end{center}
\end{figure}

Let us now discuss some results concerning $\nrle$ and $\nrlt$. 
With a two loop explanation of the electron anomaly, it follows from \refeqs{eq:deltal:2loop} and \eqref{eq:F:2loop} (see appendix \ref{appendix:dal:loops} for further details) that one could have expected that both the coupling $\nrle$ and the deviation $\delta a_{e}$ have opposite sign: this is confirmed in figure \ref{fig:03:Cs100} in the 1$\sigma$ region. However, this figure also contains regions where $\nrle$ is negative. This behavior might be understood by analysing with some detail the two loop contribution to $\delta a_e$: $F$ in \refeq{eq:F:2loop} can be decomposed as $F= F_q+F_\tau+F_\mu$, where $F_f$ is the contribution with fermion $f$ running in the closed loop. One can estimate the importance of the different contributions for different $\tb$, $\mH$ and $\mA$ ranges.
\begin{itemize}
\item For $\tb\sim1$ and $\mH,\mA>1.2$ TeV, \refeq{eq:F:2loop} gives
 \begin{equation}
  F_q>0.18\,,\quad \left|F_\tau\right|<0.13\times 10^{-4}\frac{\nrlt}{1\text{ GeV}}\,,\quad \left|F_\mu\right|<0.12\times 10^{-5}\frac{\nrlm}{1\text{ GeV}}\,.
 \end{equation}
It is clear that in this region the quark-induced contribution $F_t$ is (i) necessarily dominant and (ii) it requires $\nrle\sim 4-10$ GeV, as figure \ref{sfig:ne:mH:Cs100} illustrates, in order to reproduce $\Delta_e^{\rm Cs}\simeq -16$.

\item For $\tb>10$, $\mH\in [200, 400]$ GeV and $\mA\in [400,1000]$ GeV, \refeq{eq:F:2loop} gives
 \begin{equation}
  F_q<0.18\,,\quad F_\tau\in [0.02, 0.15]\times 10^{-2}\frac{\nrlt}{1\text{ GeV}}\,, \quad F_\mu\in [0.04, 0.25]\times 10^{-3}\frac{\nrlm}{1\text{ GeV}}\,.
 \end{equation}
In this case, large values of $\nrlt\simeq \pm 100$ GeV give $\tau$-induced contributions at the same level of, or even larger than, the quark-induced contribution. This occurs despite some cancellation among the $\tau\nH$ and $\tau\nA$ contributions in \refeq{eq:F:2loop}. This scenario would require $\nrle\lesssim-15$ GeV or $\nrle\gtrsim7$ GeV, as shown in figure \ref{sfig:ne:mH:Cs100}, in order to reproduce $\Delta_e^{\rm Cs}\simeq -16$.
\end{itemize}
From this simple estimates one can conclude that, besides the expected regions where $\delta a_e$ arises from quark-induced two loop contributions, regions where the $\tau$-induced contributions have an important role might be present. For this to occur, one might expect some peculiarities: besides light $\nH$ and large $\tb$, large values of both $|\nrlt|$ and $|\nrle|$, with $\nrlt$ and $\nrle$ having the same sign, are required. Contrary to the case with dominating quark induced contributions, one might then have allowed regions where $\nrle<0$. This is illustrated in figures \ref{sfig:ne:mH:Cs100} and \ref{sfig:nt:ne:Cs100} where one can observe how allowed $\nrle<0$ only appear for a light $\nH$, and how the regions with large $\pm\nrle$ correspond to large $\pm\nrlt$.

To close this subsection, it is worth analysing in detail the role of the lepton flavor universality constraints mentioned in section \ref{Sec:Analysis}. 
As justified later, we focus on observables involving only $\mu$'s and $e$'s.
For the ratios 
\begin{equation}
 R_{\mu e}^P=\frac{\Gamma(P^+\to \mu^+\nu)}{\Gamma(P^+\to \mu^+\nu)_{\rm SM}}\frac{\Gamma(P^+\to e^+\nu)_{\rm SM}}{\Gamma(P^+\to e^+\nu)}\,,
\end{equation}
the current constraints are \cite{ParticleDataGroup:2020ssz}
\begin{equation}\label{eq:RPmue:exp}
 R_{\mu e}^\pi=1+(4.1\pm 3.3)\times 10^{-3}\,,\qquad
 R_{\mu e}^K=1-(4.8\pm 4.7)\times 10^{-3}\,.
\end{equation}
In the present scenario,
\begin{equation}\label{eq:RPmue:01}
 R_{\mu e}^P=\frac{|1-\Delta_\mu^P|^2}{|1-\Delta_e^P|^2}\,,\qquad |1-\Delta_\ell^P|^2=\left|1-\frac{M_P^2}{\tb\mcH^2}\frac{\nrl{\ell}}{m_\ell}\right|^2\,,
\end{equation}
and thus, for $\Delta_\ell^P\ll 1$,
\begin{equation}\label{eq:RPmue:02}
 R_{\mu e}^P\simeq 1+2\frac{M_P^2}{\tb\mcH^2}\left(\frac{\nrle}{m_e}-\frac{\nrlm}{m_\mu}\right)\,.
\end{equation} 
The presence of $M_P^2$ and the lepton masses allows us to concentrate on $R_{\mu e}^K$ and neglect the $\nl{\mu}$ contribution. Therefore from \refeq{eq:RPmue:exp} we get the constraint
\begin{equation}
 \nrle < 5\frac{\tb\mcH^2}{1\text{ TeV}^2}\text{ GeV}\,.
\end{equation}
Then,
\begin{itemize}
 \item for $\tb\simeq 1$ and $\mcH\simeq 2$ TeV, $\nrle< 20$ GeV,
 \item while for $\tb\simeq 10^2$ and $\mcH\simeq 0.5$ TeV, $\nrle< 125$ GeV.
\end{itemize}
From muon decay constraints on the $\cH$ mediated contributions we also have a $\tb$ independent constraint (since the process is purely leptonic) \cite{ParticleDataGroup:2020ssz,Botella:2014ska}: 
\begin{equation}\label{eq:muenunu:01}
 \left|\frac{\nl{e} \nl{\mu}}{\mcH^2}\right|<0.035\,.
\end{equation}
This constraint is relevant for the low mass region: for $\nrlm\simeq 100$ GeV, we can rewrite
\begin{equation}\label{eq:muenunu:02}
|\nl{e}|< 87\left(\frac{\mcH}{0.5\text{ TeV}}\right)^2\text{ GeV}\,,
\end{equation}
which is more restrictive than the bound from $R_{\mu e}^K$ above. Concerning other observables involving $\tau$ leptons, semileptonic processes are not sensitive to $\nl{\tau}$ due to $\frac{m_e}{m_\tau}$ and $\frac{m_\mu}{m_\tau}$ suppressions, while purely leptonic decays have looser bounds than \refeq{eq:muenunu:01}.\\ 
This simple numerical exercise confirms that $\delta a_e^{\rm Exp}$ cannot be explained through one loop contributions.

\subsection{$|\nl{\ell}|\leq 250$ GeV}\label{sSec:Results:Cs250}
As previously motivated, perturbativity bounds on the Yukawa couplings should be studied in detail. Here we explore higher scales in $\nl{\ell}$, namely changing from  $|\nl{\ell}| \leq 100\ \rm{GeV}$ to $|\nl{\ell}| \leq 250\ \rm{GeV}$ while maintaining the same constraints of the previous section. Conversely to what one would naively expect, it is not just the allowed regions in the different $\nl{\ell}$ that might change, but it has direct consequences on other physical observables such as the scalar masses and $\tb$, among others.

Figure \ref{sfig:nmu:mH:Cs250} shows results for $\nrl{\mu}$ vs. $\mH$. It is clear that the allowed regions in parameter space are notably enlarged with respect to those in figure \ref{sfig:nmu:mH:Cs100}, which are completely embedded in the ones of this new analysis, as one could have expected. On that respect, one may realize of the appearance of a new set of intermediate values for the scalar mass, $\mH \in [0.4; 1.2]\ \rm{TeV}$, when increasing our perturbativity upper bound. It can be easily understood by tracing an horizontal line at $\nrlm_0 = -100$ GeV: we eliminate the blue region ``bridge" connecting the low and high mass solutions. Therefore, this new range of scalar masses requires large values of $|\nrl{\mu}|$.

\begin{figure}[H]
\begin{center}
\subfloat[$\nrlm$ vs. $\mH$.\label{sfig:nmu:mH:Cs250}]{\includegraphics[width=0.3\textwidth]{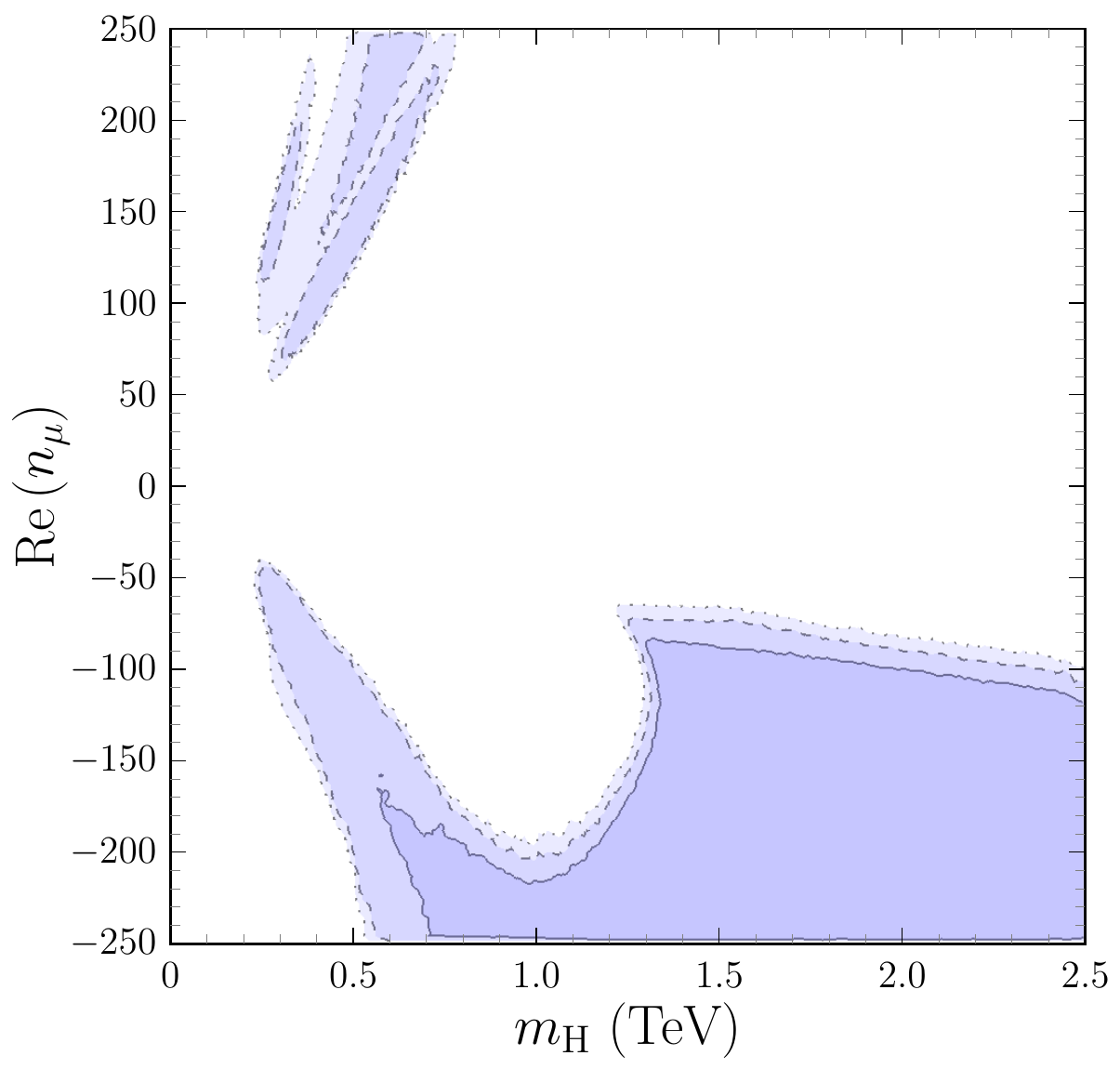}}\quad 
\subfloat[$\mH$ vs. $\tb$.\label{sfig:mH:tb:Cs250}]{\includegraphics[width=0.3\textwidth]{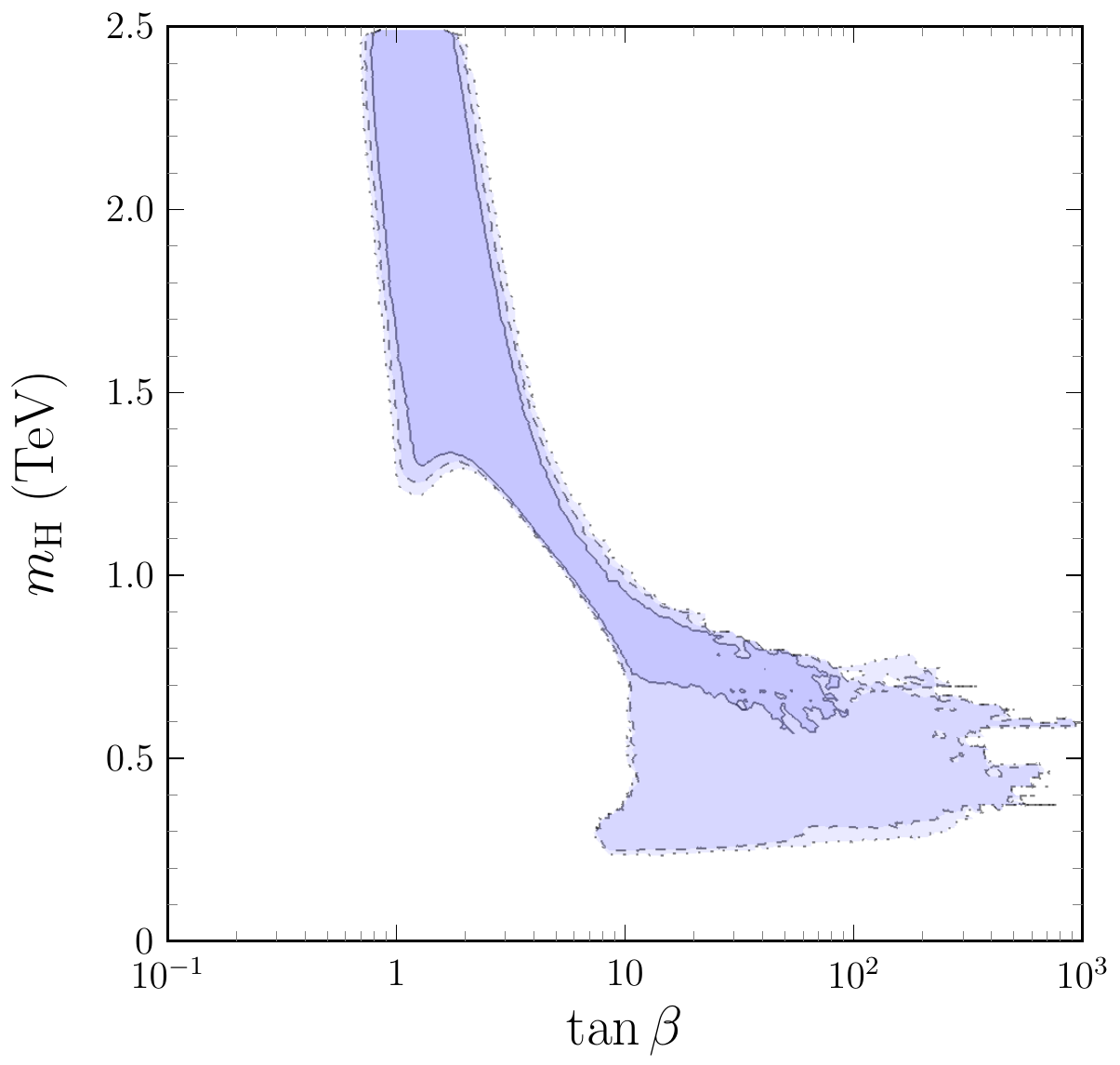}}\quad 
\subfloat[$\mH$ vs. $\mcH$.\label{sfig:mH:mCH:Cs250}]{\includegraphics[width=0.3\textwidth]{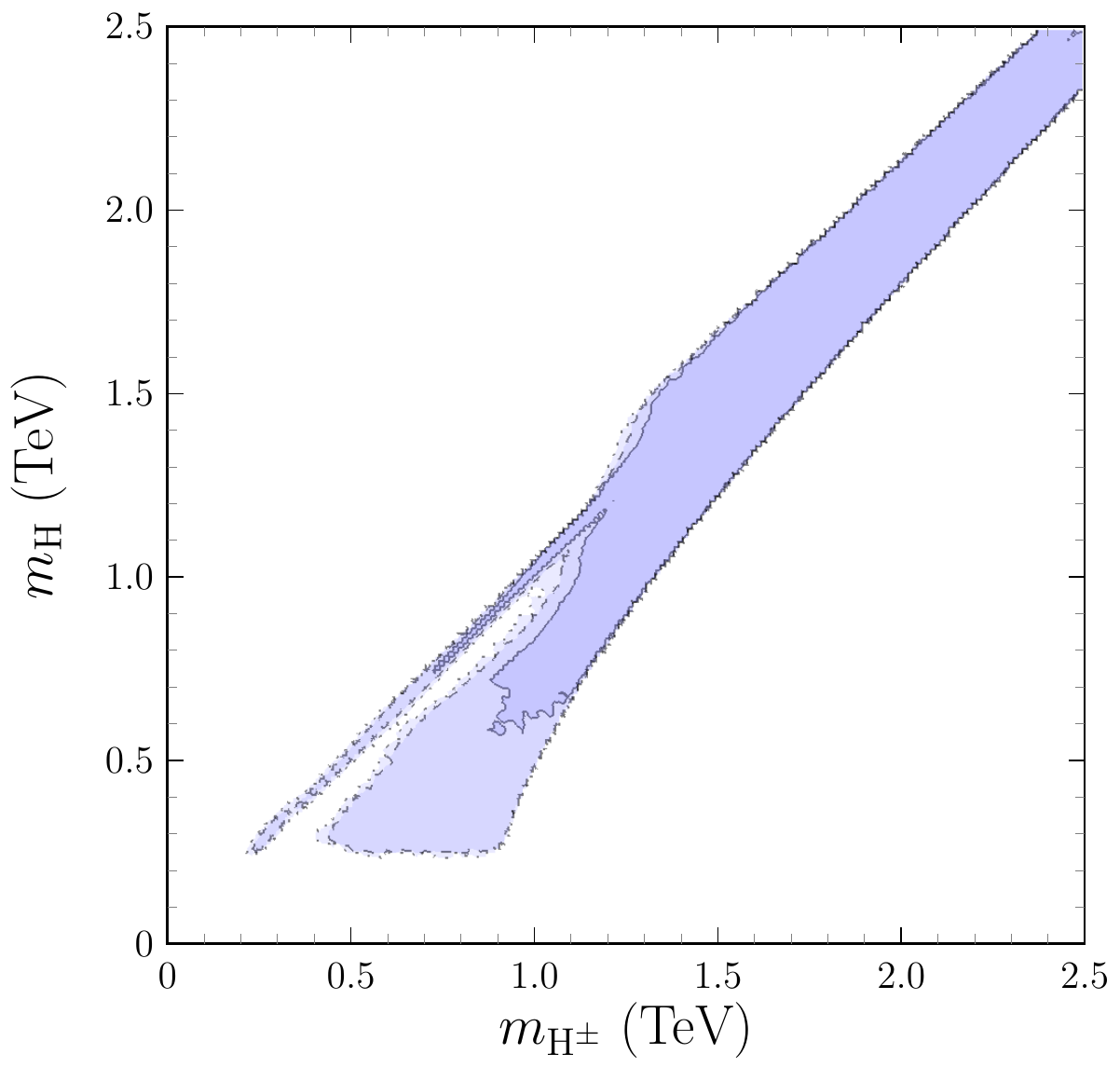}}
\caption{Allowed regions: relevant correlations involving $\mH$ with $|\nl{\ell}|\leq 250$ GeV.\label{fig:01:Cs250}}
\end{center}
\end{figure}

To fully characterize the impact of perturbativity on the allowed parameter space, figure \ref{sfig:mH:tb:Cs250} illustrates the scalar mass $\mH$ in terms of $\tb$. Taking into account the appearance of new intermediate solutions in $\mH$, one could expect that this behavior is translated into $\tb$. As figure \ref{sfig:mH:tb:Cs250} corroborates, new values of $\tb$, roughly in the range $1 \leq \tb \leq 10$, are allowed when changing the perturbativity requirement from $|\nl{\ell}| \leq 100$ GeV to $|\nl{\ell}| \leq 250$ GeV. Furthermore, one may also notice by comparing with figure \ref{sfig:mH:tb:Cs100} that the top blue region for large $\mH$ becomes wider, around a factor 2.5 in $\tb$ for each value of the scalar mass. 

Figure \ref{sfig:mH:mCH:Cs250} shows correlations among the scalar masses $\mH$ and $\mcH$. Concerning the low mass regions where $\cH$ is degenerate either with $\nH$ or $\nA$, already mentioned in figure \ref{sfig:mH:mCH:Cs100}, it can be observed that enlarging perturbativity bounds pushes the upper limit of these regions in such a way that $\mcH \in [0.2; 1.2]$ TeV for $\mcH \simeq \mH$ and $\mcH \in [0.4; 1.2]$ TeV for $\mcH \simeq \mA$, to a high degree of accuracy. Figures \ref{sfig:mA:mH:Cs250} and \ref{sfig:mA:mCH:Cs250} complete the results for the scalar masses. For instance, it is still true that $\mA > \mH$ in the low mass region, according to the general constraints presented in section \ref{Sec:General}.

\begin{figure}[H]
\begin{center}
\subfloat[$\mA$ vs. $\mH$.\label{sfig:mA:mH:Cs250}]{\includegraphics[width=0.3\textwidth]{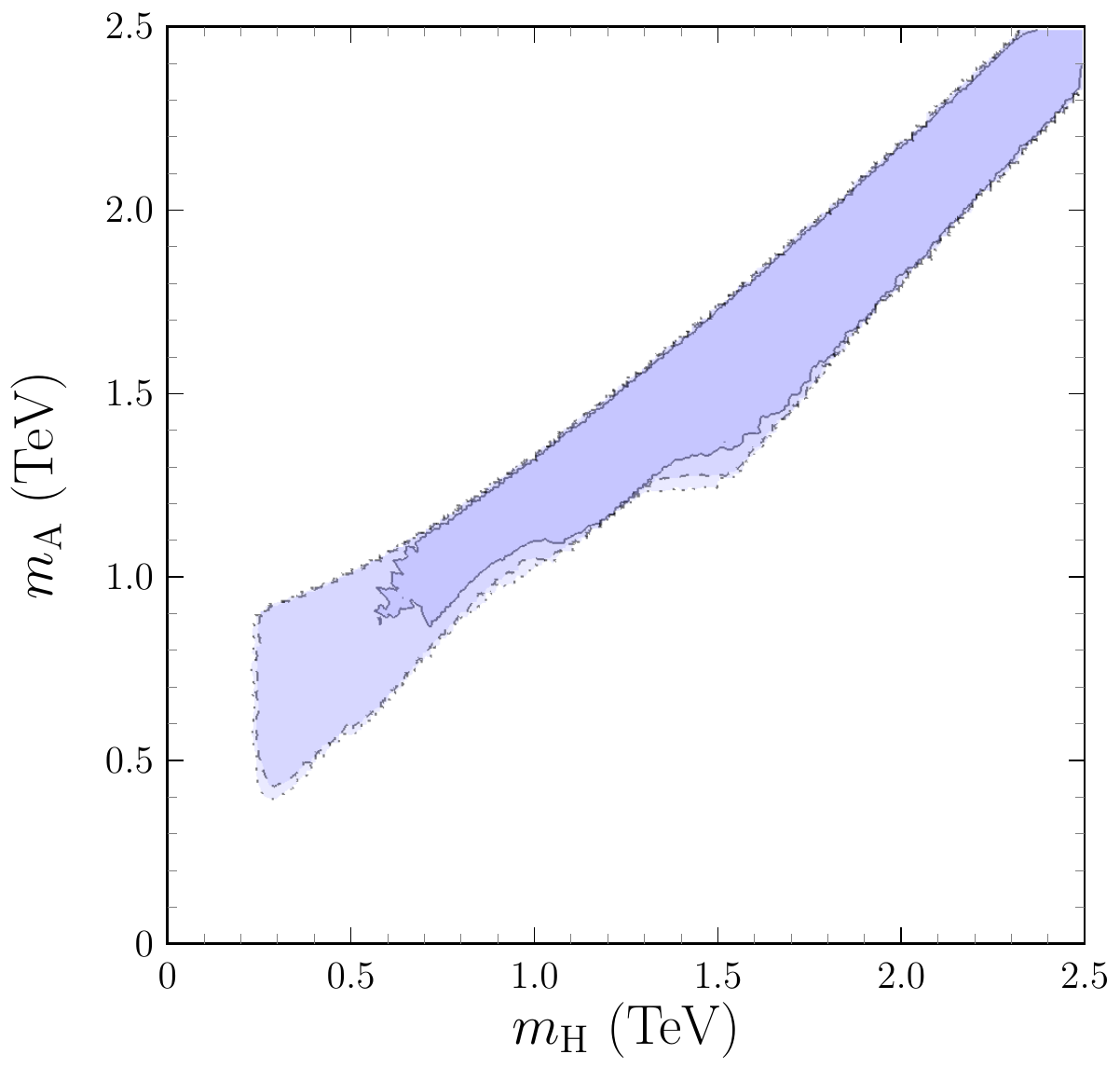}}\quad 
\subfloat[$\mA$ vs. $\mcH$.\label{sfig:mA:mCH:Cs250}]{\includegraphics[width=0.3\textwidth]{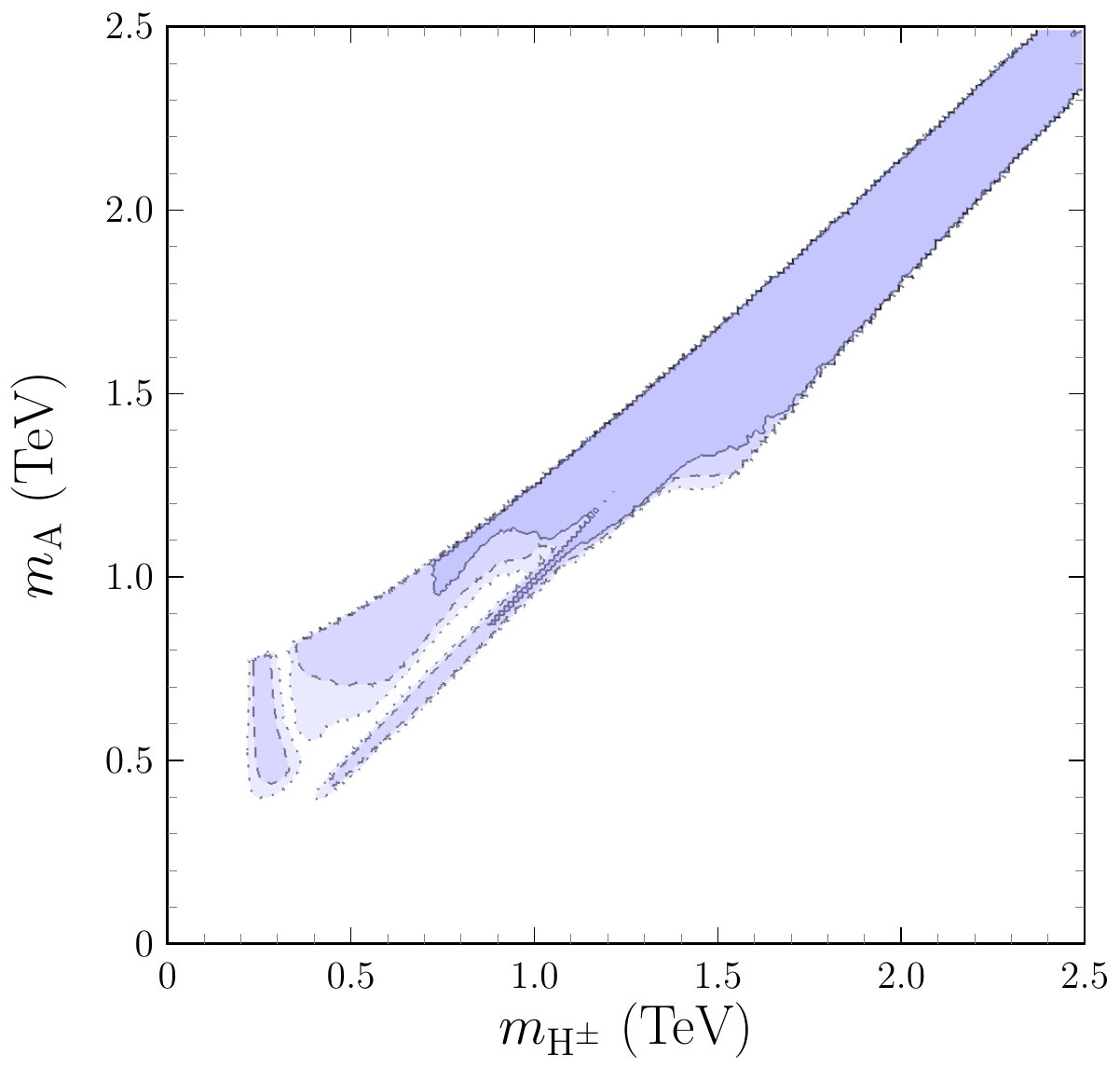}}
\caption{Allowed regions for scalar masses with $|\nl{\ell}|\leq 250$ GeV.\label{fig:03:Cs250}}
\end{center}
\end{figure}

On the other hand, figure \ref{fig:02:Cs250} shows the resonant process $[pp]_{\rm{ggF}} \to S \to \mu^{+}\mu^{-}$ as a function of the scalar mass $m_S$ for $S = \nH, \nA$, which acquire an important role since we may be entering an era of exclusion or discovery at the LHC. As disclosed above, the existence of an intermediate set of solutions, $\mH \in [0.4; 1.2]$ TeV and $\mA \in [0.9; 1.2]$ TeV, opens the possibility to detect a sizeable signal in that range of scalar masses that was not contemplated in figure \ref{fig:02:Cs100}. Moreover, it is clear that increasing $|\nl{\ell}|$ up to $|\nl{\ell}| \leq 250$ GeV modifies our expectations for $\BR{S\to\ell^{+}\ell^{-}}$ and, in particular, enlarges the allowed parameter space, as one can easily check.
\begin{figure}[h!tb]
\begin{center}
\subfloat[For $S=\nH$.\label{sfig:ppHmumu:mH:Cs250}]{\includegraphics[width=0.3\textwidth]{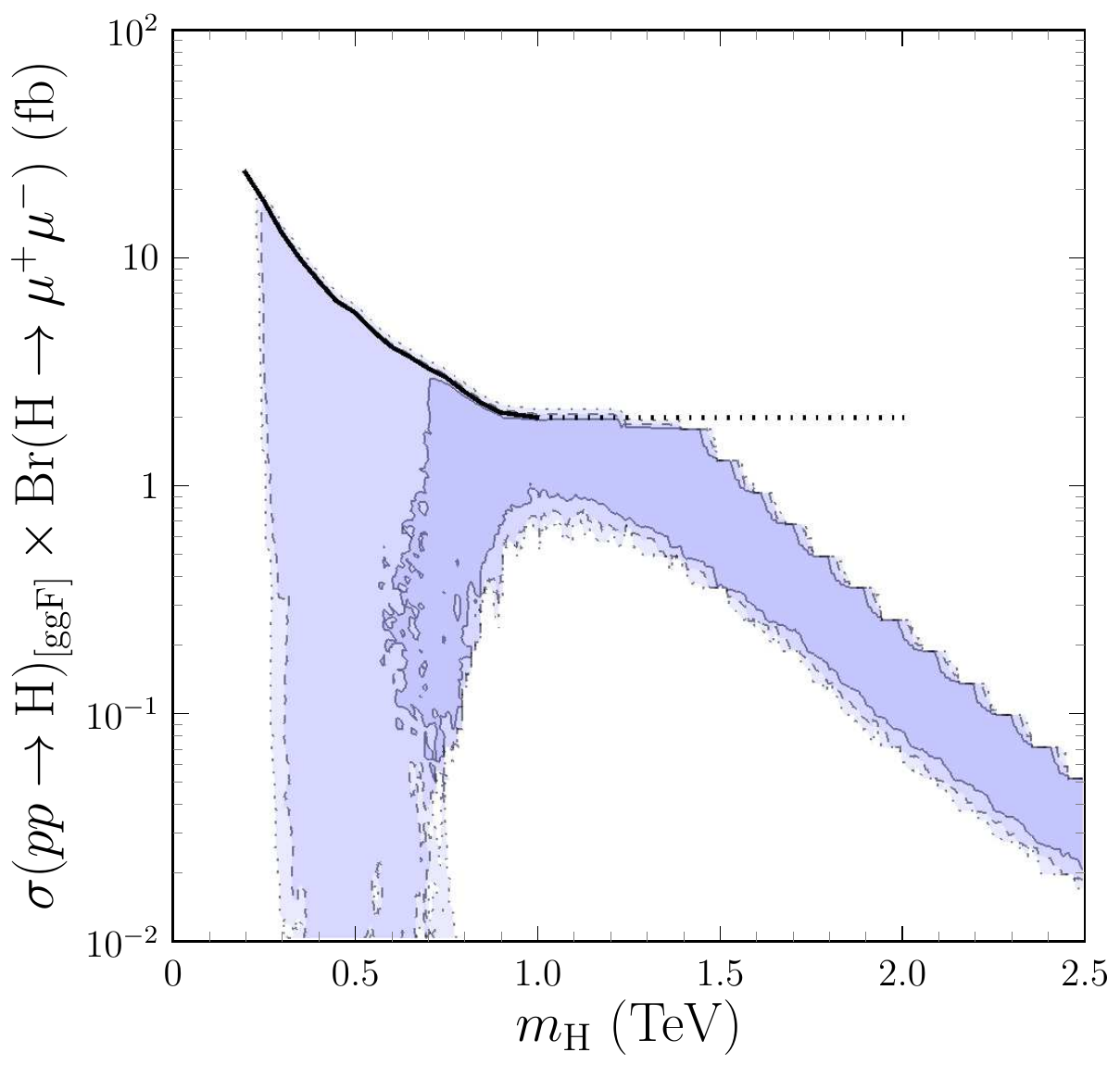}}\quad 
\subfloat[For $S=\nA$.\label{sfig:ppAmumu:mA:Cs250}]{\includegraphics[width=0.3\textwidth]{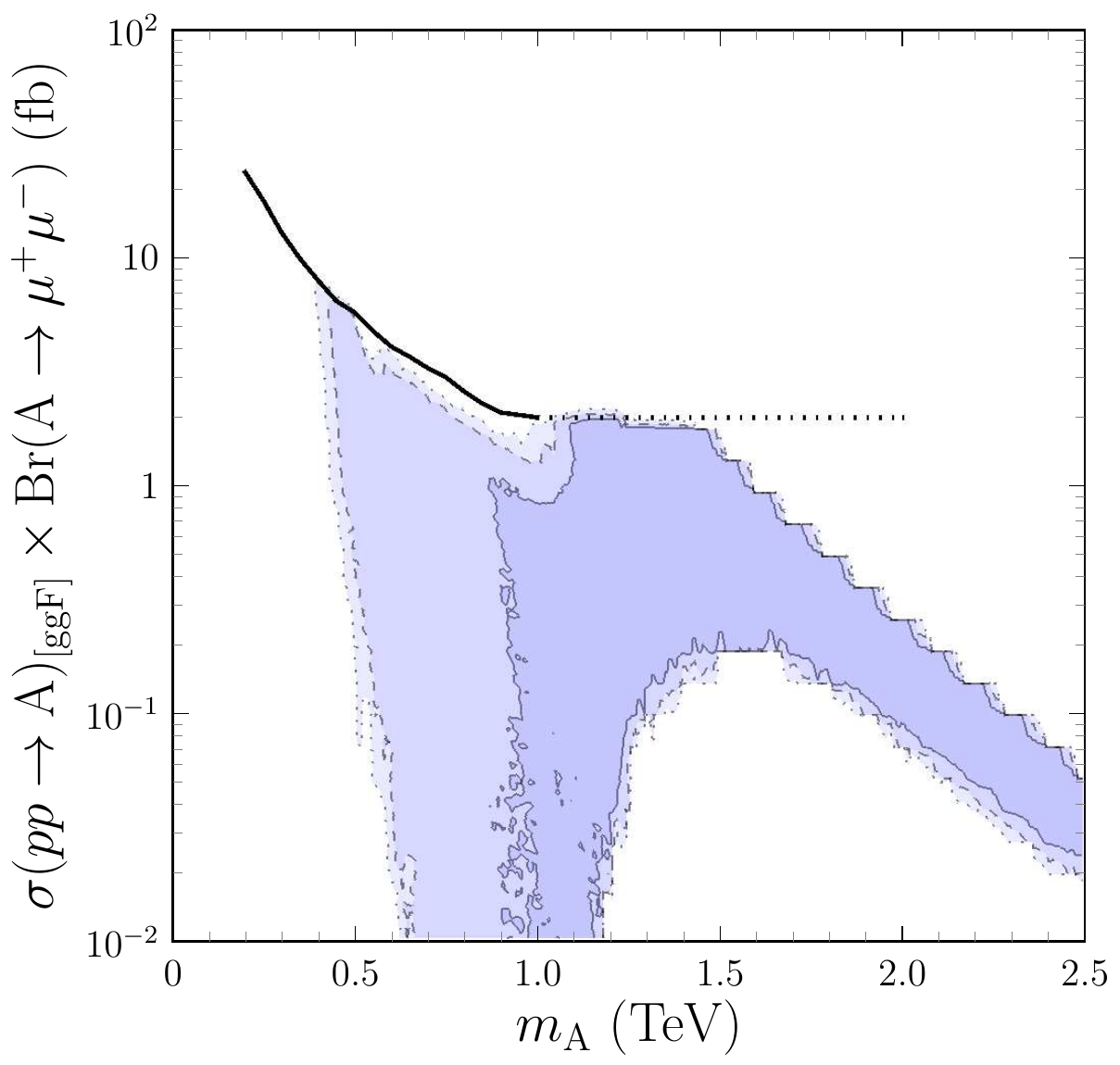}}
\caption{$\sigma(pp\to S)_{\rm [ggF]}\times\BR{S\to\mu^+\mu^-}$ vs. $m_S$ allowed regions with $|\nl{\ell}|\leq 250$ GeV.\label{fig:02:Cs250}}
\end{center}
\end{figure}

Finally, we should stress some aspects concerning $\nrle$ and $\nrlm$ from figure \ref{fig:04:Cs250}. In spite of increasing our perturbativity bound up to $|\nl{\ell}| \leq 250$ GeV, it still seems difficult to obtain a one loop explanation for the electron anomaly since it requires quite large couplings, namely $|\nl{e}| > 160$ GeV in the Cs case. Figure \ref{sfig:ne:mH:Cs250} shows that $|\nl{e}| < 150$ GeV in the relevant range of scalar masses, thus indicating that $\delta a_{e}^{\rm Exp,Cs}$ is mainly explained at two loops. This agrees with the discussion on universality constraints closing subsection \ref{sSec:Results:Cs100}.\\
As it was already explained in the discussion of figure \ref{sfig:ne:mH:Cs100}, now in figure \ref{sfig:ne:mH:Cs250} and for large scalar masses, one can easily check that the electron coupling is positive and lies in the range $\nrle \sim 4-20$ GeV. Furthermore, according to \refeq{eq:nmu:ne}, there exists a linear relation between $\nrlm$ and $\nrle$ for $\mH > 1.2$ TeV, which implies that they have opposite sign in the Cs case and therefore $\nrlm$ should be negative in this region. The region $\nrlm=-13\nrle$ can be seen in the lower part of figure \ref{sfig:nmu:ne:Cs250} inside the 1$\sigma$ region as it should. Departure from this straight line introduces an important one loop contribution to the muon anomaly lowering also the scalar masses ranges. 
On the other hand, for light scalar masses, $\nrle$ might be either positive or negative by the same arguments discussed in section \ref{sSec:Results:Cs100}. It is also important to recall that, in this low mass region, $\Delta_\mu$ receives dominant one loop contributions and thus $\nrlm$ could naturally appear with both signs. From figure \ref{sfig:nmu:ne:Cs250}, one may notice as well that $|\nl{\mu}|$ is in general larger than $|\nl{e}|$ in the whole parameter space.
\begin{figure}[h!tb]
\begin{center}
\subfloat[$\nrlm$ vs. $\nrle$.\label{sfig:nmu:ne:Cs250}]{\includegraphics[width=0.3\textwidth]{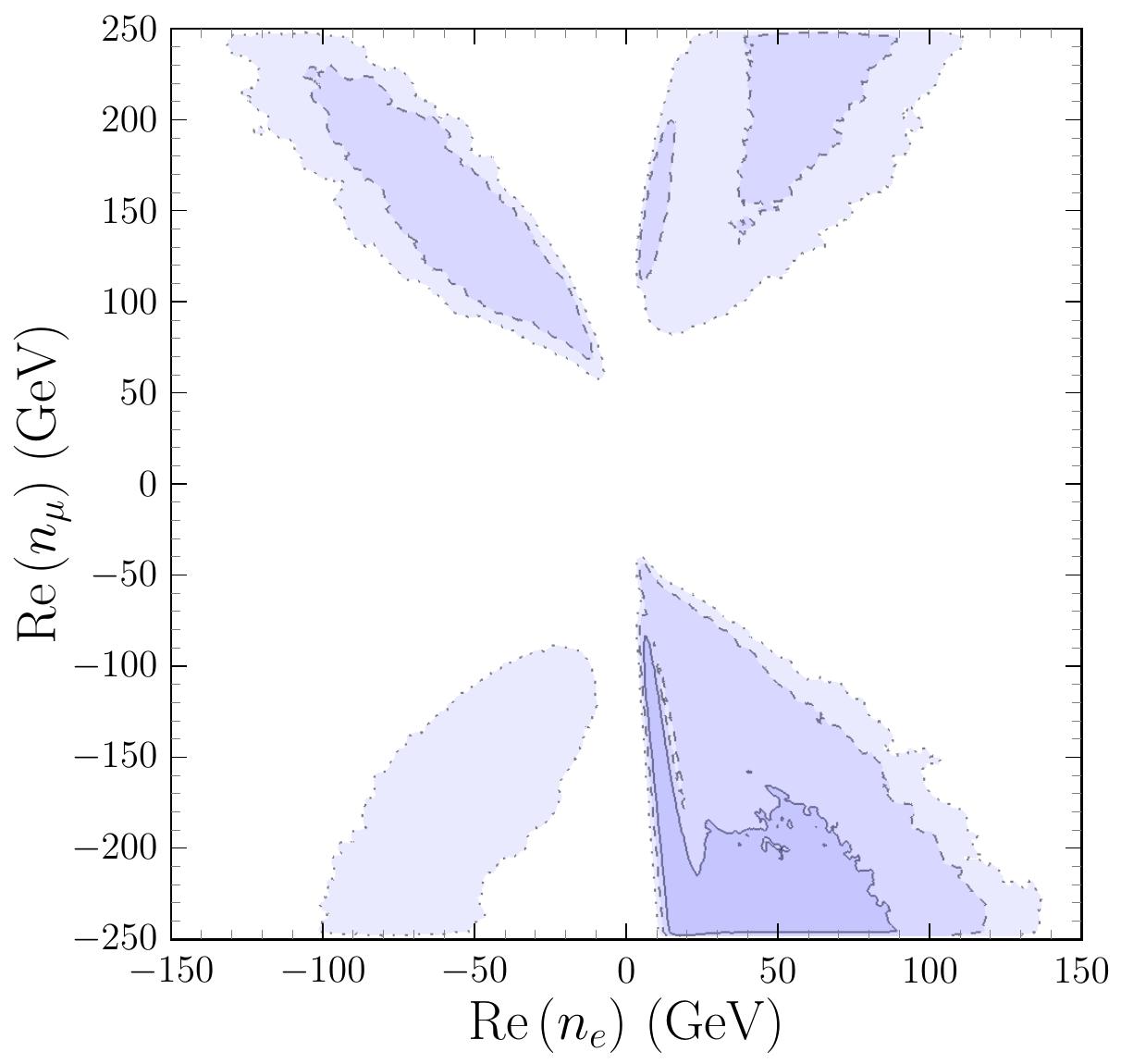}}\quad 
\subfloat[$\nrle$ vs. $\mH$.\label{sfig:ne:mH:Cs250}]{\includegraphics[width=0.3\textwidth]{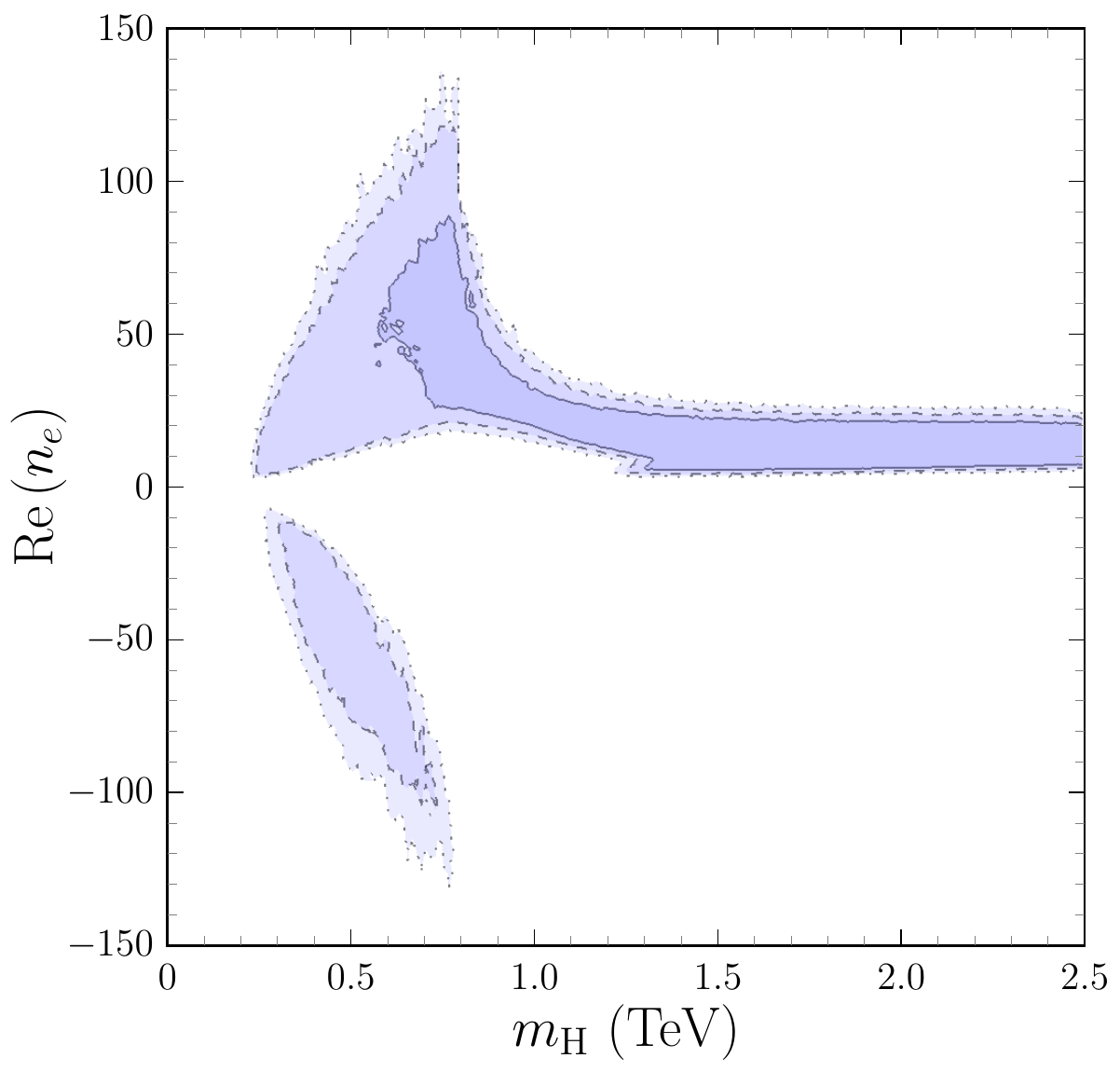}}
\caption{Allowed regions for $\nrle$ with $|\nl{\ell}|\leq 250$ GeV.\label{fig:04:Cs250}}
\end{center}
\end{figure}

\subsection{Different $\delta a_e$}\label{sSec:Results:diff:dae}
As commented in section \ref{Sec:Analysis}, the situation concerning $a_e$ is to some extent unclear. In this section we discuss the implications of different assumptions for the value of $\delta a_e^{\rm Exp}$, that is, in terms of the model, the implications of requiring different values of the new contributions $\delta a_e$. The ultimate answer is definitely provided by repeating detailed numerical analyses under the different assumptions $\delta a_e^{\rm Exp}$. However, one can anticipate part of the answer with simple considerations.
As analysed in section \ref{Sec:dal}, $\delta a_e$ arises from two loop contributions proportional to $\nrle$: this fact, together with the results of section \ref{sSec:Results:Cs250} corresponding to $\delta a_e\simeq\delta a_e^{\rm Exp,Cs}$, can give us a first insight. Consider for example an allowed point in parameter space (i.e. a point respecting all imposed constraints) which gives $\delta a_e\simeq\delta a_e^{\rm Exp,Cs}$. This point has a certain $\nrle=\nrle_{\rm Cs}$; it is straightforward that changing $\nrle\mapsto\nrle'=\nrle_{\rm Rb}=\nrle_{\rm Cs}\times\frac{\delta a_e^{\rm Exp,Rb}}{\delta a_e^{\rm Exp,Cs}}$ and no other parameter, one would obtain $\delta a_e\simeq \delta a_e^{\rm Exp,Rb}$. The question is, of course, if such a change in $\nrle$ alone still gives an allowed point. On that respect, one needs to analyse which observables constrain $\nrle$ and how those constraints work. These are the ones related to lepton flavor universality in leptonic decays $\ell_i\to\ell_j\nu\bar\nu$ and in semileptonic decays involving kaons and pions, analysed in section \ref{sSec:Results:Cs100}. In particular, attending to $\delta a_e^{\rm Exp,Cs}$, $\delta a_e^{\rm Exp,Rb}$ and $\delta a_e^{\rm Exp,Avg}$ in \refeqs{eq:dae:ExpCs}, \eqref{eq:dae:ExpRb} and \eqref{eq:dae:ExpAvg}, one is interested in the effect on those constraints of
\begin{equation}\label{eq:ne:Rb:Avg}
\begin{aligned}
 &\nrle=\nrle_{\rm Cs}\mapsto \nrle'=\nrle_{\rm Rb}=-0.55\nrle_{\rm Cs}\,,\\
 &\nrle=\nrle_{\rm Cs}\mapsto \nrle'=\nrle_{\rm Avg}=0.23\nrle_{\rm Cs}\,,
\end{aligned}
\end{equation}
when no other parameter is changed. There are two different aspects:
\begin{enumerate}
 \item since $|\nrle_{\rm Avg}|,|\nrle_{\rm Rb}|<|\nrle_{\rm Cs}|$, the constraint on $|\nl{e}|$ from $\mu\to e\nu\bar\nu$ decays in \refeq{eq:muenunu:02} is necessarily less restrictive when \refeq{eq:ne:Rb:Avg} is considered;
 \item besides the uncertainty in $R^K_{\mu e}$ in \refeq{eq:RPmue:exp}, as discussed previously, there is a ``sign'' question concerning the deviation, at the same $\simeq 5\times 10^{-3}$ level of the uncertainty, from $R^K_{\mu e}=1$. In order to obtain $\delta a_e\simeq\delta a_e^{\rm Exp,Cs}<0$, the expectation is $\nrle>0$, and that produces $R^K_{\mu e}-1>0$ in \refeq{eq:RPmue:02}, which goes ``in the wrong direction''. For both cases in \refeq{eq:ne:Rb:Avg}, that problem is alleviated.
\end{enumerate}
It is then clear that the analysis with $\delta a_e^{\rm Exp,Cs}$ is somehow a ``worst case'' scenario in terms of the dependence of the constraints on $\nrle$: besides the naive mapping of allowed regions expected from \refeq{eq:ne:Rb:Avg}, one might then expect larger allowed regions not only for $\nrle$ but also for other quantities of interest. As mentioned in section \ref{Sec:Analysis}, we also perform an analysis where $|\delta a_e|\leq 20\times 10^{-13}$ is imposed (instead of requiring some specific value, as summarized in figure \ref{fig:damu:dae}). This serves a double purpose: identifying which allowed regions are necessary in order to obtain an appropriate $\delta a_\mu$ without regard to $\delta a_e$, and identifying which regions are absolutely excluded for any value of $\delta a_e$ reasonably compatible with $\delta a_e^{\rm Exp,Cs}$ or $\delta a_e^{\rm Exp,Rb}$ that one could consider.\\
In figures \ref{fig:ae:01:RbAvgBound}, \ref{fig:ae:02:RbAvgBound} and \ref{fig:ae:01:Bound}, the color coding follows figure \ref{fig:damu:dae}. 
Figure \ref{fig:ae:01:RbAvgBound} shows $\nrlm$ vs. $\nrle$ and $\nrle$ vs. $\mH$ allowed regions: comparison with figures \ref{sfig:nmu:ne:Cs250} and \ref{sfig:ne:mH:Cs250} confirms the simple expectations of the previous discussion in terms of the position of the allowed regions and their extension. The same applies to figure \ref{fig:ae:02:RbAvgBound}, which shows $\nrlm$ vs. $\mH$ (to be compared with figure \ref{sfig:nmu:mH:Cs250}). In particular it is clear from figure \ref{sfig:nmu:mH:Bound250} that once $\delta a_\mu\simeq\delta a_\mu^{\rm Exp}$ is imposed, the allowed regions for some parameters (besides $\nrlm$, obviously) are coarsely determined and the sensitivity of the analysis on the requirement for $\delta a_e$ only concerns a finer level of detail.
\begin{figure}[H]
\begin{center}
\subfloat[\label{sfig:nmu:ne:Rb250}]{\includegraphics[width=0.3\textwidth]{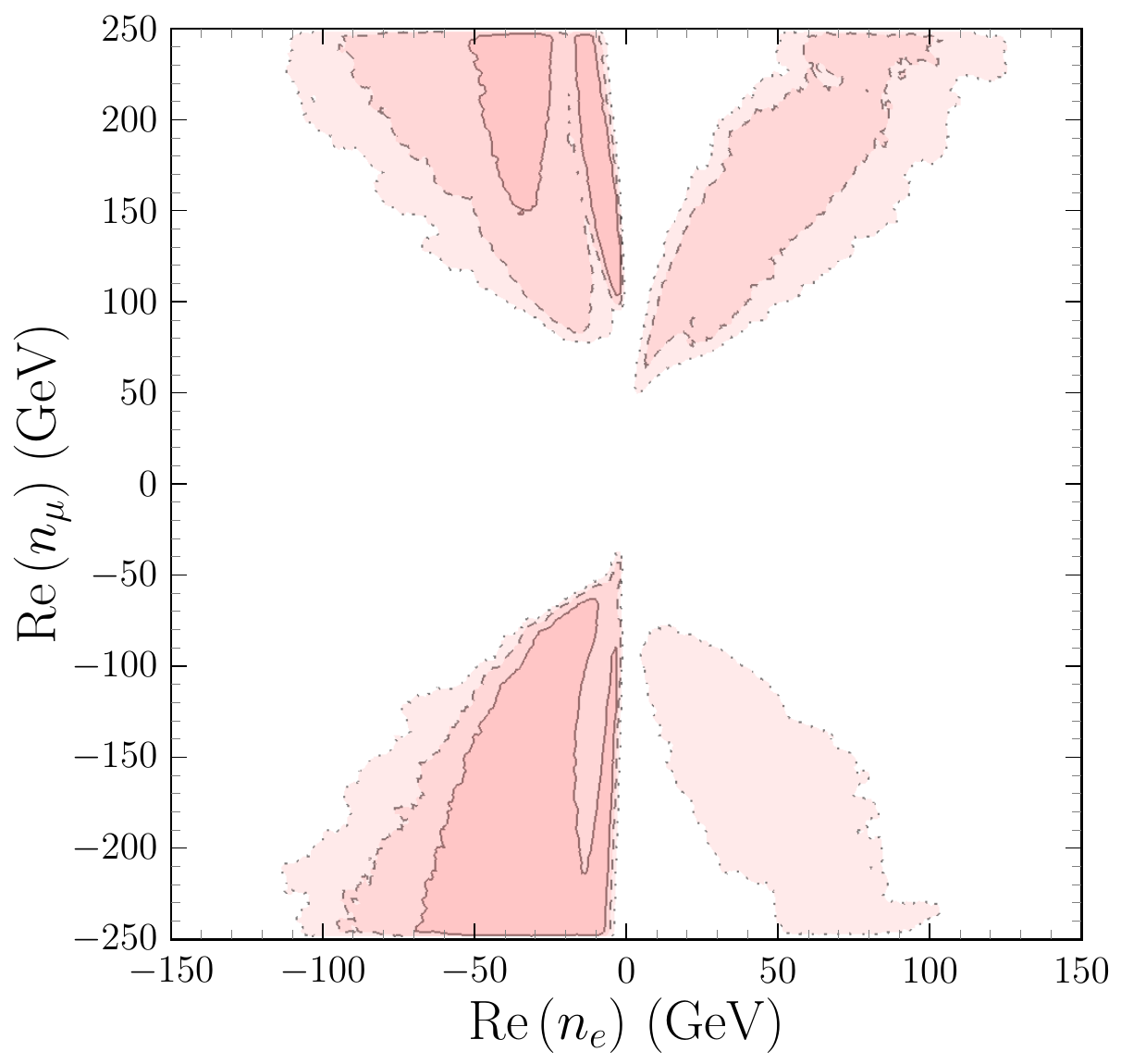}}\quad 
\subfloat[\label{sfig:nmu:ne:Avg250}]{\includegraphics[width=0.3\textwidth]{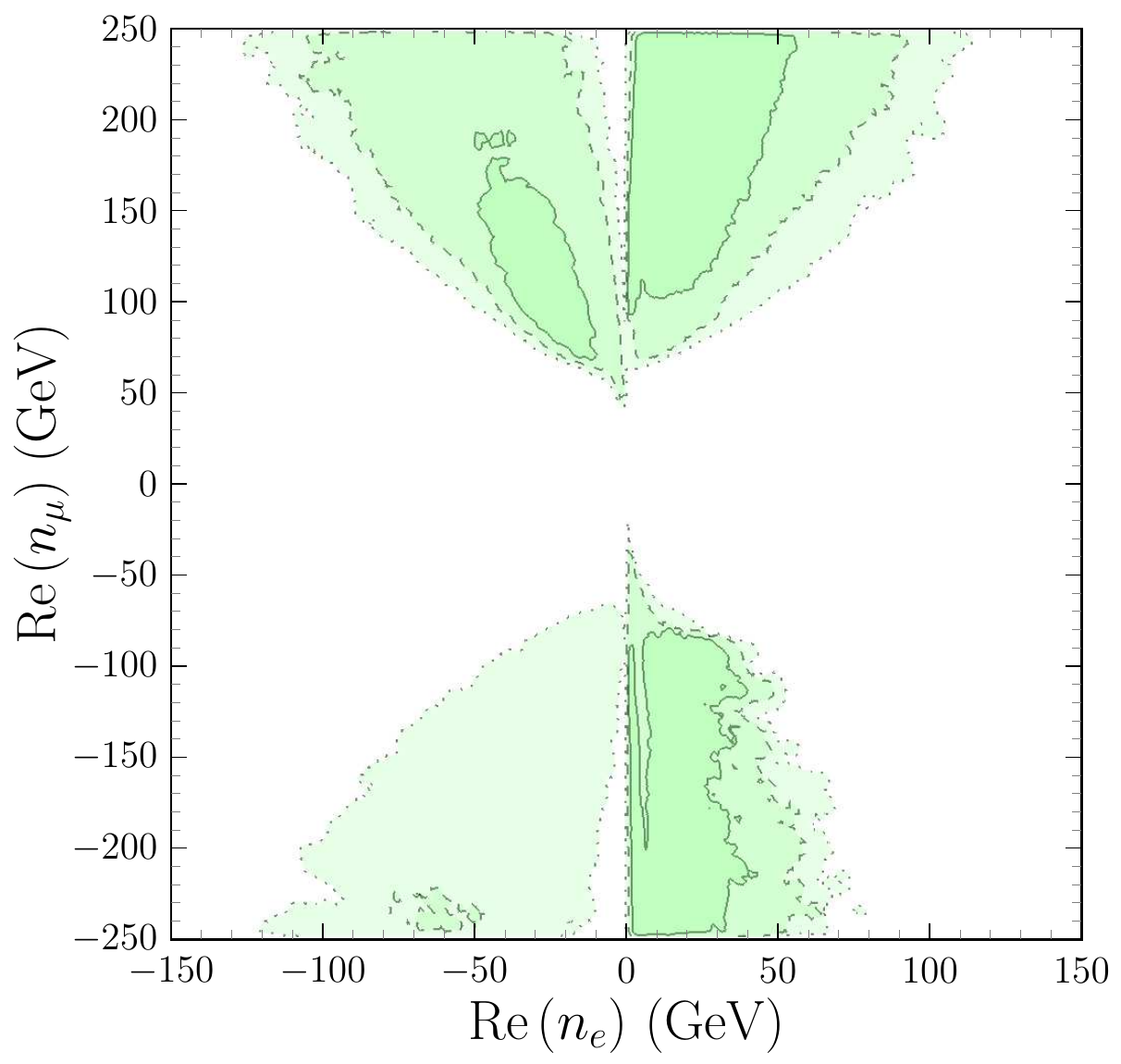}}\quad 
\subfloat[\label{sfig:nmu:ne:Bound250}]{\includegraphics[width=0.3\textwidth]{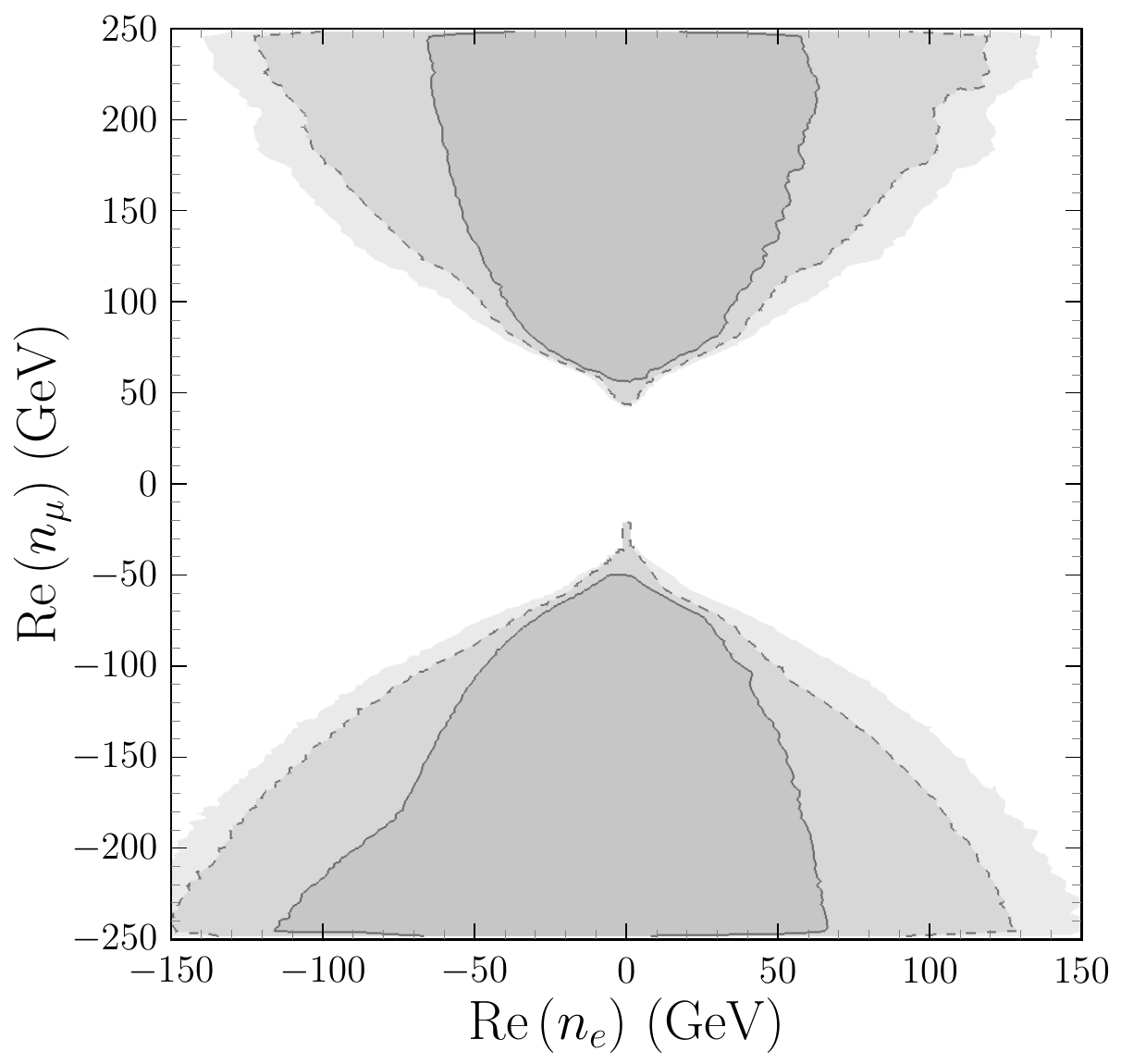}}\\ 
\subfloat[\label{sfig:ne:mH:Rb250}]{\includegraphics[width=0.3\textwidth]{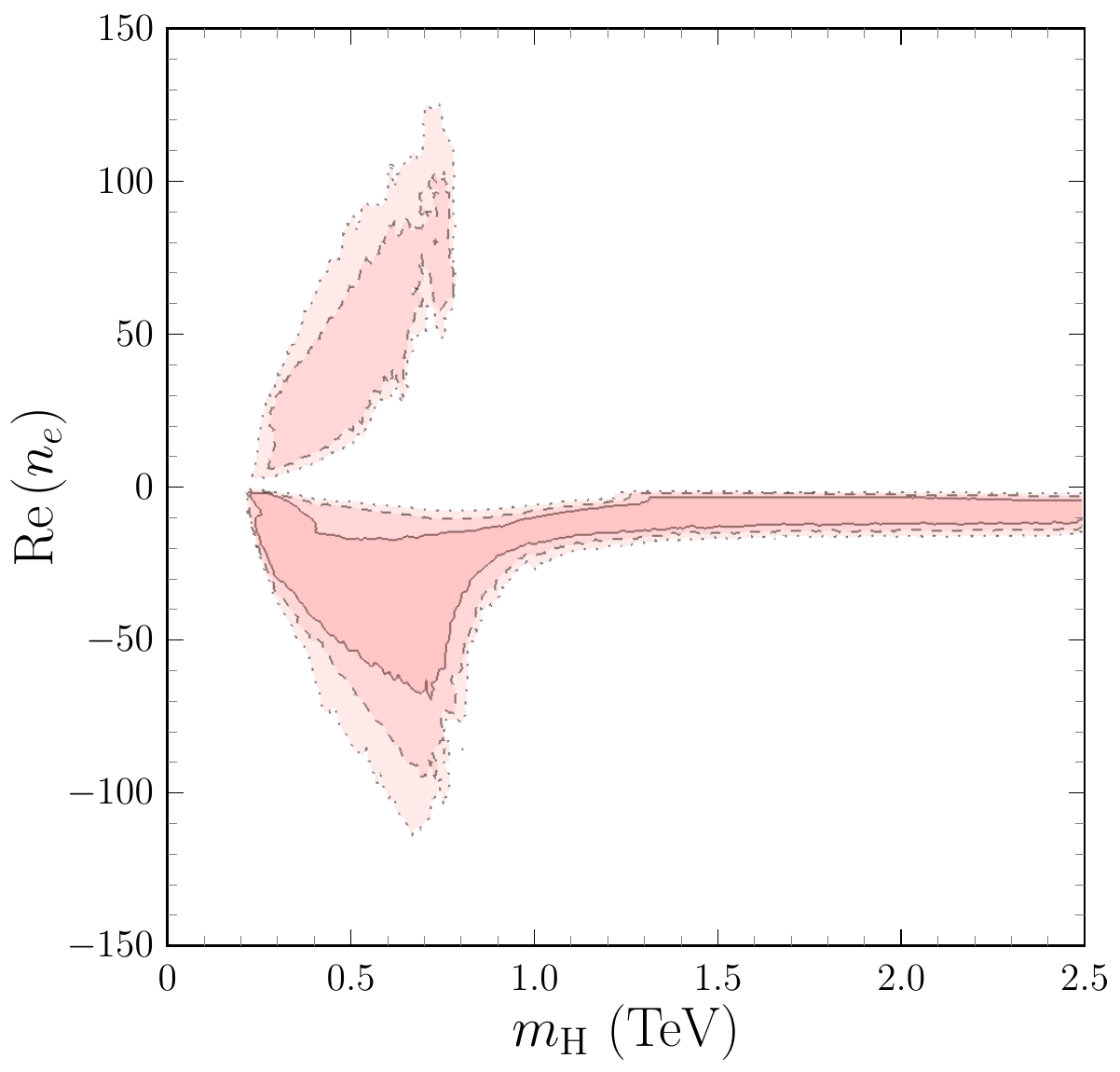}}\quad 
\subfloat[\label{sfig:ne:mH:Avg250}]{\includegraphics[width=0.3\textwidth]{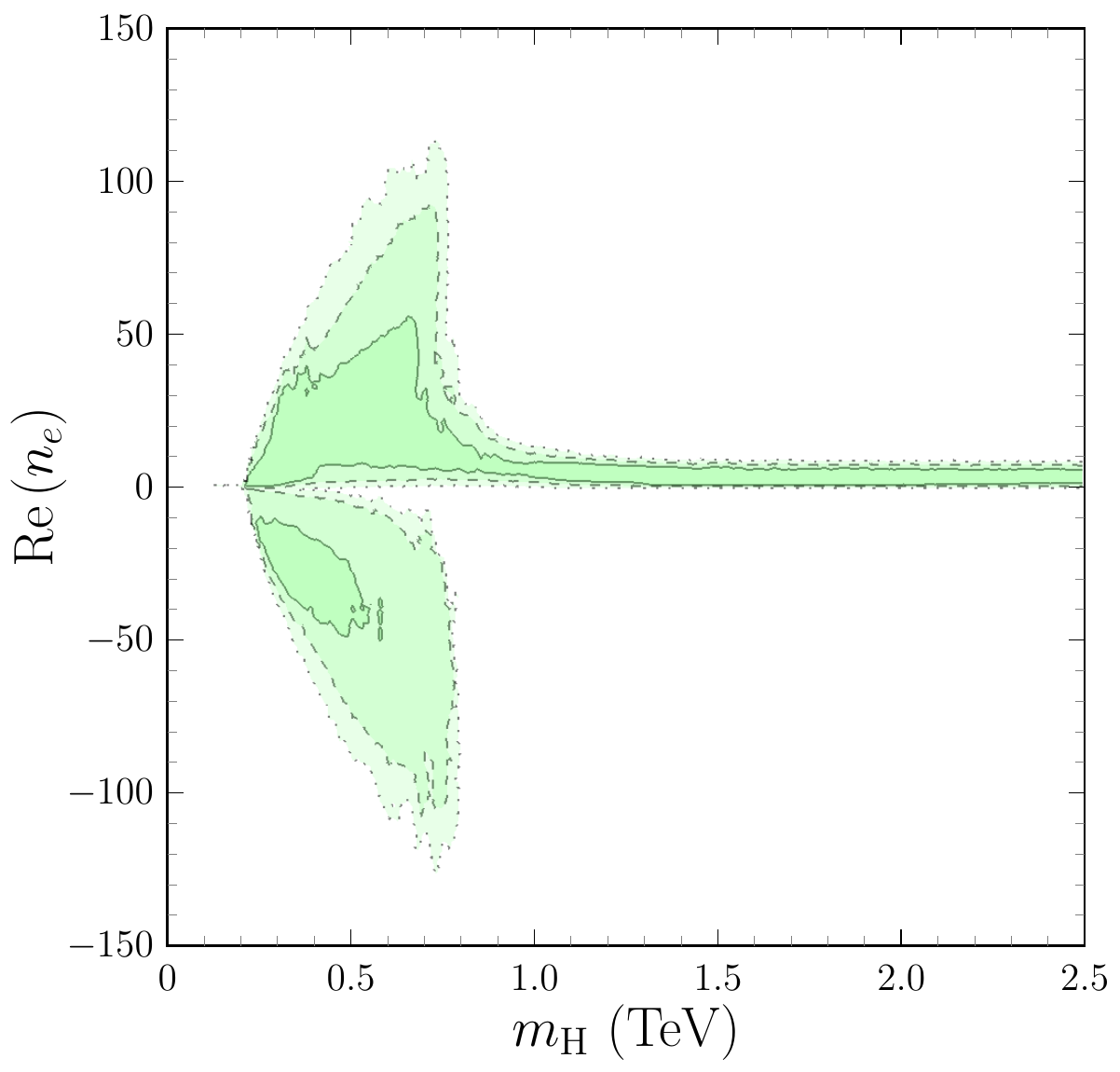}}\quad 
\subfloat[\label{sfig:ne:mH:Bound250}]{\includegraphics[width=0.3\textwidth]{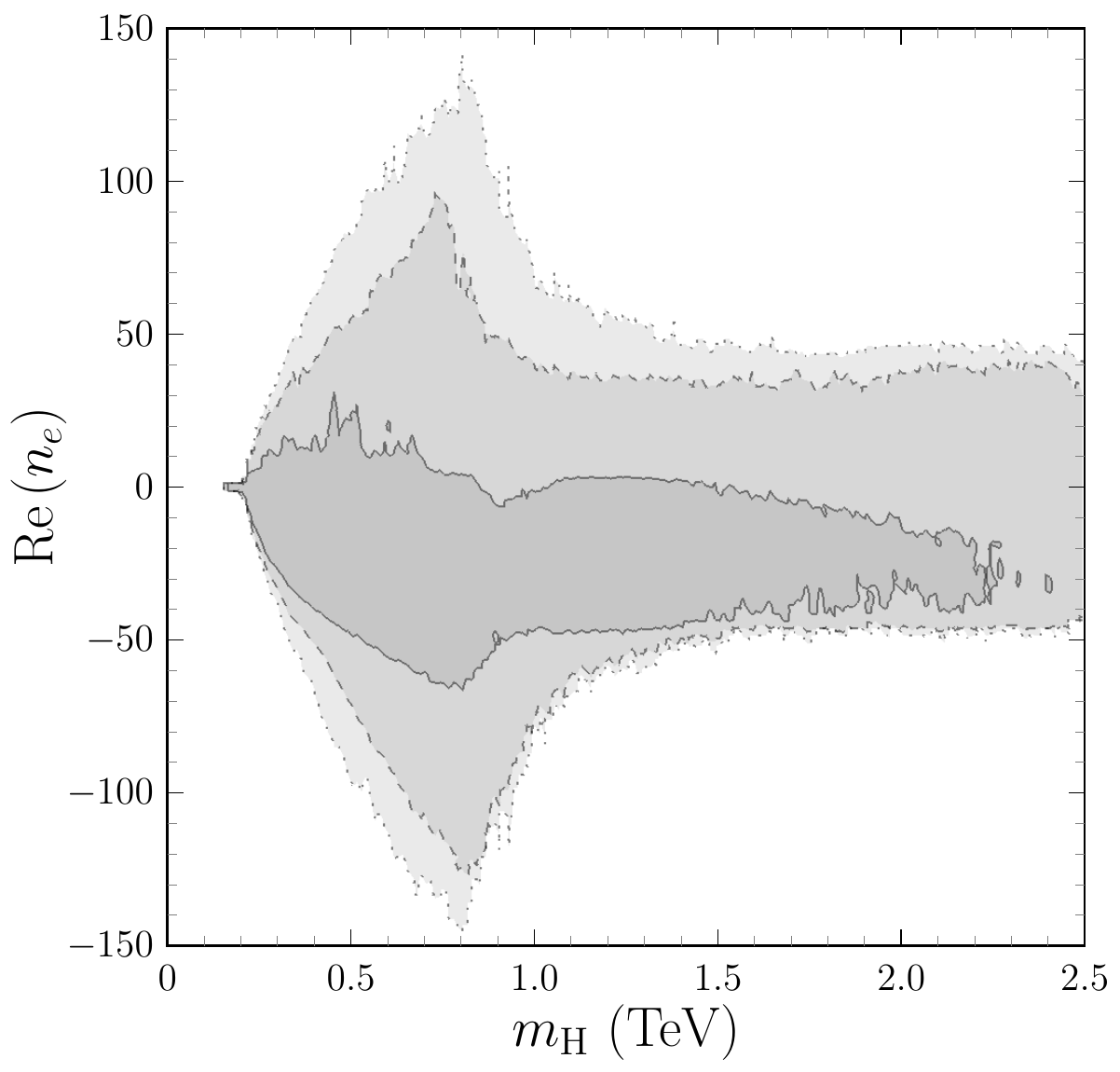}}
\caption{$\nrle$ correlations.\label{fig:ae:01:RbAvgBound}}
\end{center}
\end{figure}
\begin{figure}[H]
\begin{center}
\subfloat[\label{sfig:nmu:mH:Rb250}]{\includegraphics[width=0.3\textwidth]{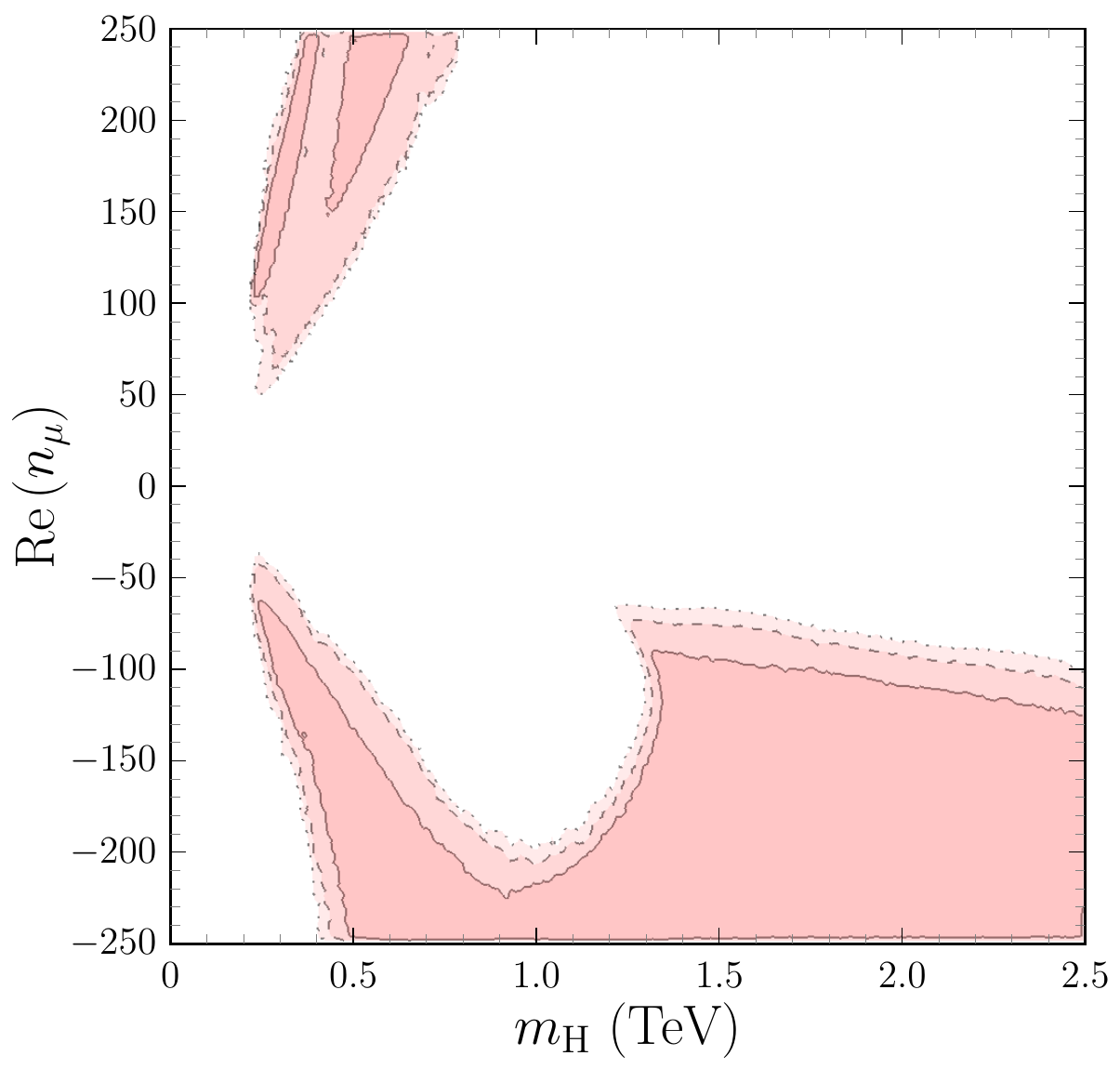}}\quad 
\subfloat[\label{sfig:nmu:mH:Avg250}]{\includegraphics[width=0.3\textwidth]{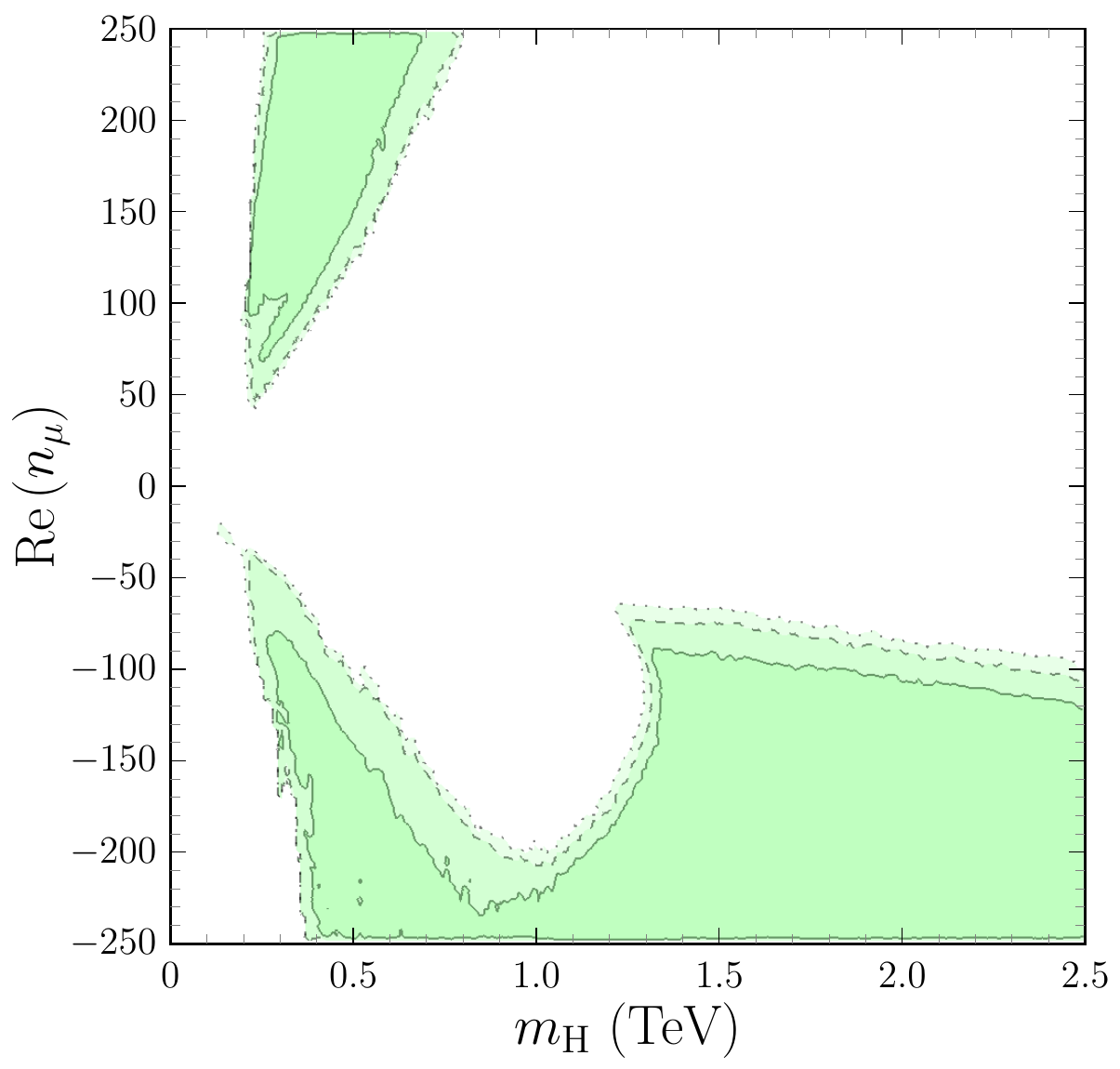}}\quad 
\subfloat[\label{sfig:nmu:mH:Bound250}]{\includegraphics[width=0.3\textwidth]{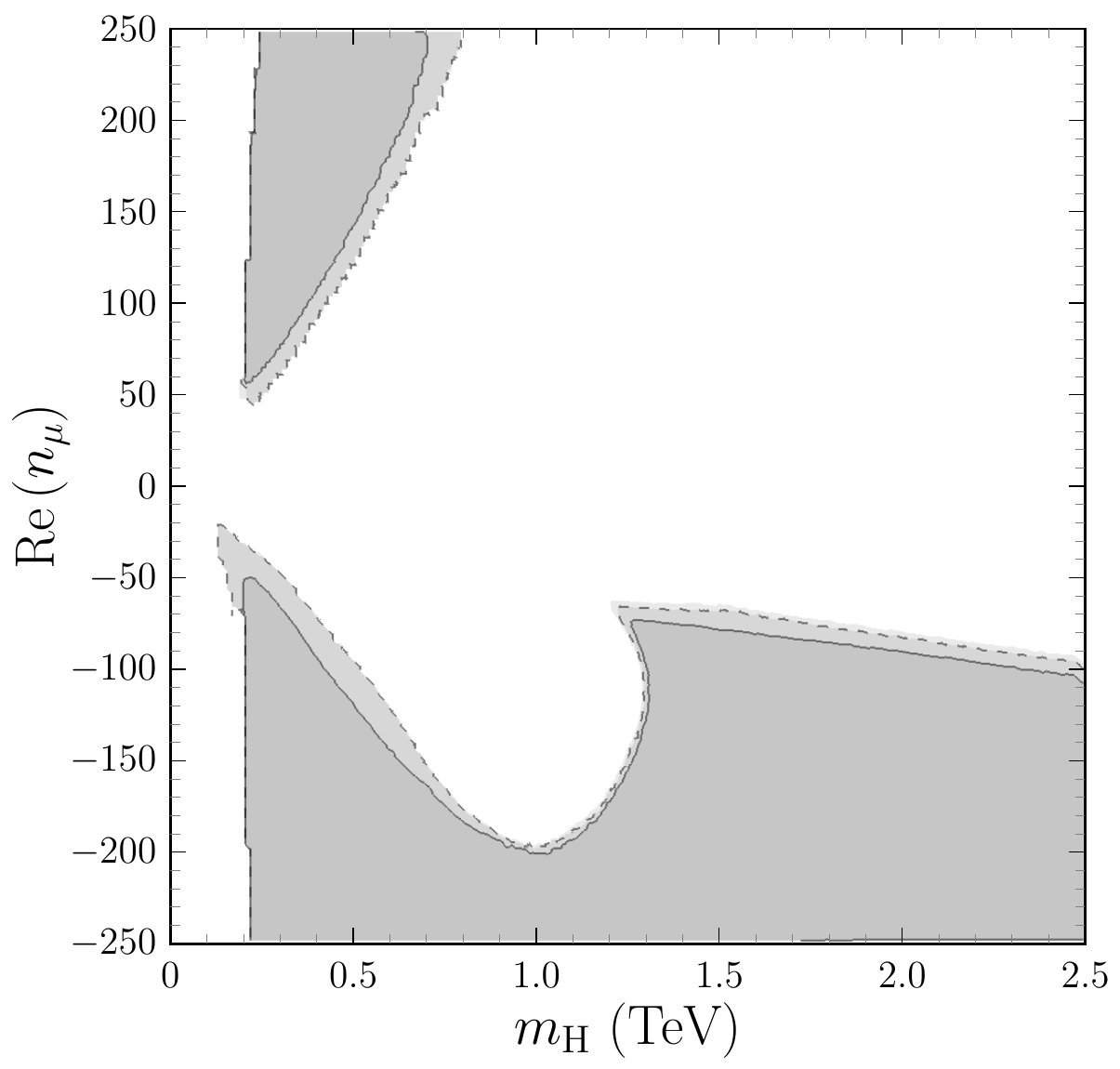}}
\caption{$\nrlm$ vs. $\mH$ correlations.\label{fig:ae:02:RbAvgBound}}
\end{center}
\end{figure}
There is a final point that the analysis with $|\delta a_e|\leq 20\times 10^{-13}$ confirms. Figure \ref{fig:ae:01:Bound} shows $\delta a_e$ vs. $\nrle$: under the simple expectations for the two loop contributions discussed in section \ref{sSec:dal:2loop}, one would have $\nrle\times \delta a_e<0$. Besides that expected region, one can observe smaller allowed regions where $\nrle\times \delta a_e>0$: they correspond to the unexpected situation in which the two loop contributions are dominated by virtual $\tau$'s in the fermion loop, and furthermore it is clear that the values of $\delta a_e$ that can be obtained in this manner are more restricted, with $|\delta a_e|<10^{-12}$.
\begin{figure}[h!tb]
\begin{center}
\includegraphics[width=0.3\textwidth]{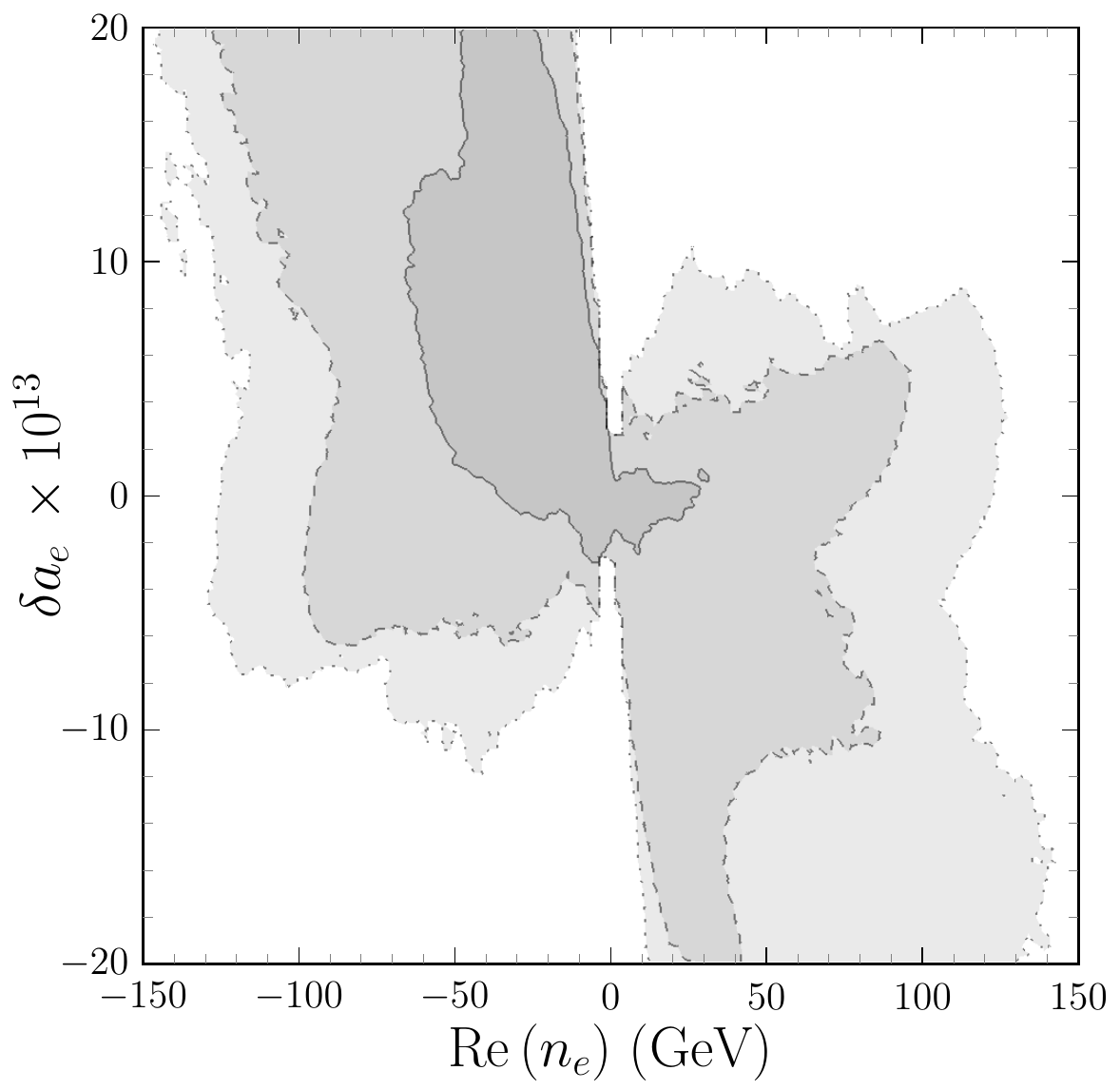}
\caption{$\delta a_e$ vs. $\nrle$.\label{fig:ae:01:Bound}}
\end{center}
\end{figure}

\subsection{The CDF $M_W$ anomaly}\label{sSec:Results:CDF}
As mentioned in section \ref{Sec:General}, one can use deviations from the SM in the oblique parameters $(\Delta S,\Delta T)\neq (0,0)$ in order to ``explain'' the CDF measurement of $M_W$ in \cite{CDF:2022hxs}: this subsection is devoted to that ``explanation''. Figures \ref{fig:Cs250W2} and \ref{fig:Cs250W1} show results analogous to the ones in section \ref{sSec:Results:Cs250} --which use \refeq{eq:obliqueval}--, except for a different $(\Delta S,\Delta T)$ constraint. Figure \ref{fig:Cs250W2} is obtained with \refeq{eq:MWST:cons} (the ``conservative'' average of \cite{deBlas:2022hdk}) and figure \ref{fig:Cs250W1} is obtained with \refeq{eq:MWST:nocons} (the results in \cite{Lu:2022bgw}). The coloring of the allowed regions corresponds, darker to lighter, to $1,2,3\sigma$ levels of a 2D-$\Delta\chi^2$. For $\Delta\chi^2$ we use the $\chi^2_{\rm Min}$ value of the analysis in section \ref{sSec:Results:Cs250} (that is, with the constraint in \refeq{eq:obliqueval} for $\Delta S$, $\Delta T$). A few comments are in order.
\begin{itemize}
 \item Besides the absence of degeneracies $\mcH\simeq\mH$ or $\mcH\simeq\mA$, masses of the new scalars larger than 2 TeV are more difficult to accommodate. This can be understood attending to the clash between the mass differences discussed in section \ref{Sec:General} that \refeqs{eq:MWST:cons} or \eqref{eq:MWST:nocons} require, and the need of near degenerate scalars that the perturbativity requirements on the scalar potential impose for new scalar masses much larger than $\vev{}$.
 \item Overall agreement with the imposed constraints is worse in several regions in figures \ref{fig:Cs250W2} and \ref{fig:Cs250W1} than it was in the analyses of section \ref{sSec:Results:Cs250} (figures \ref{fig:01:Cs250}, \ref{fig:03:Cs250} and \ref{fig:04:Cs250}). This is more dramatic in figure \ref{fig:Cs250W1}, where the agreement with constraints is worse than in figure \ref{fig:Cs250W2} to the point that several regions are beyond the represented contour levels.
\end{itemize}
Despite these changes, the main characteristics of the allowed regions discussed in the previous sections still apply and are clearly identified in both figures \ref{fig:Cs250W2} and \ref{fig:Cs250W1}.
\begin{figure}[H]
\begin{center}
\subfloat[$\mH$ vs. $\tb$.\label{sfig:mH:tb:Cs250W2}]{\includegraphics[width=0.23\textwidth]{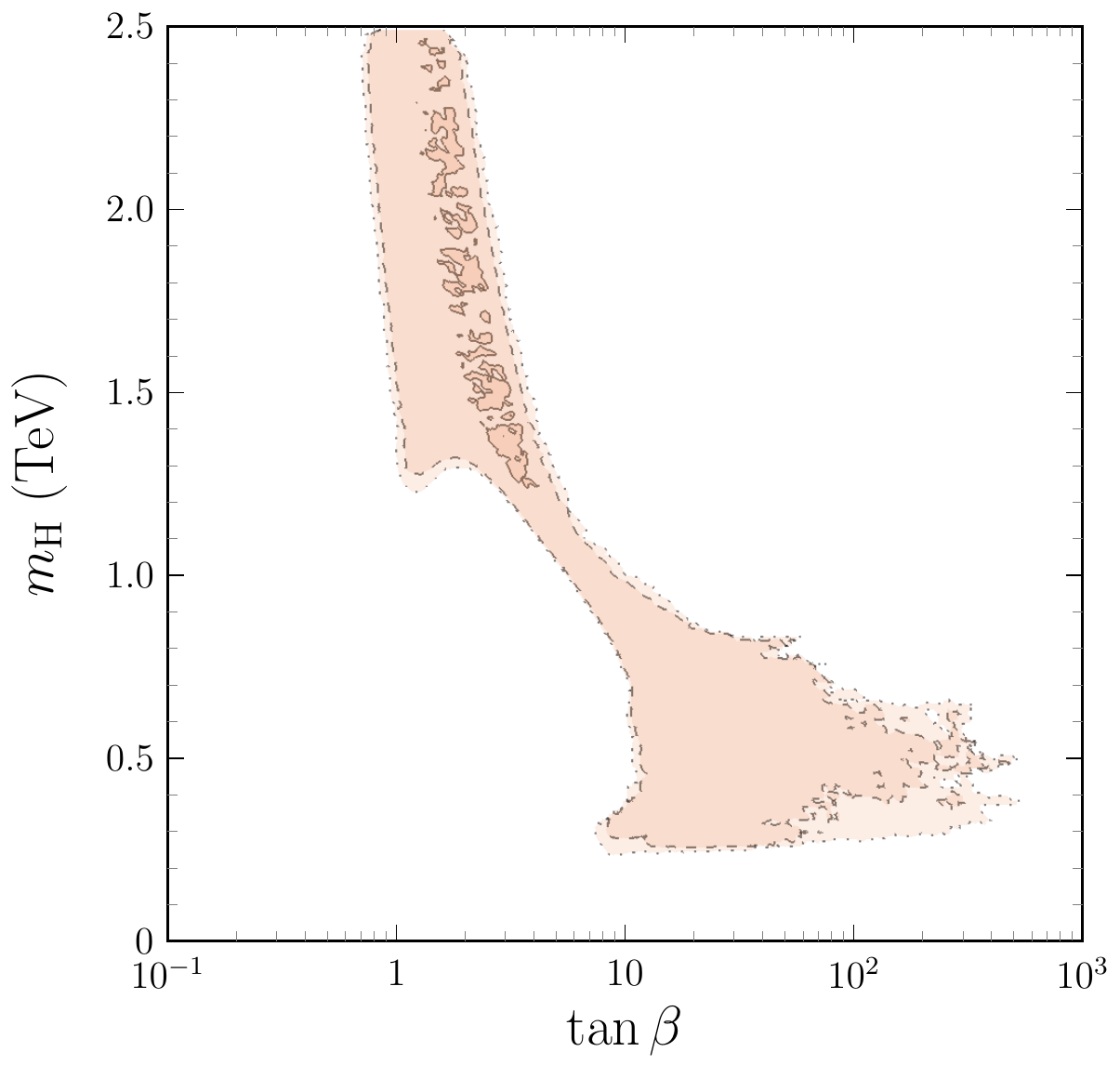}}\quad 
\subfloat[$\mA$ vs. $\mH$.\label{sfig:mA:mH:Cs250W2}]{\includegraphics[width=0.23\textwidth]{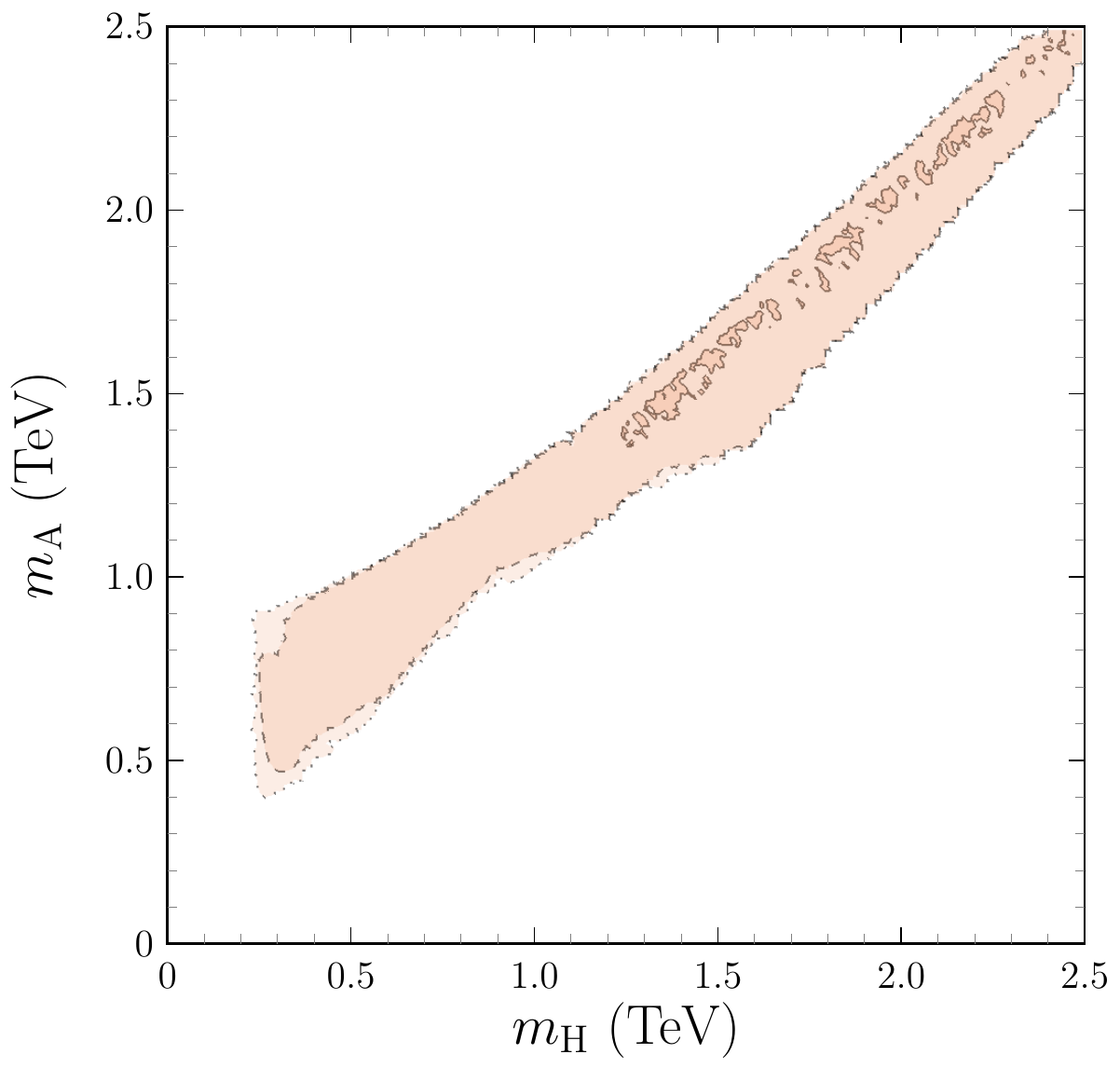}}\quad 
\subfloat[$\mA$ vs. $\mcH$.\label{sfig:mA:mcH:Cs250W2}]{\includegraphics[width=0.23\textwidth]{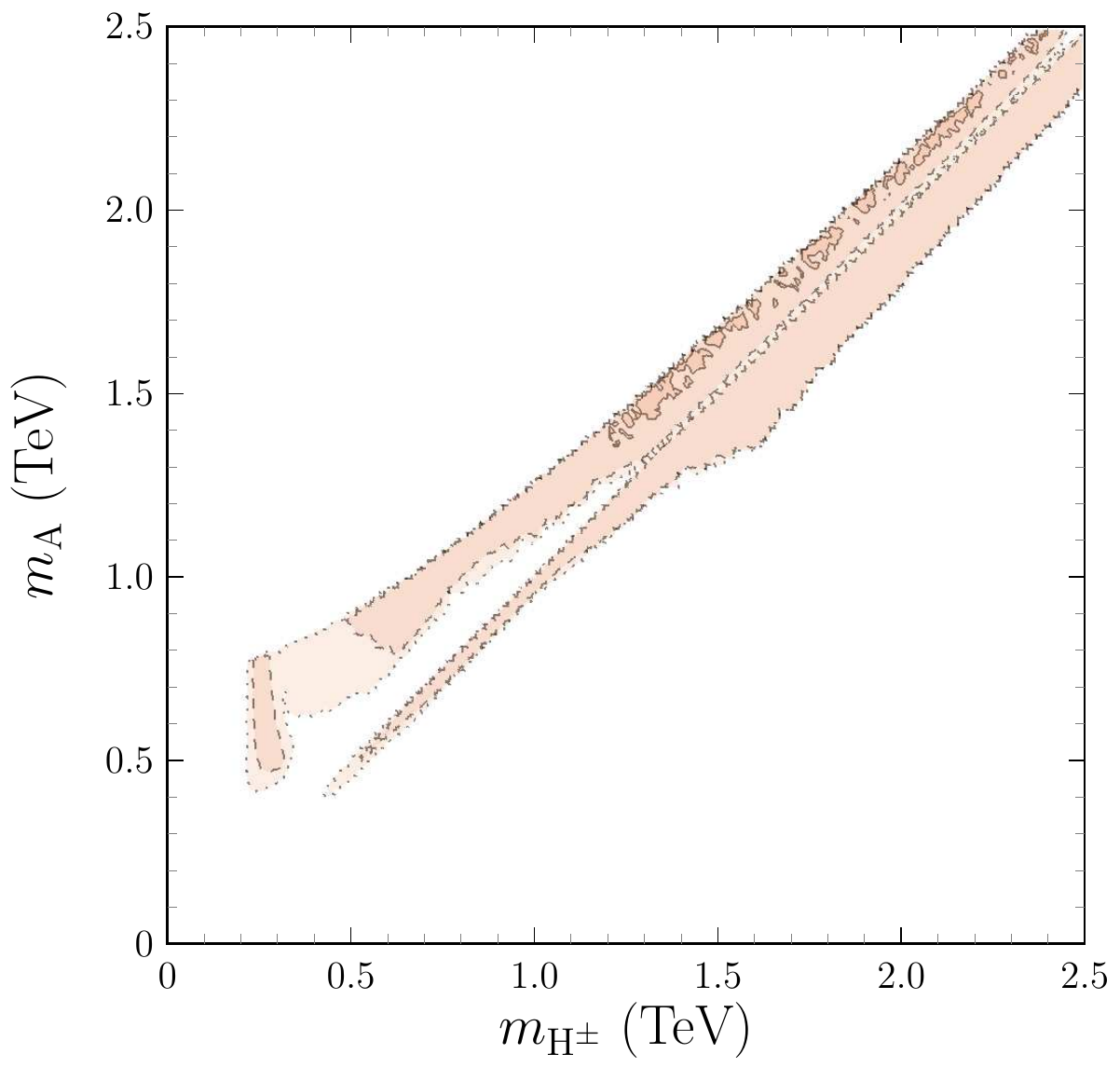}}\quad 
\subfloat[$\mH$ vs. $\mcH$.\label{sfig:mH:mcH:Cs250W2}]{\includegraphics[width=0.23\textwidth]{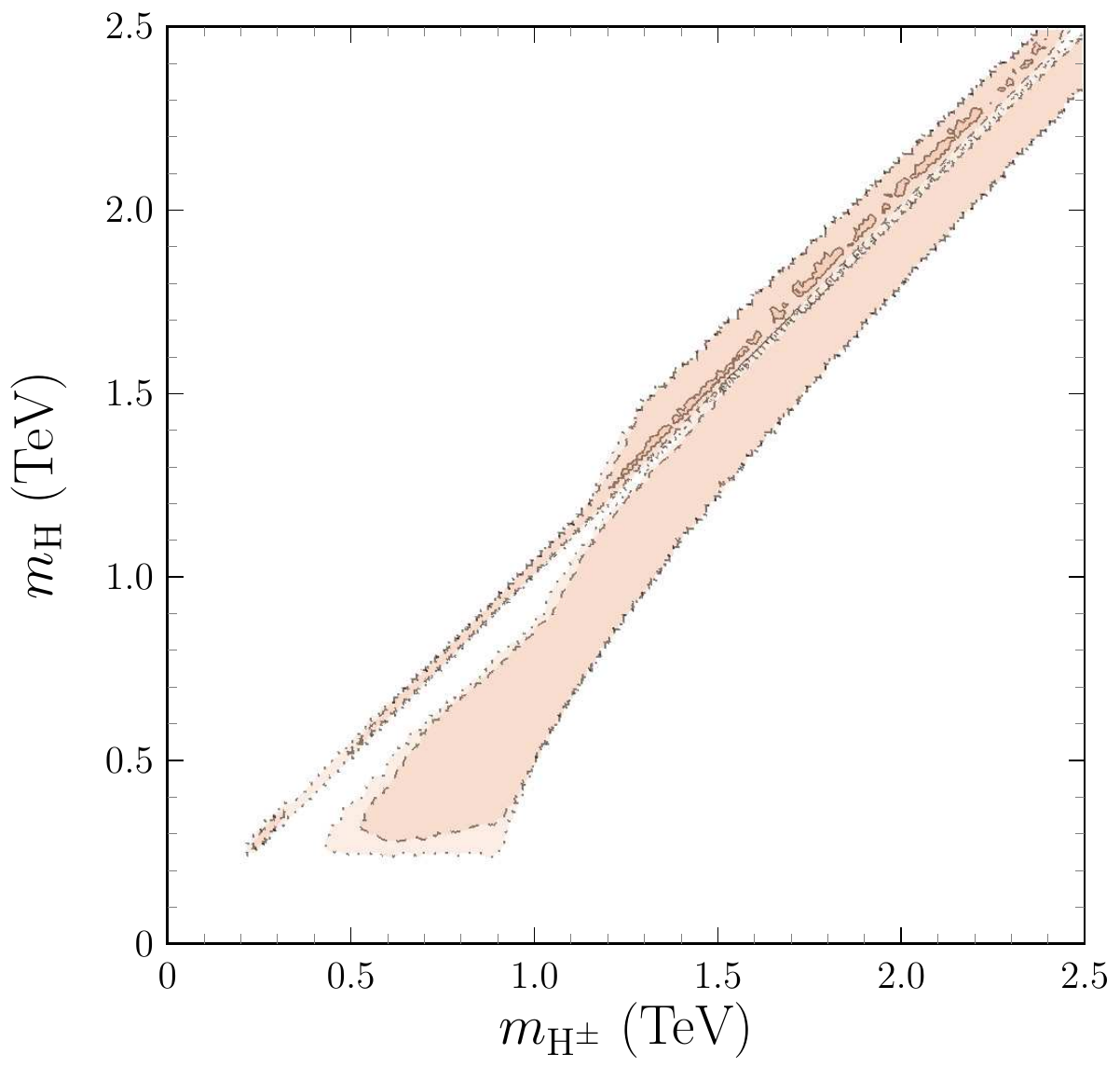}}\\
\subfloat[$\nrle$ vs. $\mH$.\label{sfig:ne:mH:Cs250W2}]{\includegraphics[width=0.3\textwidth]{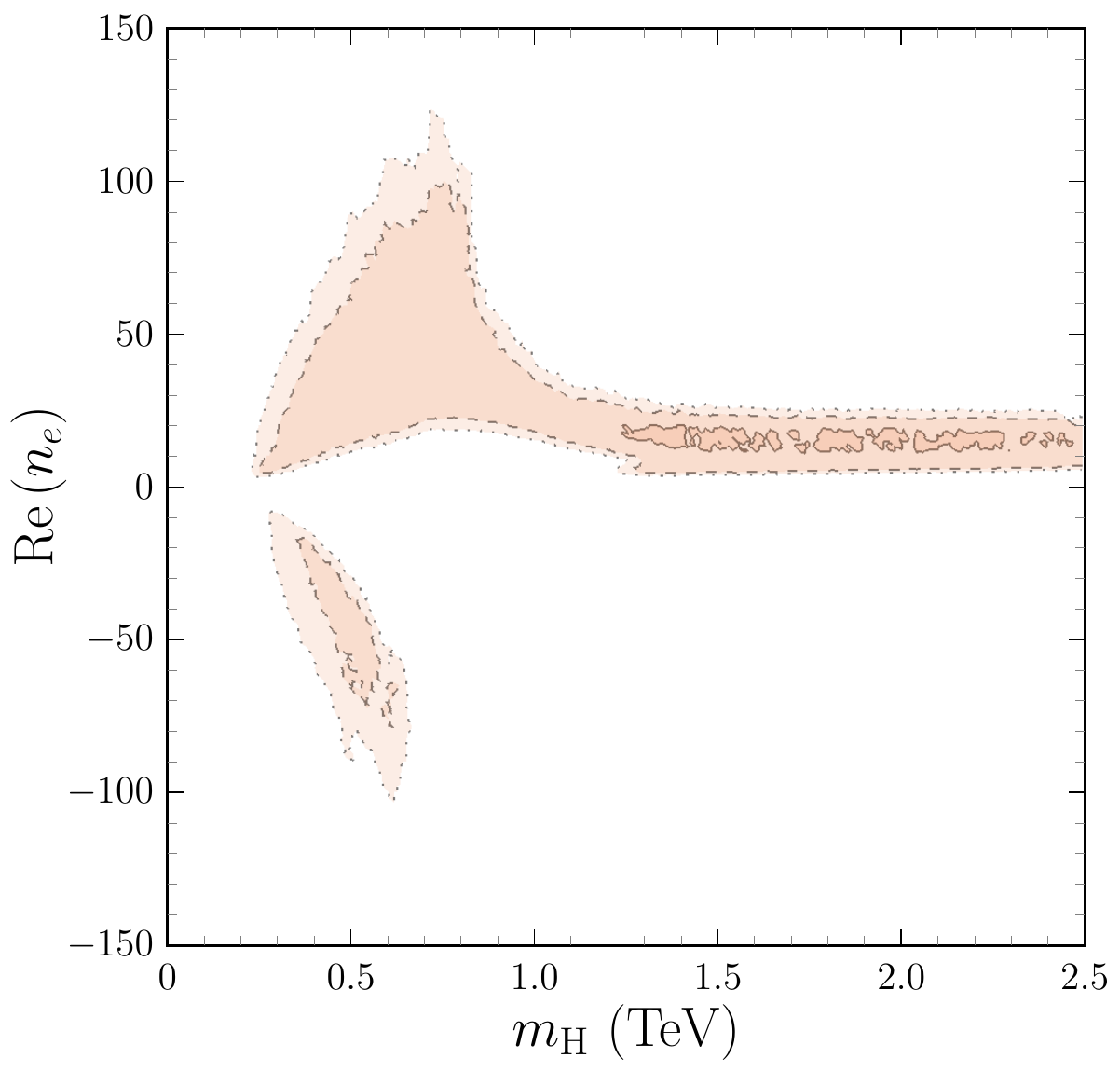}}\quad 
\subfloat[$\nrlm$ vs. $\mH$.\label{sfig:nm:mH:Cs250W2}]{\includegraphics[width=0.3\textwidth]{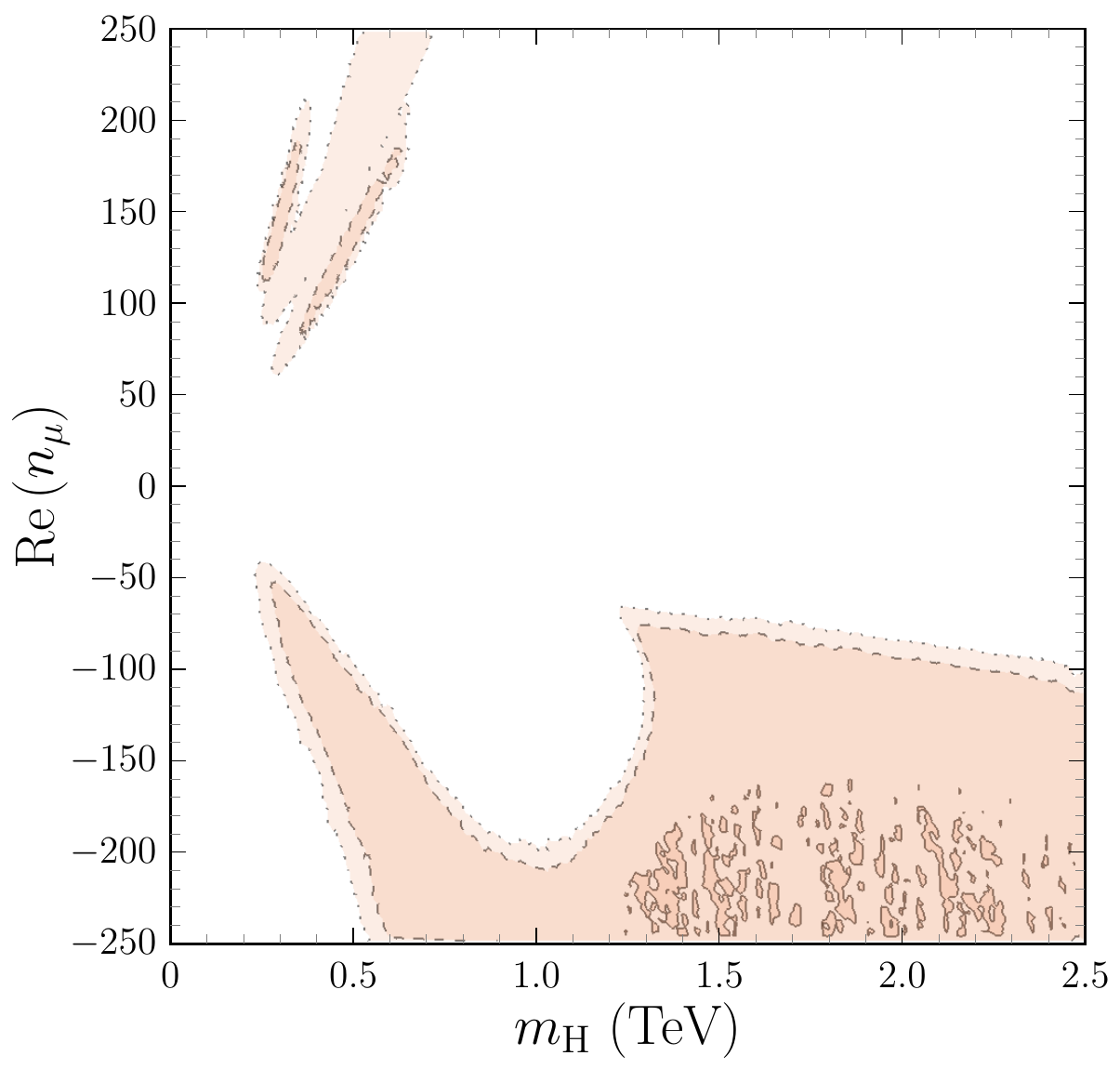}}\quad
\subfloat[$\nrlm$ vs. $\nrle$.\label{sfig:nmu:ne:Cs250W2}]{\includegraphics[width=0.3\textwidth]{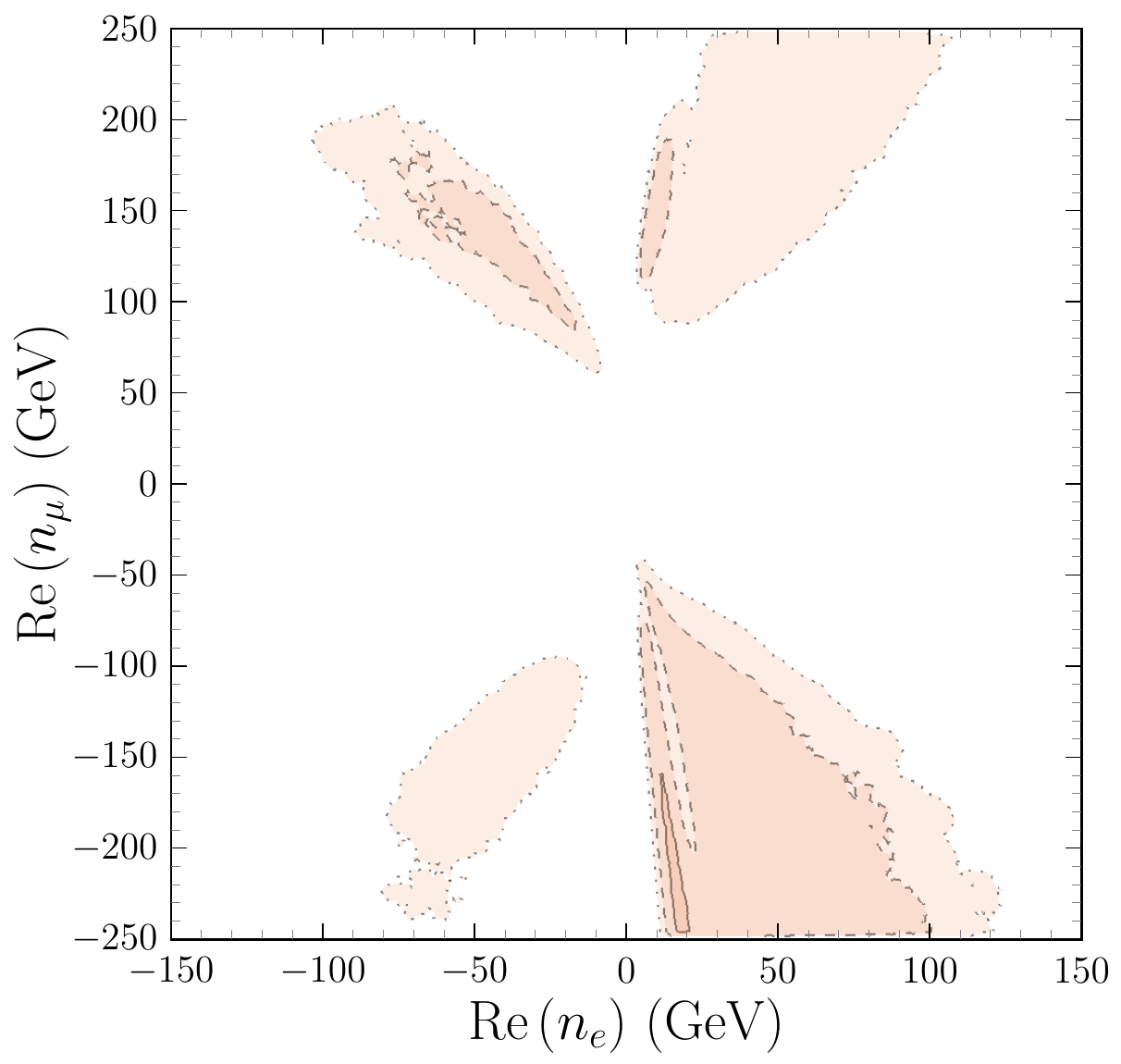}} 
\caption{Results with $(\Delta S,\Delta T)$ in \refeq{eq:MWST:cons}, ``conservative'' case in \cite{deBlas:2022hdk}.\label{fig:Cs250W2}}
\end{center}
\end{figure}

\begin{figure}[H]
\begin{center}
\subfloat[$\mH$ vs. $\tb$.\label{sfig:mH:tb:Cs250W1}]{\includegraphics[width=0.23\textwidth]{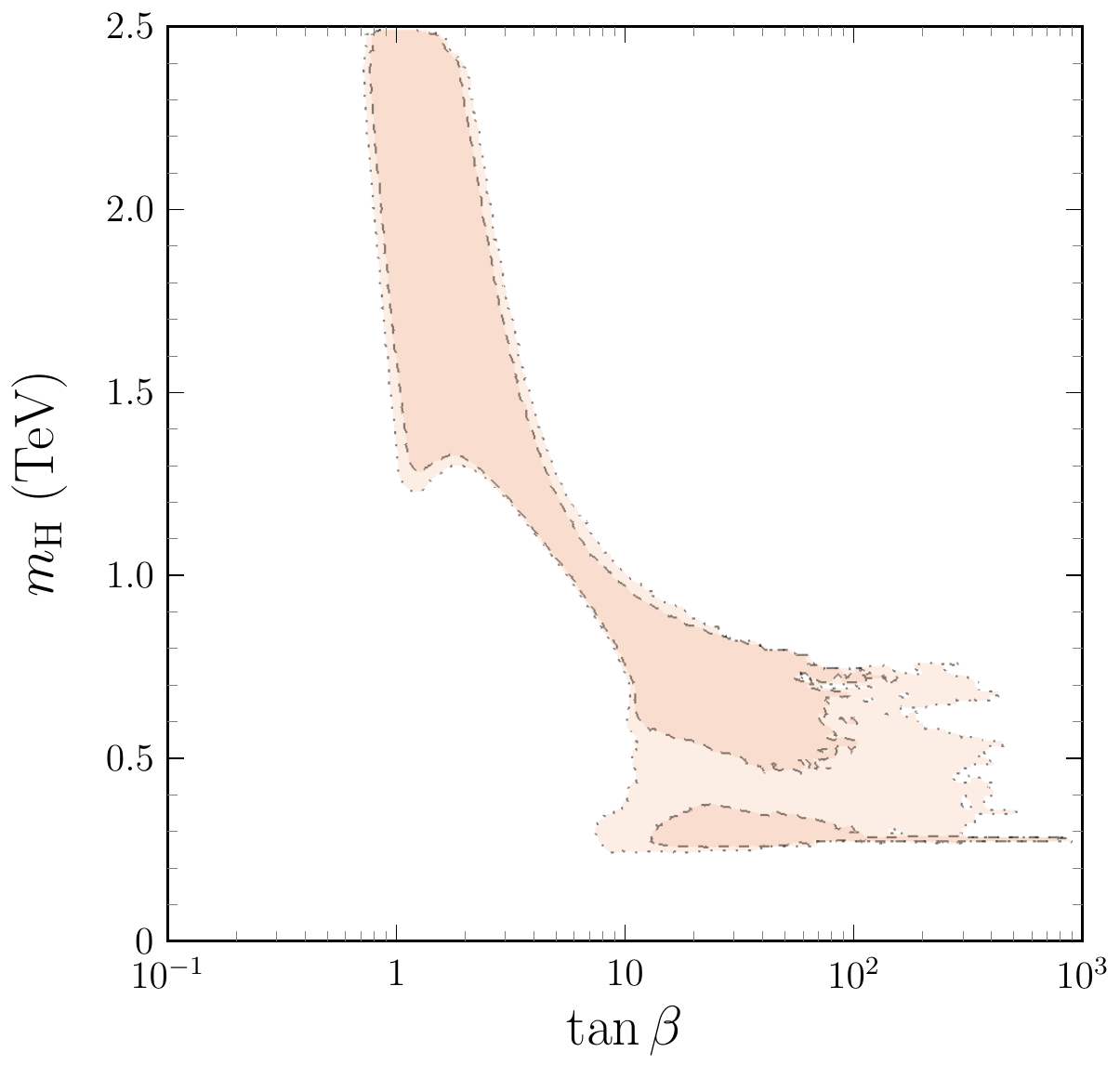}}\quad 
\subfloat[$\mA$ vs. $\mH$.\label{sfig:mA:mH:Cs250W1}]{\includegraphics[width=0.23\textwidth]{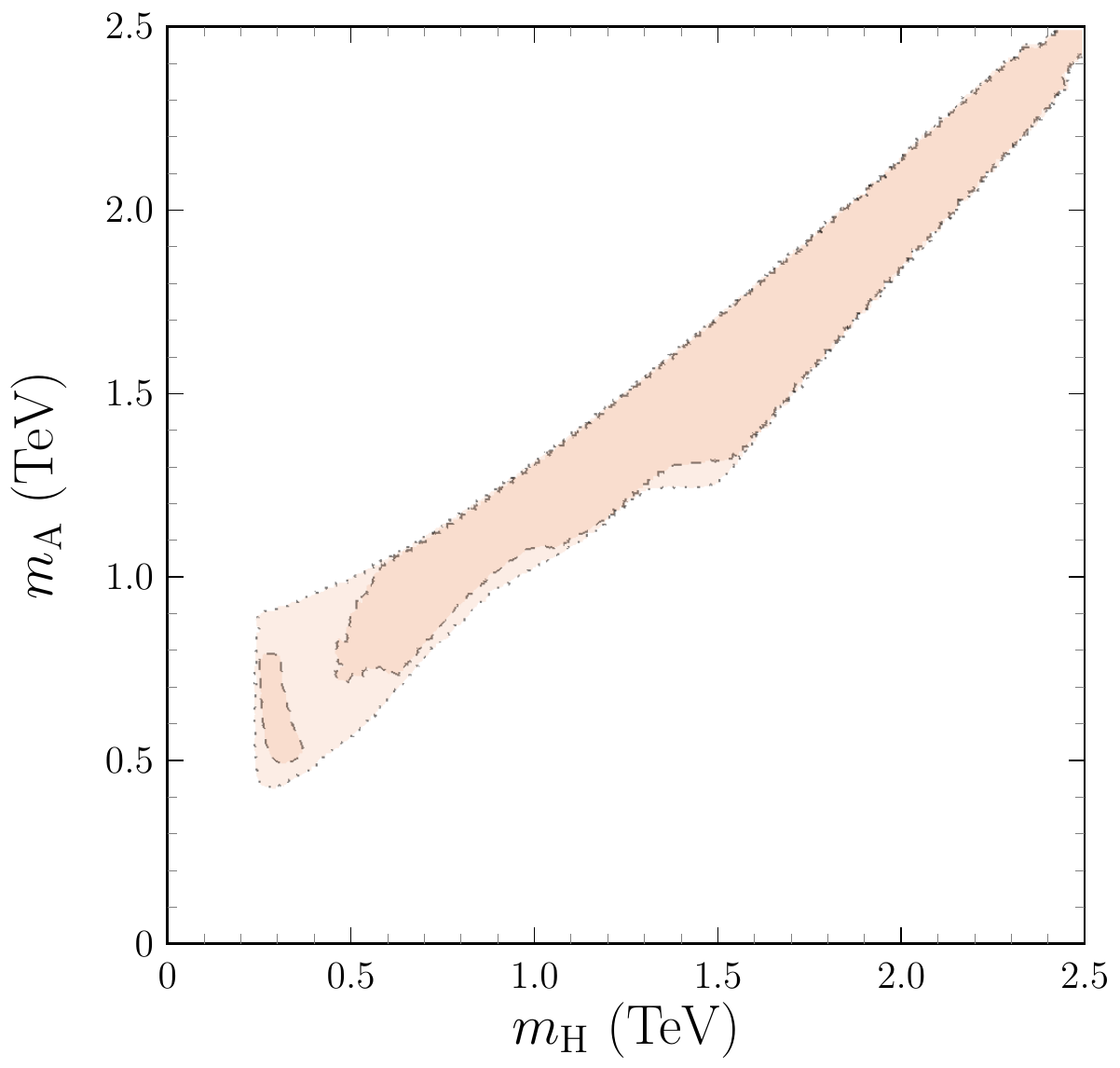}}\quad 
\subfloat[$\mA$ vs. $\mcH$.\label{sfig:mA:mcH:Cs250W1}]{\includegraphics[width=0.23\textwidth]{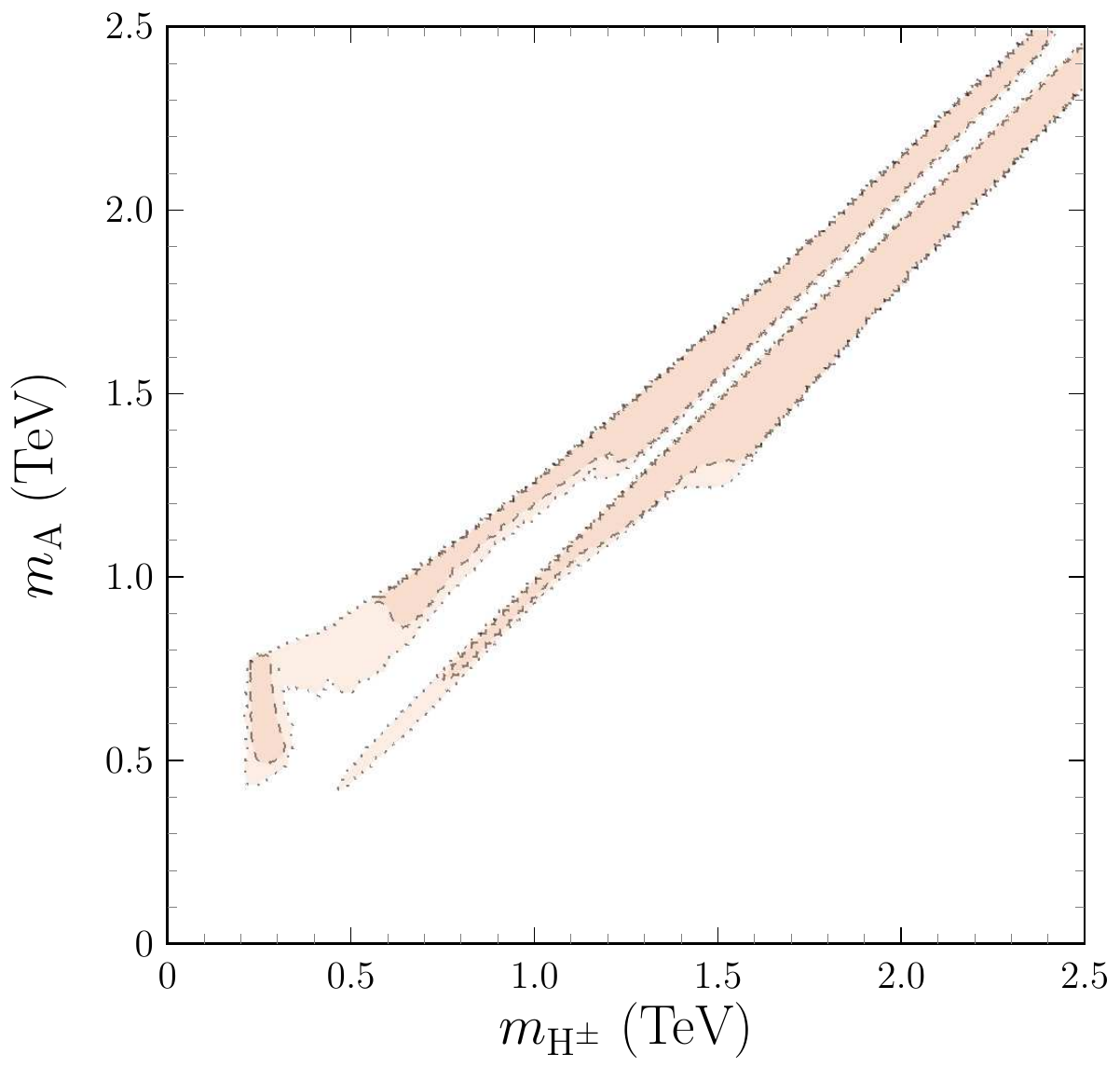}}\quad 
\subfloat[$\mH$ vs. $\mcH$.\label{sfig:mH:mcH:Cs250W1}]{\includegraphics[width=0.23\textwidth]{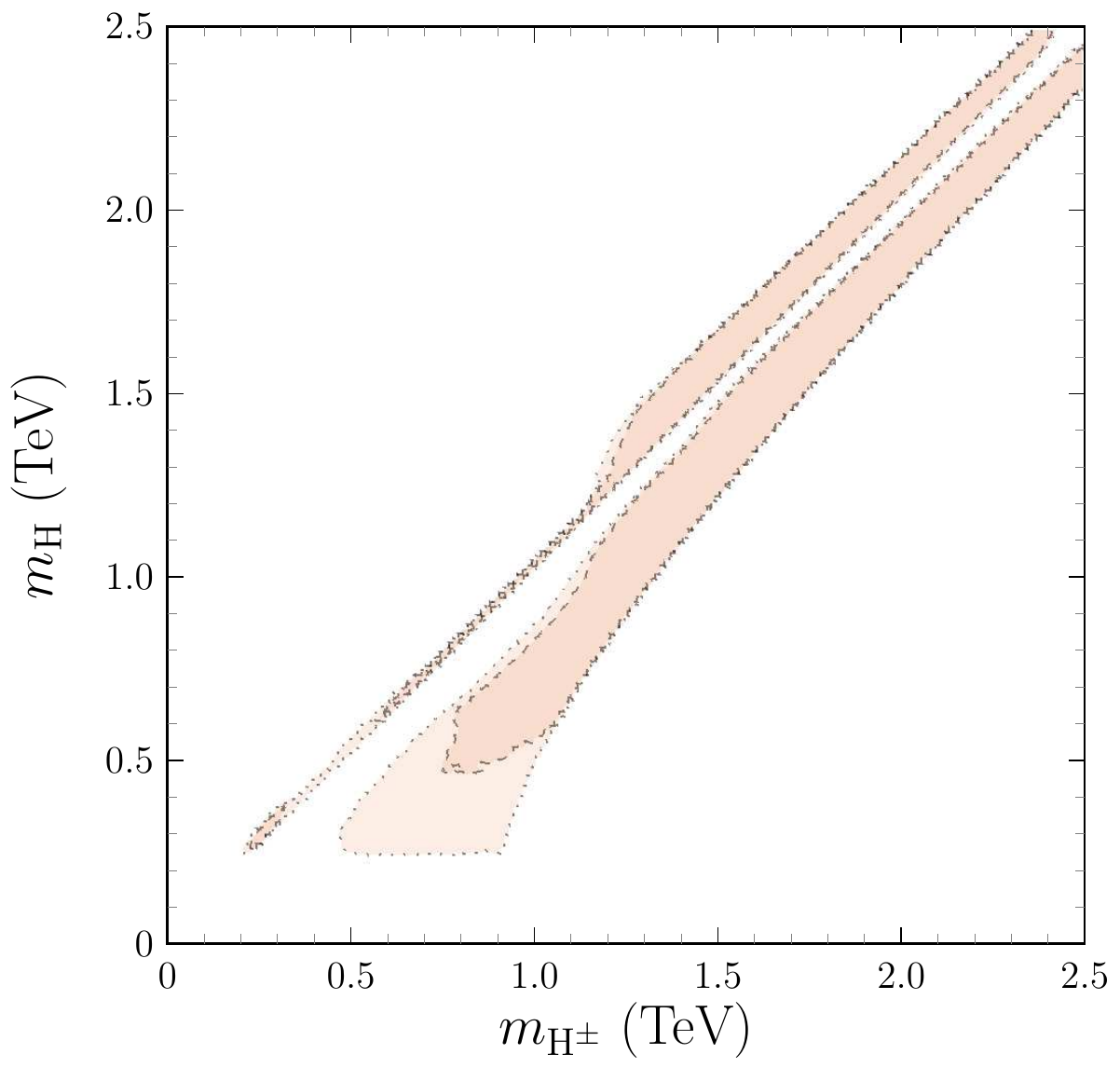}}\\
\subfloat[$\nrle$ vs. $\mH$.\label{sfig:ne:mH:Cs250W1}]{\includegraphics[width=0.3\textwidth]{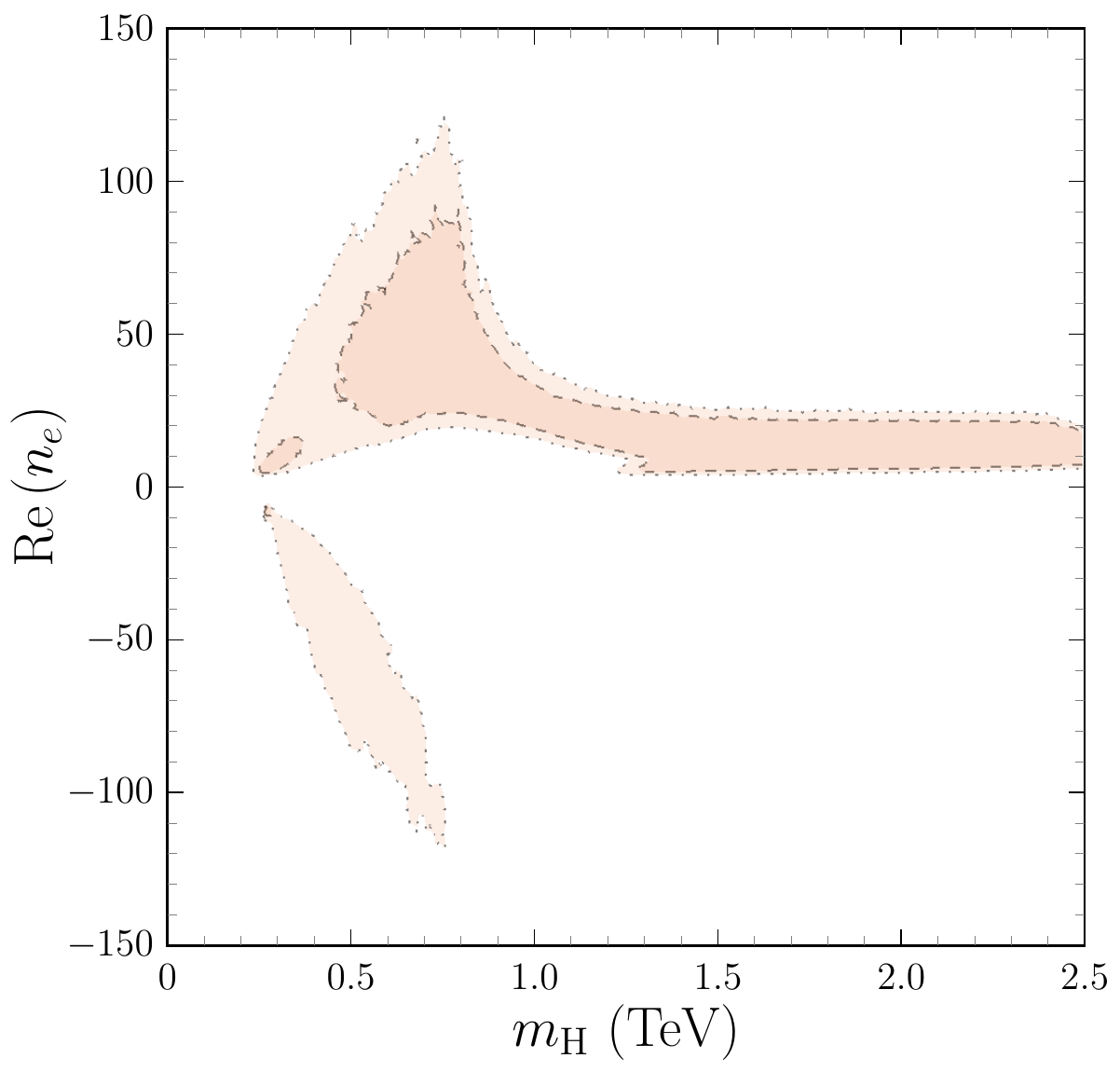}}\quad 
\subfloat[$\nrlm$ vs. $\mH$.\label{sfig:nm:mH:Cs250W1}]{\includegraphics[width=0.3\textwidth]{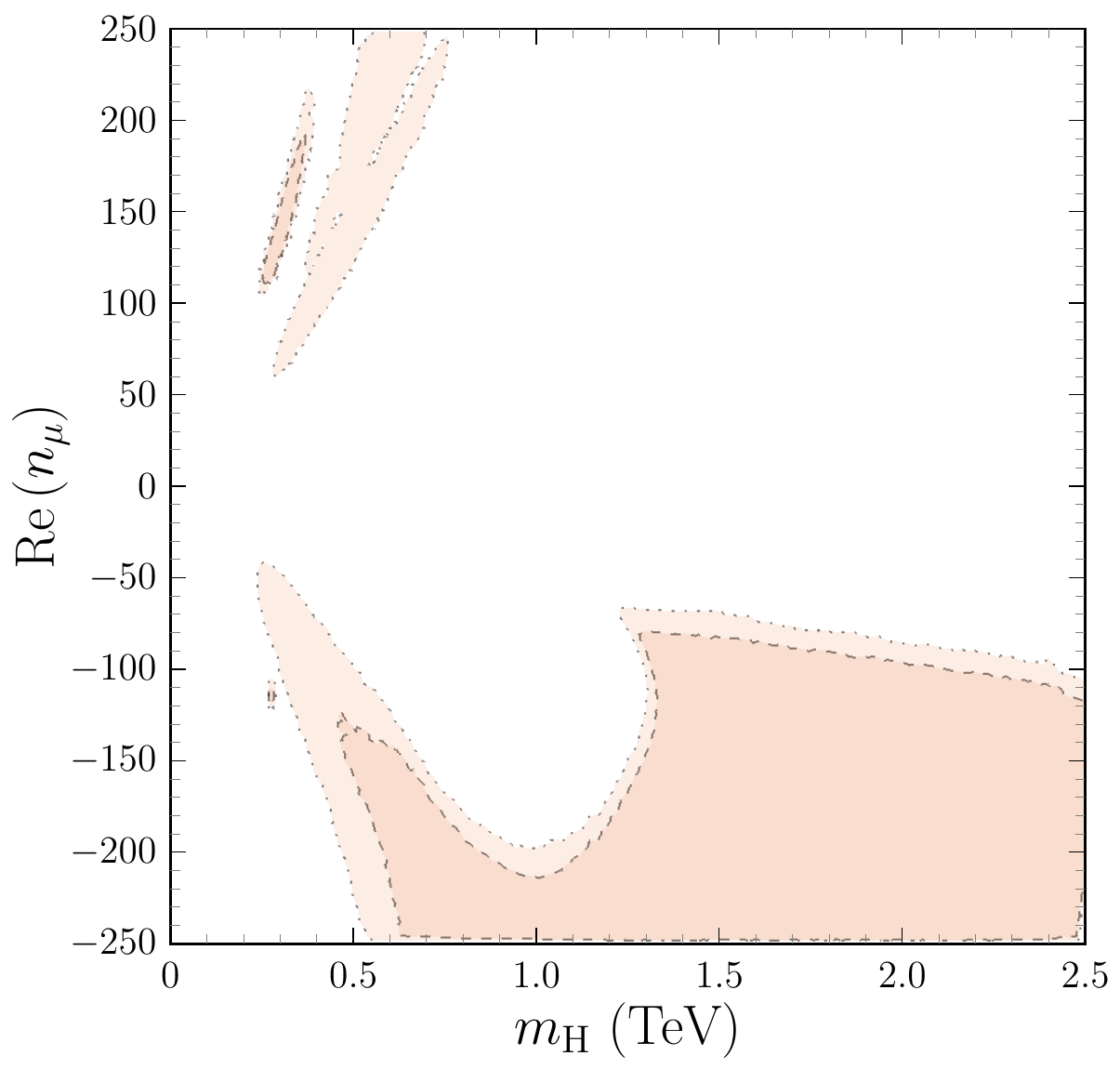}}\quad
\subfloat[$\nrlm$ vs. $\nrle$.\label{sfig:nmu:ne:Cs250W1}]{\includegraphics[width=0.3\textwidth]{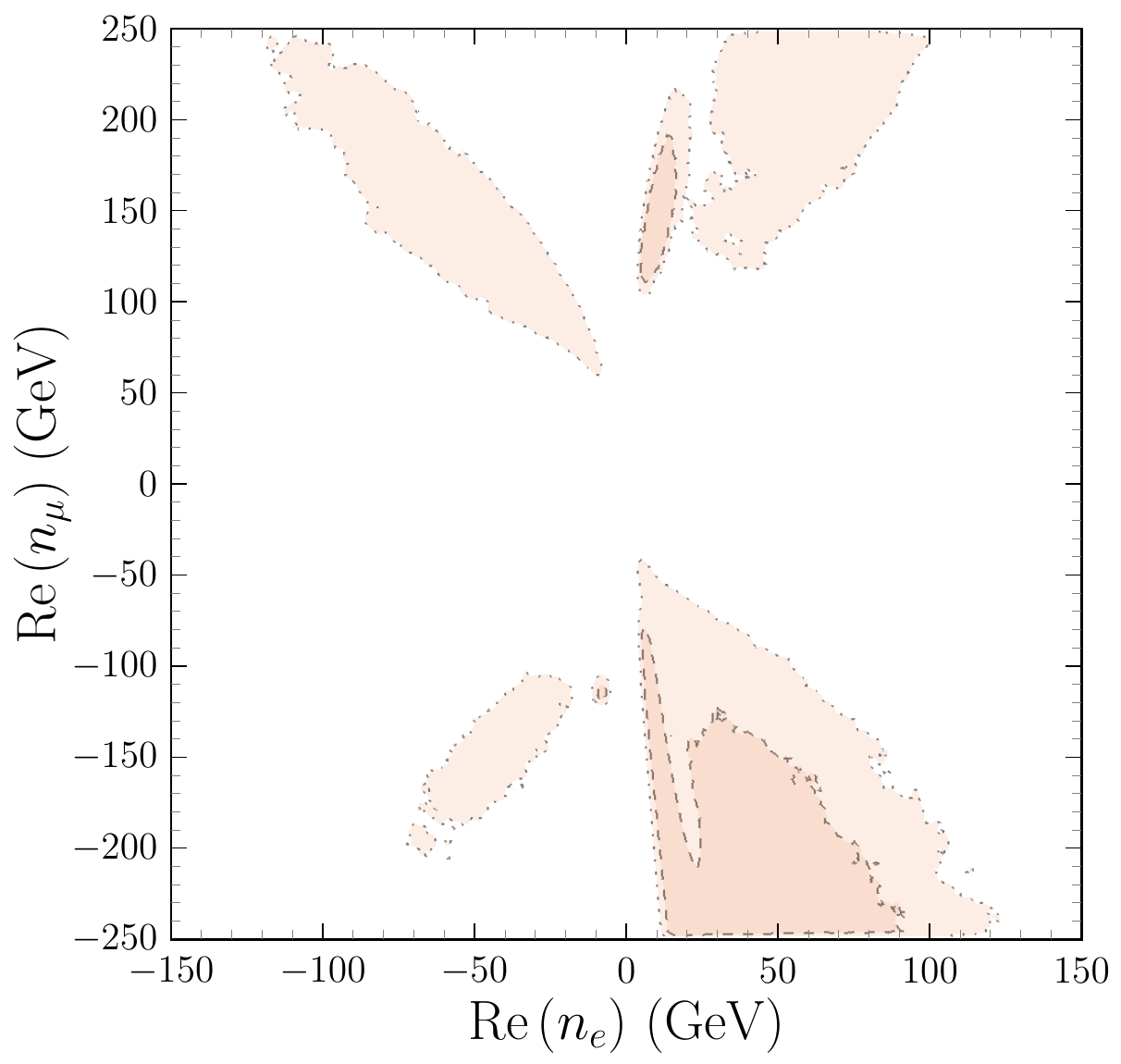}} 
\caption{Results with $(\Delta S,\Delta T)$ in \refeq{eq:MWST:nocons}, from \cite{Lu:2022bgw}.\label{fig:Cs250W1}}
\end{center}
\end{figure}
Finally, since the oblique parameters $S$ and $T$ play an important role, figure \ref{fig:DS:DT} shows allowed regions for $\Delta S$ vs. $\Delta T$ in the two scenarios considered for the CDF $M_W$ ``explanation'', together with the imposed $(\Delta T,\Delta S)$ constraint in each case. As anticipated, the constraint in \refeq{eq:MWST:nocons} appears to be more difficult to accommodate than the constraint in \refeq{eq:MWST:cons}. In fact, despite the different position of the ellipses corresponding to the $(\Delta T,\Delta S)$ constraints in figures \ref{sfig:DS:DT:Cs250W2} and \ref{sfig:DS:DT:Cs250W1}, the allowed regions are quite similar in both cases, that is, the model appears to be unable to accommodate values $\Delta T>0.22$ together with $\Delta S>0.02$. Other possible explanations of the CDF $M_W$ anomaly have been addressed in \cite{Fan:2022dck,Tang:2022pxh,Ahn:2022xeq,Han:2022juu,Chowdhury:2022moc,Ghorbani:2022vtv,Bhaskar:2022vgk,Baek:2022agi}.
\begin{figure}[H]
\begin{center}
\subfloat[$(\Delta S,\Delta T)$ in \refeq{eq:MWST:cons}.\label{sfig:DS:DT:Cs250W2}]{\includegraphics[width=0.3\textwidth]{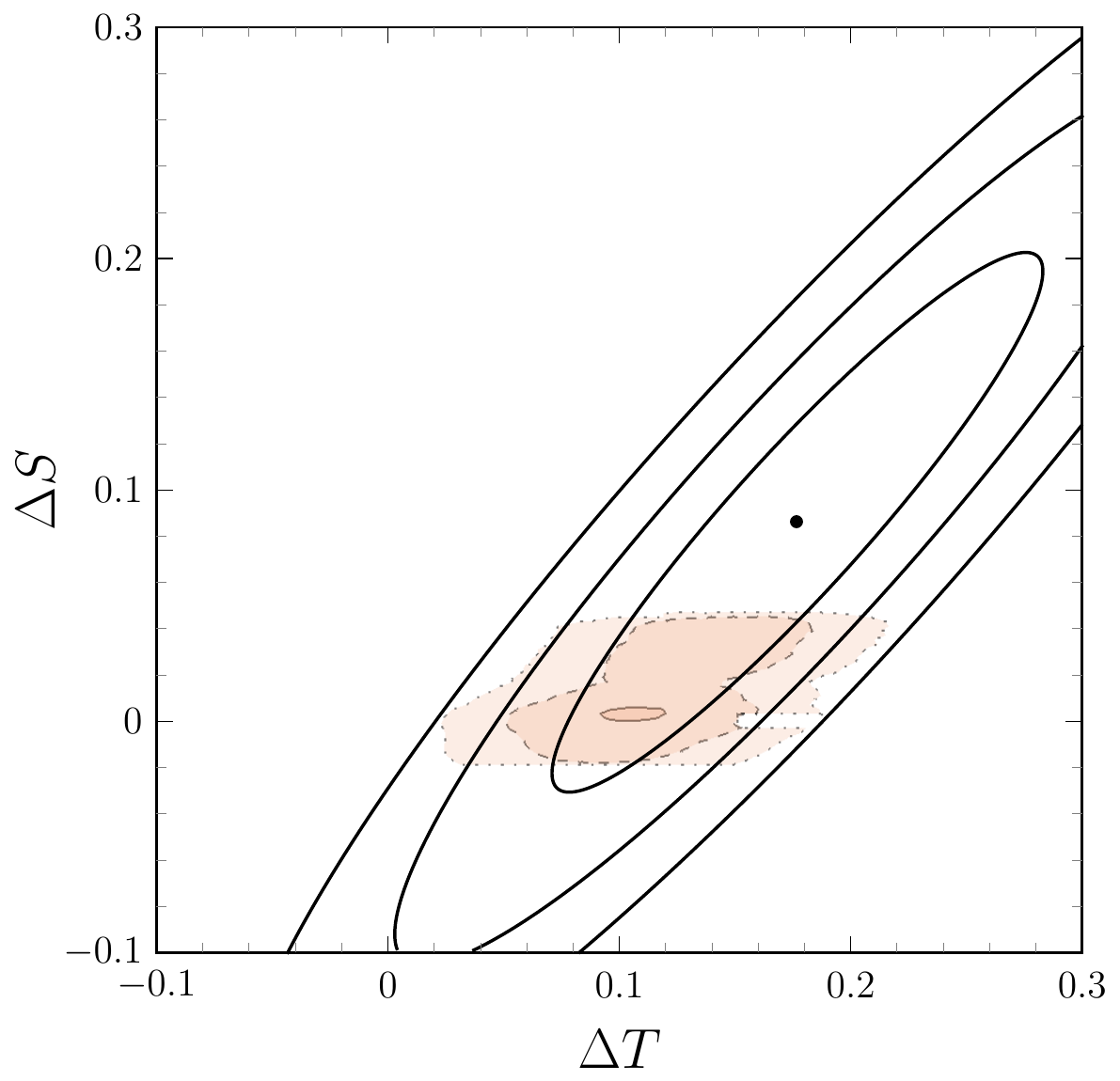}}\quad
\subfloat[$(\Delta S,\Delta T)$ in \refeq{eq:MWST:nocons}.\label{sfig:DS:DT:Cs250W1}]{\includegraphics[width=0.3\textwidth]{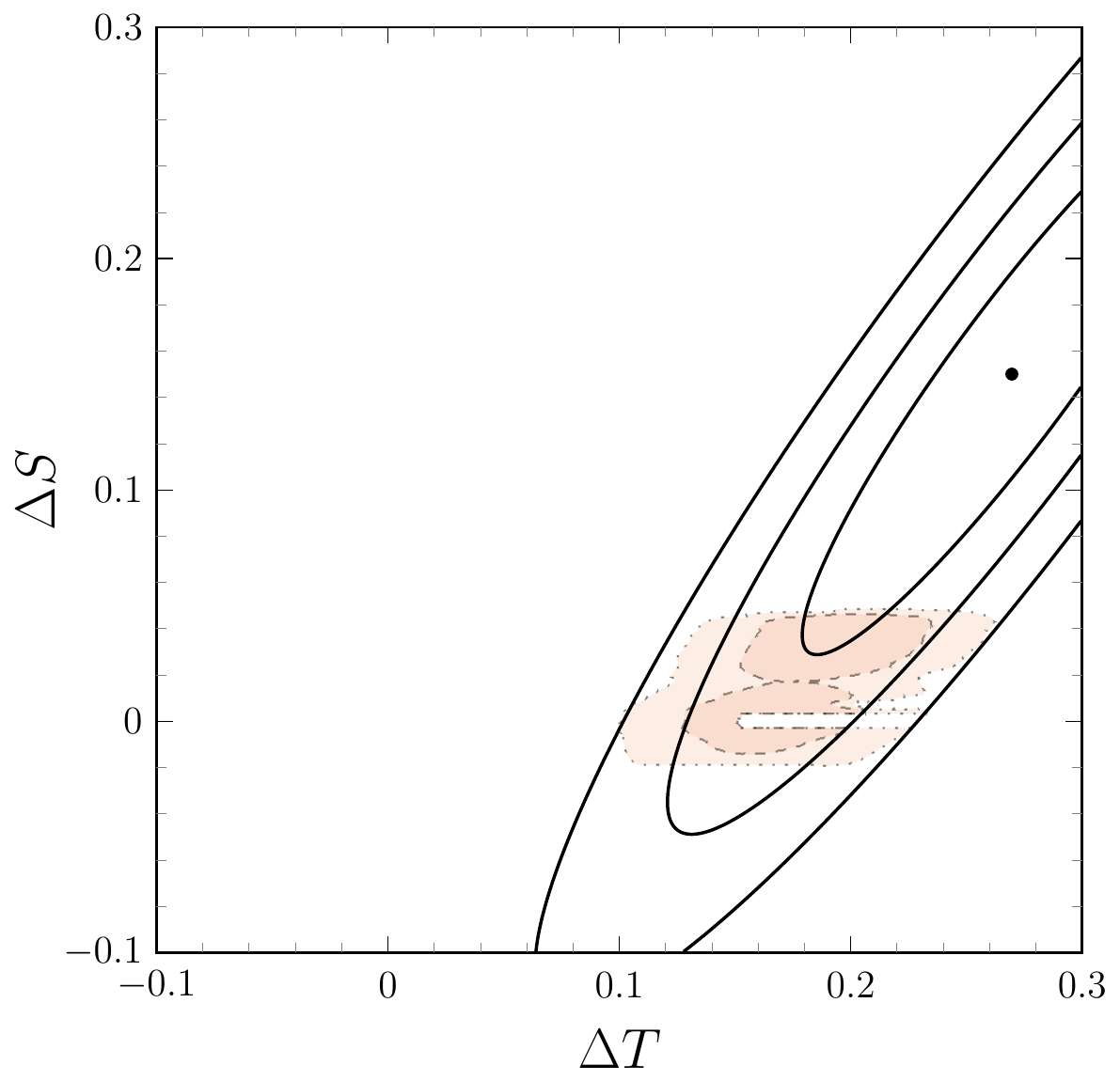}}%
\caption{$\Delta S$ vs. $\Delta T$, 1, 2 and 3$\sigma$ 2D-$\Delta\chi^2$ contours from the imposed constraint are shown together with the allowed regions.\label{fig:DS:DT}}
\end{center}
\end{figure}

\subsection{Example points}\label{sSec:Results:examples}
In this section, some example points of the allowed parameter space are presented in order to specify the behavior pointed out in the previous plots. For the sake of clarity, we only focus on the analysis with ``$a_e^{\rm Cs}$'' concerning the electron anomaly: other cases do not change substantially beyond the differences already mentioned in section \ref{sSec:Results:diff:dae}.

From table \ref{tab:masses:tb:nl}, it is clear that points 1--2 correspond to the solution with small values of $\tb$ and large scalar masses: all scalars are above $1.2\ \rm{TeV}$ and their mass differences do not exceed $\pm200$ GeV. In this region, both anomalies are explained at two loops through the top quark terms, as one can easily check in tables \ref{tab:ae:loops} and \ref{tab:amu:loops}, where all loop contributions are normalized to the total $\delta a_{\ell}$ in such a way that their sum must be 1. One may also notice that $\delta a_{\mu}$ receives a subdominant one loop contribution. The lepton couplings $\nrle$ and $\nrlm$ have opposite sign and they roughly satisfy the linear relation in \refeq{eq:nmu:ne} for the Cs case. 

Regarding the appearance of the intermediate values of the scalar masses and $\tb$ previously commented in section \ref{sSec:Results:Cs250}, our point 3 gives a perfect example of that behavior. It is important to realize that large values of $|\nrlm|$ are required in this region; in fact, they are almost reaching the perturbativity upper bound  $|\nl{\ell}| \leq 250$ GeV. On the other hand, although the top dominance still holds at two loops in the electron anomaly, the corresponding tau contributions begin to play a relevant role. This trend will continue as $\tb$ grows and the quark contributions are more suppressed.

Finally, points 4--9 belong to the low mass region corresponding to a wide range of $\tb \gg 1$ values. As we have stressed before, two possible scenarios arise: one where $\mcH \simeq \mA$ (points 4--6) and another where $\mcH \simeq \mH$ (points 7--9). In all cases, the scalar masses are below $1$ TeV and $\mA > \mH$, as anticipated. Taking into account the large values of $\tb$, the two loop contribution that dominates $\delta a_e$ is generated by the tau loop. This confirms our expectation for $\nrle$: its sign is not fixed and it could be either positive or negative (point 9). Furthermore, in this region the muon anomaly is clearly one loop dominated, albeit there exists a subdominant contribution from the tau loop as well. This in turn means that $\nrlm$ can take both signs, as one can easily check. 

For completeness, the last two points have been included to give an example of the allowed parameter space in subsection \ref{sSec:Results:CDF} considering the CDF $M_W$ anomaly. It is clear that point 10 mimics the behavior of points 1--2, while point 11 presents the same features as points 4--6. 

\begin{table}[H]
\begin{center}
 \begin{tabular}{|c|c|c|c|c|c|c|c|}\hline
  Point & $\mH$ & $\mA$ & $\mcH$ & $\tb$ & $\nrle$ & $\nrlm$ & $\nrlt$\\ \hline\hline
  1 & $1351$ & $1547$ & $1560$ & $1.34$ & $6.08$ & $-95.0$ & $-40.79$\\ \hline
  2 & $1522$ & $1567$ & $1485$ & $1.90$ & $9.09$ & $-158.9$ & $126.5$\\ \hline
  3 & $1049$ & $1322$ & $1332$ & $6.88$ & $29.86$ & $-245.0$ & $-75.75$\\ \hline
  4 & $663$ & $876$ & $888$ & $11.9$ & $22.09$ & $-172.9$ & $224.6$\\ \hline
  5 & $621$ & $938$ & $946$ & $21.5$ & $54.26$ & $238.1$ & $-74.57$\\ \hline
  6 & $350$ & $855$ & $860$ & $22.7$ & $17.31$ & $-87.22$ & $94.99$\\ \hline
  7 & $372$ & $815$ & $362$ & $34.6$ & $25.44$ & $-98.21$ & $85.64$\\ \hline
  8 & $364$ & $812$ & $355$ & $100$ & $46.18$ & $-95.14$ & $79.90$\\ \hline
  9 & $360$ & $810$ & $352$ & $186$ & $-35.52$ & $97.10$ & $-119.2$\\ \hline
  10 & $1509$ & $1604$ & $1453$ & $1.36$ & $6.55$ & $-112.8$ & $70.7$\\ \hline
  11 & $508$ & $809$ & $834$ & $89.3$ & $38.85$ & $-146.8$ & $189.1$\\ \hline
  \end{tabular}
 \caption{Example points, masses and $\nrl{\ell}$'s in GeV.}
 \label{tab:masses:tb:nl}
\end{center}
\end{table}

\begin{table}[H]
\begin{center}
 \begin{tabular}{|c|c|c|c|c|c|c|c|c|}\cline{3-9}
 \multicolumn{2}{c}{}& \multicolumn{3}{|c}{1 loop} & \multicolumn{4}{|c|}{2 loop}\\ \hline
  Point & $\delta a_e$ & $\nH$ & $\nA$ & $\cH$ & $t\nH$ & $t\nA$ & $\tau\nH$ & $\tau\nA$ \\ \hline\hline
  1 & $-7.29$ & $0$ & $0$ & $0$ & $0.469$ & $0.521$ & $-0.012$ & $0.011$ \\ \hline
  2 & $-7.06$ & $0$ & $0$ & $0$ & $0.445$ & $0.559$ & $0.047$ & $-0.051$ \\ \hline
  3 & $-8.40$ & $-0.001$ & $0.001$ & $0$ & $0.524$ & $0.526$ & $-0.145$ & $0.113$\\ \hline
  4 & $-7.53$ & $-0.002$ & $0.001$ & $0$ & $0.410$ & $0.407$ & $0.764$ & $-0.554$\\ \hline
  5 & $-7.12$ & $-0.016$ & $0.007$ & $0$ & $0.627$ & $0.541$ & $-0.734$ & $0.425$\\ \hline
  6 & $-6.96$ & $-0.005$ & $0.001$ & $0$ & $0.330$ & $0.186$ & $0.776$ & $-0.207$\\ \hline
  7 & $-7.01$ & $-0.009$ & $0.002$ & $0$ & $0.300$ & $0.187$ & $0.926$ & $-0.295$\\ \hline
  8 & $-8.41$ & $-0.027$ & $0.006$ & $0$ & $0.161$ & $0.098$ & $1.352$ & $-0.419$\\ \hline
  9 & $-7.06$ & $-0.019$ & $0.004$ & $0$ & $-0.078$ & $-0.049$ & $1.879$ & $-0.574$\\ \hline
  10 & $-7.06$ & $0$ & $0$ & $0$ & $0.453$ & $0.548$ & $0.019$ & $-0.020$\\ \hline
  11 & $-7.09$ & $-0.012$ & $0.005$ & $0$ & $0.133$ & $0.111$ & $1.863$ & $-0.996$\\ \hline
  \end{tabular}
 \caption{Example points, $\delta a_e$ values; columns 3 to 9 show the relative contributions of the different one and two loop terms to the value of $\delta a_e$ in the second column.}
 \label{tab:ae:loops}
\end{center}
\end{table}

\begin{table}[H]
\begin{center}
 \begin{tabular}{|c|c|c|c|c|c|c|c|c|}\cline{3-9}
 \multicolumn{2}{c}{}& \multicolumn{3}{|c}{1 loop} & \multicolumn{4}{|c|}{2 loop}\\ \hline
  Point & $\delta a_\mu$ & $\nH$ & $\nA$ & $\cH$ & $t\nH$ & $t\nA$ & $\tau\nH$ & $\tau\nA$ \\ \hline\hline
  1 & $2.40$ & $0.085$ & $-0.063$ & $-0.001$ & $0.459$ & $0.511$ & $-0.012$ & $0.011$\\ \hline
  2 & $2.58$ & $0.177$ & $-0.162$ & $-0.003$ & $0.439$ & $0.553$ & $0.046$ & $-0.050$\\ \hline
  3 & $2.24$ & $0.988$ & $-0.614$ & $-0.012$ & $0.334$ & $0.336$ & $-0.093$ & $0.072$\\ \hline
  4 & $2.31$ & $1.124$ & $-0.639$ & $-0.013$ & $0.216$ & $0.215$ & $0.403$ & $-0.292$\\ \hline
  5 & $2.41$ & $2.316$ & $-1.024$ & $-0.021$ & $-0.169$ & $-0.145$ & $0.197$ & $-0.114$\\ \hline
  6 & $2.51$ & $0.870$ & $-0.157$ & $-0.003$ & $0.095$ & $0.054$ & $0.224$ & $-0.060$\\ \hline
  7 & $2.43$ & $1.016$ & $-0.224$ & $-0.024$ & $0.069$ & $0.043$ & $0.213$ & $-0.068$\\ \hline
  8 & $2.21$ & $1.093$ & $-0.233$ & $-0.025$ & $0.026$ & $0.016$ & $0.219$ & $-0.068$\\ \hline
  9 & $2.37$ & $1.081$ & $-0.227$ & $-0.025$ & $-0.013$ & $-0.008$ & $0.317$ & $-0.097$\\ \hline
  10 & $2.54$ & $0.092$ & $-0.079$ & $-0.002$ & $0.448$ & $0.542$ & $0.019$ & $-0.019$\\ \hline
  11 & $2.38$ & $1.294$ & $-0.518$ & $-0.010$ & $0.031$ & $0.026$ & $0.433$ & $-0.231$\\ \hline
 \end{tabular}
 \caption{Example points, $\delta a_\mu$ values; columns 3 to 9 show the relative contributions of the different one and two loop terms to the value of $\delta a_\mu$ in the second column.}
 \label{tab:amu:loops}
\end{center}
\end{table}
\clearpage
\section{Conclusions\label{Sec:Conclusions}}
The experimental determinations of the muon and the electron anomalous magnetic moment point towards the necessity of lepton flavor non-universal New Physics. Aiming to address both leptonic anomalies simultaneously, we have considered a type I or type X 2HDM with a general flavor conserving lepton sector, one loop stable under renormalization, in which the new Yukawa couplings are completely decoupled from lepton mass proportionality. The latter turns out to be crucial in order to reproduce the $g-2$ muon anomaly together with the different scenarios one can consider for the  $g-2$ electron anomaly, related to the Cs and/or the Rb recoil measurements of the fine structure constant. A thorough analysis of the parameter space of the model has been performed including all relevant theoretical and experimental constraints. The results show that the muon anomaly receives dominant one loop contributions for light new scalar masses in the $0.2$-$1.0$ TeV range together with a significant hierarchy in the vacuum expectation values of the scalars, that is $\tb\gg 1$, while two loop Barr-Zee diagrams are also needed for heavy new scalars with masses above $1.2$ TeV together with $\tb\sim 1$. On the other hand, the electron anomaly receives dominant two loop contributions in the whole range of scalar masses. 
Furthermore, we have analysed how the perturbativity assumptions on the lepton Yukawa couplings have direct impact on relevant physical observables: intermediate values of the scalar masses and $\tb$ only arise when the perturbativity upper bound on $\nl{\ell}$ reaches the electroweak scale.
This might be relevant since we are entering an era of exclusion or discovery at the LHC, so that the allowed parameter space of the model must be fully scrutinized. 
The disagreement between the recent CDF measurement of $M_W$ and the SM expectations for electroweak precision results can be translated into deviations $(\Delta S,\Delta T)\neq (0,0)$ of the oblique parameters. We have considered two different scenarios for $(\Delta S,\Delta T)$ values which ``explain" the CDF disagreement. Both scenarios require a scalar spectrum where near degeneracies $\mcH\simeq\mH$ or $\mcH\simeq\mA$ are now disfavored, and where masses larger than 2 TeV are more difficult to accommodate. However, concerning the $\nl{\ell}$ couplings and $\tb$, the allowed regions have the same characteristics as in the analyses compatible with $(\Delta S,\Delta T)=(0,0)$.

\section*{Acknowledgments}
The authors acknowledge support from Spanish \textit{Agencia Estatal de Investigaci\'on}-\textit{Ministerio de Ciencia e Innovaci\'on} (AEI-MICINN) under grants PID2019-106448GB-C33 and PID2020-113334GB-I00/AEI/10.13039/501100011033 (AEI/FEDER, UE) and from \textit{Generalitat Valenciana} under grant PROMETEO 2019-113. 
The work of FCG is funded by MICINN, Spain (grant BES-2017-080070). 
CM is funded by \textit{Conselleria de Innovación, Universidades, Ciencia y Sociedad Digital} from \textit{Generalitat Valenciana} (grant ACIF/2021/284). 
MN is supported by the \textit{GenT Plan} from \textit{Generalitat Valenciana} under project CIDEGENT/2019/024.


\begin{appendices}
\section{One loop stability under RGE}\label{appendix:RGE:stability}
The evolution of the Yukawa couplings under one loop RGE \cite{Cvetic:1997zd,PhysRevD.58.116003,Ferreira:2010xe,Grimus:2004yh,PhysRevD.9.2259} is given by:
\begin{equation}\label{eq:RGE:Yd:00}
\begin{multlined}
\der\matYukD{\alpha}=
a_d\matYukD{\alpha}+\sum_{\rho=1}^{2}T^{d}_{\alpha,\rho}\matYukD{\rho}\\
+\sum_{\rho=1}^{2}\left(-2\matYukU{\rho}\matYukUd{\alpha}\matYukD{\rho}+\matYukD{\alpha}\matYukDd{\rho}\matYukD{\rho}+\frac{1}{2}\matYukU{\rho}\matYukUd{\rho}\matYukD{\alpha}+\frac{1}{2}\matYukD{\rho}\matYukDd{\rho}\matYukD{\alpha}\right)\\
\text{with }T^{d}_{\alpha,\rho}\equiv 3\,\tr{\matYukD{\alpha}\matYukDd{\rho}+\matYukUd{\alpha}\matYukU{\rho}}+\tr{\matYukL{\alpha}\matYukLd{\rho}},
\end{multlined}
\end{equation}
\begin{equation}\label{eq:RGE:Yu:00}
\begin{multlined}
\der\matYukU{\alpha}=
a_u\matYukU{\alpha}+\sum_{\rho=1}^{2}T^{u}_{\alpha,\rho}\matYukU{\rho}\\
+\sum_{\rho=1}^{2}\left(-2\matYukD{\rho}\matYukDd{\alpha}\matYukU{\rho}+\matYukU{\alpha}\matYukUd{\rho}\matYukU{\rho}+\frac{1}{2}\matYukD{\rho}\matYukDd{\rho}\matYukU{\alpha}+\frac{1}{2}\matYukU{\rho}\matYukUd{\rho}\matYukU{\alpha}\right)\\
\text{with }T^{u}_{\alpha,\rho}\equiv 3\,\tr{\matYukU{\alpha}\matYukUd{\rho}+\matYukDd{\alpha}\matYukD{\rho}}+\tr{\matYukLd{\alpha}\matYukL{\rho}}=T^{d\,\ast}_{\alpha,\rho}\,,
\end{multlined}
\end{equation}
\begin{equation}\label{eq:RGE:Yl:00}
\begin{multlined}
\der\matYukL{\alpha}=
a_\ell\matYukL{\alpha}+\sum_{\rho=1}^{2}T^{\ell}_{\alpha,\rho}\matYukL{\rho}+\sum_{\rho=1}^{2}\left(\matYukL{\alpha}\matYukLd{\rho}\matYukL{\rho}+\frac{1}{2}\matYukL{\rho}\matYukLd{\rho}\matYukL{\alpha}\right)\\
\text{with }T^{\ell}_{\alpha,\rho}\equiv T^{d}_{\alpha,\rho},
\end{multlined}
\end{equation}
where $\der\equiv 16\pi^2\frac{d}{d\ln\mu}$, $\mu$ is the renormalization scale and
\begin{equation}
a_d=-8g_c^2-\frac{9}{4}g^2-\frac{5}{12}g^{\prime 2},\qquad a_u=a_d-g^{\prime 2},\qquad a_\ell=-\frac{9}{4}g^2-\frac{15}{4}g^{\prime 2},
\label{eq:RGEa:00}
\end{equation}
with $g_c$, $g$, $g^\prime$ the corresponding gauge coupling constants of $SU(3)_c$, $SU(2)_L$ and $U(1)_Y$, respectively.

The alignment condition in the quark sector
\begin{equation}\label{eq:alignment:dquarks}
\matYukD{2} = d \matYukD{1},
\end{equation}
\begin{equation}\label{eq:alignment:uquarks}
\matYukU{2} = u \matYukU{1},
\end{equation}
together with the existence of two unitary matrices $W_{L,R}$ in the lepton sector such that
\begin{equation}\label{eq:gFC:leptons}
L_i \equiv W_L^\dagger \matYukL{i} W_R
\end{equation}
are diagonal, guarantee the absence of SFCNC at tree level. 

In order to ensure that \refeqs{eq:alignment:dquarks} and \eqref{eq:alignment:uquarks} in the quark sector hold at one loop, it is sufficient to impose \cite{Botella:2015yfa}
\begin{equation}
\der\matYukD{2}=\der(d)\matYukD{1}+d\der\matYukD{1},
\end{equation}
\begin{equation}
\der\matYukU{2}=\der(u)\matYukU{1}+u\der\matYukU{1},
\end{equation}
or equivalently
\begin{equation}
\der\matYukD{2}-d\der\matYukD{1}=\der(d)\matYukD{1} \propto \matYukD{1},
\end{equation}
\begin{equation}
\der\matYukU{2}-u\der\matYukU{1}=\der(u)\matYukU{1} \propto \matYukU{1},
\end{equation}
where the proportionality constants are precisely the running of the parameters $d$ and $u$ in \refeqs{eq:alignment:dquarks} and \eqref{eq:alignment:uquarks}. It is easy to check that
\begin{equation}
\begin{multlined}
\der\matYukD{2}-d\der\matYukD{1}\\
=\left \{ 3(u^* - d)(1 + ud) \tr{\matYukUd{1}\matYukU{1}} + \tr{(\matYukL{2} - d\matYukL{1})(\matYukLd{1}+ d\matYukLd{2})} \right \}\matYukD{1}\\ 
+ 2(d - u^*)(1 + ud) \matYukU{1}\matYukUd{1}\matYukD{1},
\end{multlined}
\end{equation}
\begin{equation}
\begin{multlined}
\der\matYukU{2}-u\der\matYukU{1}\\
=\left \{ 3(d^* - u)(1 + ud) \tr{\matYukDd{1}\matYukD{1}} + \tr{(\matYukLd{2} - u\matYukLd{1})(\matYukL{1} + u\matYukL{2})} \right \}\matYukU{1}\\ 
+ 2(u - d^*)(1 + ud) \matYukD{1}\matYukDd{1}\matYukU{1}.
\end{multlined}
\end{equation}
Then we should have
\begin{equation}
(d - u^*)(1 + ud) = 0,
\end{equation}
and, in particular, we are interested in the solution $d = u^*$. Therefore, the relation $\der(d) = \der(u^*)$ needs to be checked for the sake of consistency. Taking into account that, in our case,
\begin{equation}
\der(d) = \tr{(\matYukL{2} - d\matYukL{1})(\matYukLd{1} + d\matYukLd{2})},
\end{equation}
\begin{equation}
\der(u) = \tr{(\matYukLd{2} - u\matYukLd{1})(\matYukL{1} + u\matYukL{2})},
\end{equation}
it is clear that
\begin{equation}
\begin{aligned}
\der(u)^* &= \tr{(\matYukLd{1} + u^*\matYukLd{2})(\matYukL{2} - u^*\matYukL{1})}\\
&= \tr{(\matYukL{2} - u^*\matYukL{1})(\matYukLd{1} + u^*\matYukLd{2})}\\
&= \tr{(\matYukL{2} - d\matYukL{1})(\matYukLd{1} + d\matYukLd{2})} = \der(d),
\end{aligned}
\end{equation}
as it should. Hence, the quark sector is stable under RGE.

Concerning the lepton sector, one loop stability requires that
\begin{equation}
L_i + \der(L_i) \equiv W_L^\dagger (\matYukL{i} + \der(\matYukL{i})) W_R
\end{equation}
remain simultaneously diagonal. In this sense, the only apparently problematic term in $\der(\matYukL{i})$ has the structure $\matYukL{a}\matYukLd{b}\matYukL{c}$, but
\begin{equation}
W_L^\dagger (\matYukL{a}\matYukLd{b}\matYukL{c}) W_R = W_L^\dagger \matYukL{a} W_R W_R^\dagger \matYukLd{b} W_L W_L^\dagger \matYukL{c} W_R = L_a L_b^\dagger L_c,
\end{equation}
that is obviously diagonal \cite{Botella:2018gzy}. Therefore, the lepton sector is also stable under RGE.

\section{$\delta a_\ell$ loops}\label{appendix:dal:loops}
\subsection{One loop contributions}
The interaction Lagrangian of neutral scalars $S$ with charged leptons given by
\begin{equation}\label{eq:YukSLep:generic}
\mathscr{L}_{S\ell\ell}=-\frac{m_\ell}{\vev{}}S\bar\ell(A_\ell^S+iB_\ell^S\gamma_5)\ell\,
\end{equation}
generates the following one loop contribution to the anomalous magnetic moment of lepton $\ell$
\begin{equation}
\delta a_\ell^{(1)}=\frac{1}{8\pi^2}\left(\frac{m_\ell}{v}\right)^2\sum_{S}\left\{[A_\ell^S]^2[2I_1(x_{\ell S})-I_2(x_{\ell S})]-[B_\ell^S]^2I_2(x_{\ell S})\right\},
\end{equation}
where $x_{\ell S}\equiv {m_\ell^2}/{m_{S}^2}$ and
\begin{equation}
I_1(x)=1+\frac{1-2x}{2x\sqrt{1-4x}}\ln\left(\frac{1+\sqrt{1-4x}}{1-\sqrt{1-4x}}\right)+\frac{1}{2x}\ln x,
\end{equation}
\begin{equation}
I_2(x)=\frac{1}{2}+\frac{1}{x}+\frac{1-3x}{2x^2\sqrt{1-4x}}\ln\left(\frac{1+\sqrt{1-4x}}{1-\sqrt{1-4x}}\right)+\frac{1-x}{2x^2}\ln x.
\end{equation}
Taking into account that in the limit $x\ll 1$
\begin{equation}
I_1(x)\simeq x\left(-\frac{3}{2}-\ln x\right)+x^2\left(-\frac{16}{3}-4\ln x\right)+\mathcal O(x^3),
\end{equation}
\begin{equation}
I_2(x)\simeq x\left(-\frac{11}{6}-\ln x\right)+x^2\left(-\frac{89}{12}-5\ln x\right)+\mathcal O(x^3),
\end{equation}
one can write for $m_\ell\ll m_S$
\begin{equation}
\delta a_\ell^{(1)}=\frac{1}{8\pi^2}\left(\frac{m_\ell}{v}\right)^2\frac{m_\ell^2}{m_S^2}\left\{-[A_\ell^S]^2\left[\frac{7}{6}+\ln\left(\frac{m_\ell^2}{m_S^2}\right)\right]+[B_\ell^S]^2\left[\frac{11}{6}+\ln\left(\frac{m_\ell^2}{m_S^2}\right)\right]\right\}\,.
\end{equation}

On the other hand, the interaction Lagrangian of charged scalars $C^\pm$ with leptons written as
\begin{equation}
\mathscr{L}_{C\ell\nu}=-C^-\bar\ell(A_\ell^C+iB_\ell^C\gamma_5)\nu-C^+\bar\nu(A_\ell^{C\ast}+iB_\ell^{C\ast}\gamma_5)\ell\,,
\end{equation}
gives rise to one loop contributions to the anomalous magnetic moment of lepton $\ell$ of the form
\begin{equation}
\delta a_\ell^{(1)}=-\frac{1}{8\pi^2}\sum_{C}\left\{\abs{A_\ell^C}^2+\abs{B_\ell^C}^2\right\}\,I_3(x_{\ell C})\,,
\end{equation}
with $x_{\ell C}=m_\ell^2/m_{C^\pm}^2$ and
\begin{equation}
I_3(x)=-\frac{1}{2}+\frac{1}{x}+\frac{1-x}{x^2}\ln(1-x).
\end{equation}
For $x \ll 1$,
\begin{equation}
I_3(x)\simeq \frac{x}{6}+\frac{x^2}{12}+\mathcal O(x^3).
\end{equation}
\begin{figure}[h]
\subfloat[Neutral mediated one loop contribution.\label{sfig:feynman:1loop:S}]{
\includegraphics[width=0.3\textwidth]{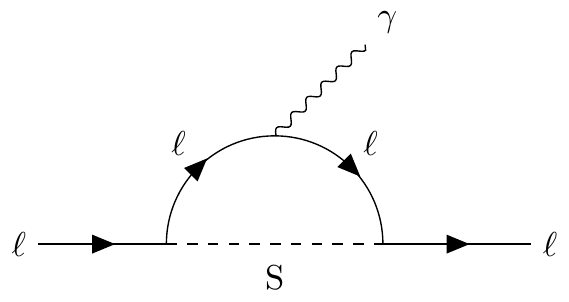}
}
\hfill
\subfloat[Charged mediated one loop contribution.\label{sfig:feynman:1loop:cH}]{
\includegraphics[width=0.3\textwidth]{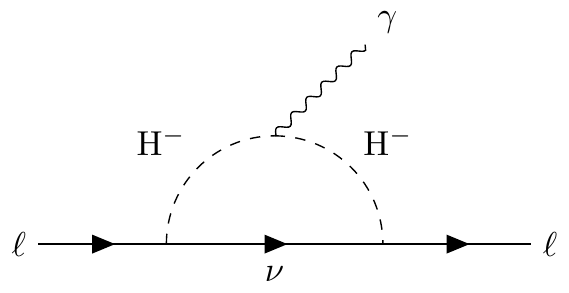}
}
\hfill
\subfloat[Barr-Zee two loop contribution.\label{sfig:feynman:2loop}]{
\includegraphics[width=0.3\textwidth]{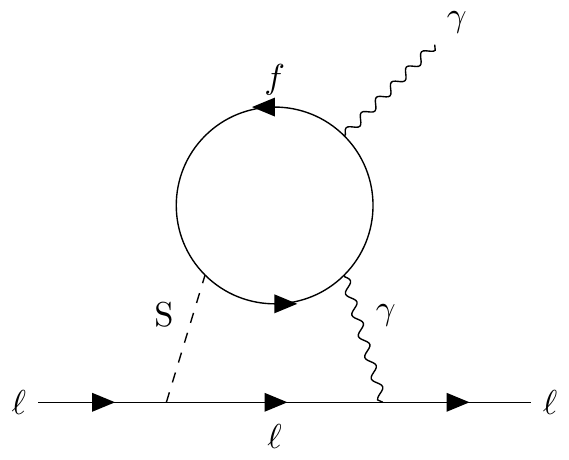}
}
\caption{Illustrative one and two loop contributions to $\delta a_\ell$.\label{fig:feynman}}
\end{figure}

\subsection{Two loop contributions}
Together with \refeq{eq:YukSLep:generic}, the interactions
\begin{equation}
\mathscr{L}_{S\bar ff}=-\frac{m_f}{\vev{}}S\bar f(\alpha_f^S+i\beta_f^S\gamma_5)f\,
\end{equation}
generate two loop Barr-Zee contributions to the anomalous magnetic moment of lepton $\ell$: 
\begin{equation}
\delta a_\ell^{(2)}=-\frac{2\alpha}{\pi}\frac{1}{8\pi^2}\left(\frac{m_\ell}{v}\right)^2\sum_{f}\sum_{S}N_c^fQ_f^2\left\{A_\ell^S\alpha_f^Sf(z_{fS})-B_\ell^S\beta_f^Sg(z_{fS})\right\}\,,
\end{equation}
where $N_c^f$ and $Q_f$ are the number of colours and the electric charge of the fermion running in the closed loop of figure \ref{sfig:feynman:2loop}, respectively, and $z_{fS}=m_f^2/m_S^2$. The two loop functions $f(z)$ and $g(z)$ are
\begin{equation}
f(z)=\frac{z}{2}\int_0^1 dx\,\frac{1-2x(1-x)}{x(1-x)-z}\,\ln\left(\frac{x(1-x)}{z}\right)\,,
\end{equation}
\begin{equation}
g(z)=\frac{z}{2}\int_0^1 dx\,\frac{1}{x(1-x)-z}\,\ln\left(\frac{x(1-x)}{z}\right)\,.
\end{equation}
We refer to \cite{Cherchiglia:2016eui} to see other two loop contributions.
\end{appendices}

}

\bibliographystyle{JHEP}
\bibliography{refs.bib}
\end{document}